# Corrections to Off-Axis $\Delta v$ Measurements from Event Data Recorders

Bob Scurlock, Ph.D., ACTAR, Andrew Rich, BSME, ACTAR, and Kyle Poe

**Introduction**

In this article, we derive a mathematical transformation which corrects $\Delta v$ measurements from event data recorders at arbitrary positions in a vehicle to the equivalent values at the center-of-gravity. The method is illustrated using staged collision data. We also demonstrate the method's consistency with simulation.

**Use of EDR Data**

It has become increasingly common for event data recorders (EDRs) to play a central role in accident reconstruction analyses. Both pre-crash speed data as well as acceleration and change-in-velocity (CG) data can provide extremely valuable constraints for the analyst's calculations and corresponding opinions. Though it is common for event data recorders to be located very near a vehicle's CG, this is not always the case. The analyst must be aware of how an EDR's distance from a vehicle's CG can cause inaccuracies to be introduced into an analysis if not properly corrected for [1,2]. Even when near or at the center-of-gravity, it is important for the analyst to be aware of how EDR-based results may be affected by issues such as large rotational velocities. Below, we develop a mathematical transformation to correct for EDR displacement from the CG.

**Mathematical Development of Transformation Equations**

We begin with a rigorous derivation of the equations needed for our inverse transformation from EDR measured $\Delta\bar{v}$ to equivalent $\Delta\bar{v}$ at the center-of-gravity based on classical mechanics. For a thorough review of classical mechanics, we refer the reader to reference [3].

*Position of Points in a Moving Reference Frame*

The position of an arbitrary point, $P$, can be specified with respect to an inertial frame (Earth frame), $O$, as the vector sum of $P$'s position with respect to a moving reference frame, $O'$, and the position of the moving reference frame's origin with respect to the inertial frame. That is,

$$\bar{R}^P = \bar{R}^{O'} + \bar{r}^P \tag{1}$$

where $\bar{R}^{O'}$ is the position vector of the moving reference frame's origin with respect to the inertial frame, $\bar{r}^P$ is the position of point $P$ with respect to the moving reference frame, and $\bar{R}^P$ is the position of point $P$ with respect to the inertial frame (Figure 1).

*Velocity of Points in a Moving Reference Frame*

Taking the time derivative of (1), we can calculate the linear velocity of point $P$ in the inertial frame. This is given by:

$$\dot{\bar{R}}^P = \bar{v}^{P,O} = \bar{V}^{O'} + \dot{\bar{r}}^P \tag{2}$$

where $\bar{v}^{P,O}$ indicates the linear velocity of point $P$ evaluated in the inertial reference frame $O$, and the linear velocity of frame $O'$ with respect to frame $O$ is given by $\bar{V}^{O'} = \dot{\bar{R}}^{O'}$.

Suppose we know the position of point $P$ with respect to frame $O'$, given by $\bar{r}^P$. That is, we can write:

$$\bar{r}^P = r^P_{x'}\hat{x}' + r^P_{y'}\hat{y}' + r^P_{z'}\hat{z}' \tag{3}$$

where $\hat{x}', \hat{y}',$ and $\hat{z}'$ are the ortho-normal basis vectors for the moving frame $O'$, whose orientations can change with time with respect to frame $O$, and $r^P_{x'}$, $r^P_{y'}$, and $r^P_{z'}$ are the time-dependent components along those basis vectors.

The components of $\bar{r}^P$ in frame $O'$ can be related to the components in frame $O$ by a 3×3 rotation matrix **M**:

$$\begin{pmatrix} r^P_x \\ r^P_y \\ r^P_z \end{pmatrix} = \mathbf{M} \cdot \begin{pmatrix} r^P_{x'} \\ r^P_{y'} \\ r^P_{z'} \end{pmatrix}$$

where $\bar{r}^P$ can be expressed with respect to frame $O$ as:

$$\bar{r}^P = r^P_x\hat{x} + r^P_y\hat{y} + r^P_z\hat{z}$$

Here $\hat{x}, \hat{y},$ and $\hat{z}$ are the ortho-normal basis vectors for the inertial frame $O$, whose orientations we take as fixed (time-independent), and $r^P_x$, $r^P_y$, and $r^P_z$ are the time-dependent components along those basis vectors. The rotation matrix is given by the direction cosines:

$$\mathbf{M} = \begin{pmatrix} \hat{x}'\cdot\hat{x} & \hat{y}'\cdot\hat{x} & \hat{z}'\cdot\hat{x} \\ \hat{x}'\cdot\hat{y} & \hat{y}'\cdot\hat{y} & \hat{z}'\cdot\hat{y} \\ \hat{x}'\cdot\hat{z} & \hat{y}'\cdot\hat{z} & \hat{z}'\cdot\hat{z} \end{pmatrix}$$

Taking the derivative of both sides of (3) with respect to time, and applying the product rule, we have:

$$\dot{\bar{r}}^P = \left(\dot{r}^P_{x'}\hat{x}' + \dot{r}^P_{y'}\hat{y}' + \dot{r}^P_{z'}\hat{z}'\right) + \left(r^P_{x'}\dot{\hat{x}}' + r^P_{y'}\dot{\hat{y}}' + r^P_{z'}\dot{\hat{z}}'\right) \tag{4}$$

where the term in the first set of parentheses represents the contribution to the time rate of change of vector $\bar{r}^P$ due to time-dependent components evaluated with respect to frame $O'$, and the second term represents the contribution due to the frame $O'$ time-dependent orientation with respect to frame $O$.

For an infinitesimal rotation in frame O about an arbitrary axis $\delta\bar{\Theta}$, a vector of fixed magnitude, $\bar{\lambda}$, is transformed to:

$$\bar{\lambda} \to \bar{\lambda} + \delta\bar{\lambda} \tag{5}$$

where,

$$\delta\bar{\lambda} = \delta\bar{\Theta} \times \bar{\lambda} \tag{6}$$

This implies:

$$\frac{\delta\bar{\lambda}}{\delta t} = \frac{\delta\bar{\Theta}}{\delta t} \times \bar{\lambda} \tag{7}$$

or in the limit, $\delta t \to 0$,

$$\dot{\bar{\lambda}} = \bar{\omega} \times \bar{\lambda} \tag{8}$$

With (8), we can now evaluate the time derivatives of the frame $O'$ basis vectors:

$$\begin{aligned} \dot{\hat{x}}' &= \bar{\omega} \times \hat{x}' \\ \dot{\hat{y}}' &= \bar{\omega} \times \hat{y}' \\ \dot{\hat{z}}' &= \bar{\omega} \times \hat{z}' \end{aligned} \tag{9}$$

where $\bar{\omega}$ is the instantaneous angular velocity vector of the reference frame $O'$ as measured in frame $O$.

Thus,

$$r^P_{x'}\dot{\hat{x}}' + r^P_{y'}\dot{\hat{y}}' + r^P_{z'}\dot{\hat{z}}' = \bar{\omega} \times \left(r^P_{x'}\hat{x}' + r^P_{y'}\hat{y}' + r^P_{z'}\hat{z}'\right) \tag{10}$$

We can now rewrite (4) as:

$$\dot{\bar{r}}^P = \bar{v}^{P,O'} + \bar{\omega} \times \bar{r}^P \tag{11}$$

Here $\bar{v}^{P,O'}$ is the linear velocity of point $P$ whose components are evaluated in the moving reference frame, $O'$:

$$\bar{v}^{P,O'} = \dot{r}^P_{x'}\hat{x}' + \dot{r}^P_{y'}\hat{y}' + \dot{r}^P_{z'}\hat{z}' \tag{12}$$

Finally, combining (2) and (11), we have our final expression for the linear velocity vector of point $P$ with respect to the inertial frame $O$:

$$\bar{v}^{P,O} = \bar{V}^{O'} + \bar{v}^{P,O'} + \bar{\omega} \times \bar{r}^P \tag{13}$$

*Acceleration of Points in a Moving Reference Frame*

Taking the time derivative of (13), we can now find an expression for the acceleration of point *P*. This is given by:

$$\bar{a}^{P,O} = \dot{\bar{v}}^{P,O} = \dot{\bar{V}}^{O'} + \dot{\bar{v}}^{P,O'} + \frac{d}{dt}(\bar{\omega} \times \bar{r}^P) \quad (14)$$

The first term is simply the linear acceleration of the moving reference frame $O'$ with respect to the inertial frame $O$:

$$\dot{\bar{V}}^{O'} = \bar{a}^{O'} \quad (15)$$

For the second term, we have:

$$\begin{aligned}\dot{\bar{v}}^{P,O'} &= \frac{d}{dt}\left(\dot{r}^P_{x'}\hat{x}' + \dot{r}^P_{y'}\hat{y}' + \dot{r}^P_{z'}\hat{z}'\right) \\ &= \ddot{r}^P_{x'}\hat{x}' + \ddot{r}^P_{y'}\hat{y}' + \ddot{r}^P_{z'}\hat{z}' \\ &\quad + \dot{r}^P_{x'}\dot{\hat{x}}' + \dot{r}^P_{y'}\dot{\hat{y}}' + \dot{r}^P_{z'}\dot{\hat{z}}'\end{aligned} \quad (16)$$

where $\ddot{r}^P_{x'}\hat{x}' + \ddot{r}^P_{y'}\hat{y}' + \ddot{r}^P_{z'}\hat{z}'$ is the linear acceleration of point *P* in frame $O'$, which we denote $\bar{a}^{P,O'}$. Using (9), the second term in (16) becomes

$$\begin{aligned}&\dot{r}^P_{x'}\dot{\hat{x}}' + \dot{r}^P_{y'}\dot{\hat{y}}' + \dot{r}^P_{z'}\dot{\hat{z}}' = \\ &\bar{\omega} \times \left(\dot{r}^P_{x'}\hat{x}' + \dot{r}^P_{y'}\hat{y}' + \dot{r}^P_{z'}\hat{z}'\right) \\ &= \bar{\omega} \times \bar{v}^{P,O'}\end{aligned} \quad (17)$$

We can thus rewrite (16) as:

$$\dot{\bar{v}}^{P,O'} = \bar{a}^{P,O'} + \bar{\omega} \times \bar{v}^{P,O'} \quad (18)$$

The final term in (14) is:

$$\frac{d}{dt}(\bar{\omega} \times \bar{r}^P) = \dot{\bar{\omega}} \times \bar{r}^P + \bar{\omega} \times \dot{\bar{r}}^P \quad (19)$$

where $\dot{\bar{\omega}}$ is the angular acceleration of reference $O'$ frame as measured in $O$.

Using (11), we can rewrite the second term in (19) as:

$$\begin{aligned}\bar{\omega} \times \dot{\bar{r}}^P &= \bar{\omega} \times \left(\bar{v}^{P,O'} + \bar{\omega} \times \bar{r}^P\right) \\ &= \bar{\omega} \times \bar{v}^{P,O'} + \bar{\omega} \times (\bar{\omega} \times \bar{r}^P)\end{aligned} \quad (20)$$

Putting this together, we can rewrite (14) as[1]:

$$\begin{aligned}\bar{a}^{P,O} &= \bar{a}^{O'} + \bar{a}^{P,O'} \\ &\quad + \dot{\bar{\omega}} \times \bar{r}^P + \bar{\omega} \times (\bar{\omega} \times \bar{r}^P) \\ &\quad + 2\bar{\omega} \times \bar{v}^{P,O'}\end{aligned} \quad (21)$$

*Acceleration of Accelerometers in a Moving Reference Frame*

Let us now suppose we have two vehicles undergoing a collision. We can assign to each vehicle its own moving reference frame $O'_k$, where $k$ is an index used to label the vehicle, and each vehicle frame origin is placed at the CG. Let us also suppose, at a given point *P*, vehicle $k$ has an accelerometer. Let's now denote the accelerometer's position with the superscript *A*. Suppose that the accelerometer is sufficiently far away from the volume of crush damage, that we can regard its position as fixed and stationary with respect to the vehicle's reference frame. That is, we have:

$$\begin{aligned}\bar{a}^{A,O'} &= 0 \\ \bar{v}^{A,O'} &= 0\end{aligned} \quad (22)$$

Using (21), we can now write an expression for the expected linear acceleration at the accelerometer position as a function of time, in the inertial frame $O$:

$$\begin{aligned}\bar{a}^A_k(t) &= \bar{a}^{CG}_k(t) + \dot{\bar{\omega}}_k(t) \times \bar{r}^A_k \\ &\quad + \bar{\omega}_k(t) \times (\bar{\omega}_k(t) \times \bar{r}^A_k)\end{aligned} \quad (23)$$

We can re-express the vectors $\bar{r}^A_k$ and $\bar{\omega}_k$ in cylindrical coordinates by:

$$\bar{r}^A_k = |\bar{r}^A_k| \cdot \hat{r}^A_k(t) \quad (24)$$

and

$$\bar{\omega}_k(t) = |\bar{\omega}_k(t)| \cdot \hat{\omega}(t) \quad (25)$$

where we define an instantaneous right-handed cylindrical coordinate system whose axes are centered at the accelerometer, where the unit vector $\hat{r}^A_k(t)$ points from the CG to accelerometer position, $\hat{\omega}(t)$ defines the axis of rotation, and $\hat{\theta}^A_k(t)$ points in the direction of rotation (Figure 2).

Next, let's redefine $\bar{r}^A_k$ in terms of its components parallel and perpendicular to the unit vector $\hat{\omega}(t)$ (Figure 2). That is,

$$\bar{r}^A_k = \left|\bar{r}^A_{k,\perp}\right| \cdot \hat{r}^A_{k,\perp}(t) + \left|\bar{r}^A_{k,\parallel}\right| \cdot \hat{\omega}(t) \quad (26)$$

This of course implies:

$$\hat{\omega}(t) \times \hat{r}^A_k(t) = \hat{\omega}(t) \times \hat{r}^A_{k,\perp}(t) \quad (27)$$

Let's now examine the special case where (1) rotation occurs only about the inertial frame's $\hat{z}$ axis and (2) $\bar{r}^A_k$ lies on the $\hat{x}' - \hat{y}'$ plane, thereby reducing our model to two dimensions (Figure 3). Note, (1) ensures $\dot{\bar{\omega}}_k(t)$ and $\bar{\omega}_k(t)$ are both aligned with the $\hat{z}$ axis and (2) ensures $\bar{r}^A_{k,\parallel} = 0$.[2] With these simplifying assumptions, we have:

$$\bar{\omega}_k(t) = |\bar{\omega}_k(t)| \cdot \hat{z} \quad (28)$$

and the angular acceleration becomes:

$$\dot{\bar{\omega}}_k(t) = |\dot{\bar{\omega}}_k(t)| \cdot \hat{z} \quad (29)$$

Our cross products are thus given by:

$$\begin{aligned}\hat{\omega} \times \hat{r}^A_k(t) &= \hat{z} \times \hat{r}^A_k(t) = \hat{\theta}^A_k(t) \\ \hat{\theta}^A_k(t) \times \hat{\omega} &= \hat{\theta}^A_k(t) \times \hat{z} = \hat{r}^A_k(t) \\ \hat{r}^A_k(t) \times \hat{\theta}^A_k(t) &= \hat{\omega} = \hat{z}\end{aligned} \quad (30)$$

We can now re-express the first cross-product in (23):

$$\begin{aligned}&\dot{\bar{\omega}}_k(t) \times \bar{r}^A_k(t) \\ &= |\dot{\bar{\omega}}_k(t)| \cdot |\bar{r}^A_k| \cdot \left(\hat{z} \times \hat{r}^A_k(t)\right) \\ &= |\dot{\bar{\omega}}_k(t)| \cdot |\bar{r}^A_k| \cdot \hat{\theta}^A_k(t)\end{aligned} \quad (31)$$

We can re-express $(\bar{\omega}_k(t) \times \bar{r}^A_k)$ by:

$$\begin{aligned}&\bar{\omega}_k(t) \times \bar{r}^A_k \\ &= |\bar{\omega}_k(t)| \cdot |\bar{r}^A_k| \cdot \left(\hat{z} \times \hat{r}^A_k(t)\right) \\ &= |\bar{\omega}_k(t)| \cdot |\bar{r}^A_k| \cdot \hat{\theta}^A_k(t)\end{aligned} \quad (32)$$

The second cross-product in (23) is therefore:

$$\begin{aligned}&\bar{\omega}_k(t) \times (\bar{\omega}_k(t) \times \bar{r}^A_k) \\ &= |\bar{\omega}_k(t)|^2 \cdot |\bar{r}^A_k| \cdot \left(\hat{z} \times \hat{\theta}^A_k(t)\right) \\ &= -|\bar{\omega}_k(t)|^2 \cdot |\bar{r}^A_k| \cdot \hat{r}^A_k(t)\end{aligned} \quad (33)$$

We can now write (23) as:

$$\begin{aligned}\bar{a}^A_k(t) &= \bar{a}^{CG}_k(t) \\ &\quad + |\dot{\bar{\omega}}_k(t)| \cdot |\bar{r}^A_k| \cdot \hat{\theta}^A_k(t) \\ &\quad - |\bar{\omega}_k(t)|^2 \cdot |\bar{r}^A_k| \cdot \hat{r}^A_k(t)\end{aligned} \quad (34)$$

With this, we can now relate the acceleration at the CG to the measured acceleration at the accelerometer position.

*$\Delta v$ in Continuous Time*

Let us now take the dot-product of (34) with an arbitrary unit-vector, $\hat{c}$, which we define as time-independent in the inertial frame, and calculate the time integral of both sides from the start of the crash pulse at time = 0 to the end of the crash pulse at time = $\Delta t$:

$$\begin{aligned}&\int_0^{\Delta t} dt \cdot \bar{a}^A_k(t) \cdot \hat{c} \\ &= \int_0^{\Delta t} dt \cdot \bar{a}^{CG}_k(t) \cdot \hat{c} \\ &+ \int_0^{\Delta t} dt \cdot |\dot{\bar{\omega}}_k(t)| \cdot |\bar{r}^A_k| \cdot \hat{\theta}^A_k(t) \cdot \hat{c} \\ &- \int_0^{\Delta t} dt \cdot |\bar{\omega}_k(t)|^2 \cdot |\bar{r}^A_k| \cdot \hat{r}^A_k(t) \cdot \hat{c}\end{aligned} \quad (35)$$

Note, the projections of $\bar{a}^A_k(t)$, $\bar{a}^{CG}_k(t)$, $\hat{\theta}^A_k(t)$, and $\hat{r}^A_k(t)$ along the $\hat{c}$ axis are all changing as a function of time, and therefore the dot-products cannot simply be factored outside of the time-integrals. Let's now simplify (35).

We first want to evaluate the second integral using integration by parts:

$$\int_0^{\Delta t} dt \cdot |\dot{\bar{\omega}}_k(t)| \cdot |\bar{r}^A_k| \cdot \hat{\theta}^A_k(t) \cdot \hat{c} \quad (36)$$

Let us define the function $u(t)$ by:

$$u(t) = |\bar{r}^A_k| \cdot \hat{\theta}^A_k(t) \cdot \hat{c} \quad (37)$$

We can thus express $du$ as:

$$du = dt \cdot |\bar{r}^A_k| \cdot \dot{\hat{\theta}}^A_k(t) \cdot \hat{c} \quad (38)$$

[1] For a thorough derivation of these equations, see chapter 10 of reference [3].

[2] Note, typically this vector points from the vehicle CG to accelerometer (or EDR), which in general can have a non-zero z-component. To use the formalism presented in the rest of this work, simply ignore the z component, using only the projection on the vehicle's local *x-y* plane to define this vector.

Using (8), we know:

$$\dot{\hat{\theta}}_k^A(t) = \bar{\omega}(t) \times \hat{\theta}_k^A(t) \\ = |\bar{\omega}_k(t)| \cdot \left(\hat{\omega} \times \hat{\theta}_k^A(t)\right) \tag{39}$$

Using (30), (39) becomes:

$$\dot{\hat{\theta}}_k^A(t) = \bar{\omega}(t) \times \hat{\theta}_k^A(t) \\ = -|\bar{\omega}_k(t)| \cdot \hat{r}_k^A(t) \tag{40}$$

With this, we can rewrite (38):

$$du = -dt \cdot |\bar{r}_k^A| \cdot |\bar{\omega}_k(t)| \cdot \hat{r}_k^A(t) \cdot \hat{c} \tag{41}$$

Let us now define the function

$$g(t) = |\bar{\omega}_k(t)| \tag{42}$$

and its differential:

$$dg = dt \cdot |\dot{\bar{\omega}}_k(t)| \tag{43}$$

Using integration by parts, we have:

$$\int_0^{\Delta t} u(t) dg \\ = \int_0^{\Delta t} d(u(t) \cdot g(t)) - \int_0^{\Delta t} g(t) du \tag{44}$$

Using the above, we can rewrite (36) as:

$$\int_0^{\Delta t} dt \cdot |\dot{\bar{\omega}}_k(t)| \cdot |\bar{r}_k^A| \cdot \hat{\theta}_k^A(t) \cdot \hat{c} \\ = \int_0^{\Delta t} d\left(|\bar{r}_k^A| \cdot |\bar{\omega}_k(t)| \cdot \hat{\theta}_k^A(t) \cdot \hat{c}\right) \\ + \int_0^{\Delta t} dt \cdot |\bar{r}_k^A| \cdot |\bar{\omega}_k(t)|^2 \cdot \hat{r}_k^A(t) \cdot \hat{c} \tag{45}$$

Here we see the last term in (45) is equal and opposite to the last term in (35). (35) therefore simplifies to:

$$\int_0^{\Delta t} dt \cdot \bar{a}_k^A(t) \cdot \hat{c} \\ = \int_0^{\Delta t} dt \cdot \bar{a}_k^{CG}(t) \cdot \hat{c} \\ + \int_0^{\Delta t} d\left(|\bar{r}_k^A| \cdot |\bar{\omega}_k(t)| \cdot \hat{\theta}_k^A(t) \cdot \hat{c}\right) \tag{46}$$

or

$$\int_0^{\Delta t} dt \cdot \bar{a}_k^A(t) \cdot \hat{c} \\ = \int_0^{\Delta t} dt \cdot \bar{a}_k^{CG}(t) \cdot \hat{c} \\ + \left(|\bar{\omega}_k(\Delta t)| \cdot \hat{\theta}_k^A(\Delta t) \cdot \hat{c} \cdot |\bar{r}_k^A| \right. \\ \left. - |\bar{\omega}_k(0)| \cdot \hat{\theta}_k^A(0) \cdot \hat{c} \cdot |\bar{r}_k^A|\right) \tag{47}$$

Using (32), (47) can also be written in the equivalent form:

$$\int_0^{\Delta t} dt \cdot \bar{a}_k^A(t) \cdot \hat{c} = \int_0^{\Delta t} dt \cdot \bar{a}_k^{CG}(t) \cdot \hat{c} \\ + \left(\bar{\omega}_k(\Delta t) \times \bar{r}_k^A(\Delta t) \right. \\ \left. - \bar{\omega}_k(0) \times \bar{r}_k^A(0)\right) \cdot \hat{c} \tag{48}$$

With (48), we can now write our equations to estimate $\Delta\bar{v}_k^{CG}$ components in the inertial frame. These are given by:

$$\Delta\bar{v}_{k,x}^{CG} = \int_0^{\Delta t} dt \cdot \bar{a}_k^{CG}(t) \cdot \hat{x} \\ = \int_0^{\Delta t} dt \cdot \bar{a}_k^A(t) \cdot \hat{x} \\ -\left(\bar{\omega}_{k,f} \times \bar{r}_{k,f}^A - \bar{\omega}_{k,i} \times \bar{r}_{k,i}^A\right) \cdot \hat{x} \tag{49}$$

and

$$\Delta\bar{v}_{k,y}^{CG} = \int_0^{\Delta t} dt \cdot \bar{a}_k^{CG}(t) \cdot \hat{y} \\ = \int_0^{\Delta t} dt \cdot \bar{a}_k^A(t) \cdot \hat{y} \\ -\left(\bar{\omega}_{k,f} \times \bar{r}_{k,f}^A - \bar{\omega}_{k,i} \times \bar{r}_{k,i}^A\right) \cdot \hat{y} \tag{50}$$

where the $i$ and $f$ subscripts denote initial and final values. Note, the dot products in the time integrals $\int_0^{\Delta t} dt \cdot \bar{a}_k^A(t) \cdot \hat{x}$ and $\int_0^{\Delta t} dt \cdot \bar{a}_k^A(t) \cdot \hat{y}$ must be evaluated time-step by time-step as the acceleration vector at the accelerometer position, $\bar{a}_k^A(t)$, will likely rotate as the collision unfolds. Also, recall $\bar{a}_k^A(t)$ is the acceleration measured at the accelerometer position in the vehicle. $\bar{a}_k^A(t)$ itself is, of course, the acceleration at the accelerometer position, evaluated in the inertial frame. This means to properly evaluate (49) and (50), one must first transform the acceleration vector components, typically given in the moving vehicle frame of reference, to the Earth-fixed inertial frame. This transformation typically requires vehicle yaw angle versus time data which defines the vehicle frame $\hat{x}'(t)$ and $\hat{y}'(t)$ behavior with respect to the inertial frame.

*$\Delta\boldsymbol{v}$ in the Instantaneous Limit*

Let us now approximate the collision as occurring instantly in time, thus in the limit $\Delta t \to 0$. With this, our integrals become:

$$\int_0^{\Delta t} dt \cdot \bar{a}_k^{CG}(t) \cdot \hat{c} = \Delta\bar{v}_k^{CG} \cdot \hat{c} \tag{51}$$

and

$$\int_0^{\Delta t} dt \cdot \bar{a}_k^A(t) \cdot \hat{c} = \Delta\bar{v}_k^A \cdot \hat{c} \tag{52}$$

Our cross-products become:

$$\left(\bar{\omega}_{k,f} \times \bar{r}_{k,f}^A - \bar{\omega}_{k,i} \times \bar{r}_{k,i}^A\right) \cdot \hat{c} \\ = (\Delta\bar{\omega}_k \times \bar{r}_k^A) \cdot \hat{c} \tag{53}$$

We can therefore rewrite (52) as:

$$\int_0^{\Delta t} dt \cdot \bar{a}_k^A(t) \cdot \hat{c} \to \Delta\bar{v}_k^A \cdot \hat{c} \\ = \Delta\bar{v}_k^{CG} \cdot \hat{c} + (\Delta\bar{\omega}_k \times \bar{r}_k^A) \cdot \hat{c} \tag{54}$$

Because there is no rotation in the instantaneous limit, we can now factor out our dot-product, and simplify (52) by:

$$\Delta\bar{v}_k^A = \Delta\bar{v}_k^{CG} + (\Delta\bar{\omega}_k \times \bar{r}_k^A) \tag{55}$$

Note, because we are assuming $\Delta t \to 0$ (the instantaneous collision condition), this will result in some inaccuracies because this approximation assumes there is no component rotation in $\bar{a}_k^A$ during vehicle contact. For collision events in which there are large rotations as forces are being exchanged, this assumption may lead to large inaccuracies.

Equation (55) gives us a convenient way to transform between $\Delta\bar{v}_k^A$ and $\Delta\bar{v}_k^{CG}$; however, this transformation requires knowledge of $\Delta\bar{\omega}_k$. One can model expected $\Delta\bar{\omega}_k$ with the aid of computer simulation, but this isn't necessary. Below we complete the development of our mathematical transformation between $\Delta\bar{v}_k^A$ and $\Delta\bar{v}_k^{CG}$ that is independent of $\Delta\bar{\omega}_k$ through the use of a closed-form impulse-based collision model.

*Impulse*

Let us define the total impulse imparted to vehicle *k* by:

$$\bar{J}_k = \int_0^{\Delta t} dt \cdot \bar{F}_k \tag{56}$$

where $\bar{F}_k$ is the total force versus time acting on vehicle *k* during the duration $\Delta t$. Using Newton's 2nd Law, we can rewrite (56) as:

$$\bar{J}_k = \int_0^{\Delta t} dt \cdot \left[\frac{d}{dt}(m_k \bar{v}_k)\right] \\ = m_k \cdot \Delta\bar{v}_k = \Delta\bar{p}_k \tag{57}$$

*Torque*

We can write an expression for the total torque on vehicle *k* caused by the application of force $\bar{F}_k$:

$$\bar{\Gamma}_k = I_k \bar{\alpha}_k = \bar{r}_k \times \bar{F}_k \tag{58}$$

where $\bar{\alpha}_k = d\bar{\omega}_k/dt$ is the angular acceleration about the center-of-gravity of object *k*, and $\bar{r}_k$ is the lever-arm extending from the CG to the point of contact, and $I_k$ is the moment-of-inertia for rotation about the $\hat{\Gamma}_k$ axis.

Taking the time integral of the total torque over interaction duration $\Delta t$, we have:

$$\int_0^{\Delta t} dt \cdot \bar{\Gamma}_k = \int_0^{\Delta t} dt \cdot (I_k \bar{\alpha}_k) \\ = \int_0^{\Delta t} dt \cdot \left[I_k \frac{d\bar{\omega}_k}{dt}\right] \\ = I_k \Delta\bar{\omega}_k = \Delta\bar{L}_k \tag{59}$$

Therefore, the torque delivered over time $\Delta t$ is associated with a change in angular momentum $\Delta L_k$, where the angular momentum is given by $\bar{L}_k = I_k \bar{\omega}_k$.

Therefore using (58) and (59), we have:

$$\Delta\bar{L}_k = I_k \Delta\bar{\omega}_k = \int_0^{\Delta t} dt \cdot (\bar{r}_k \times \bar{F}_k) \\ = \bar{r}_k \times \int_0^{\Delta t} dt \cdot \bar{F}_k \\ = \bar{r}_k \times \bar{J}_k = m_k \cdot (\bar{r}_k \times \Delta\bar{v}_k) \tag{60}$$

*Change in Angular Velocity*

Using (60), we can now write an expression for $\Delta\bar{\omega}_k$ in terms of $\Delta\bar{v}_k$:

$$\Delta\bar{\omega}_k = \frac{\bar{r}_k \times \bar{J}_k}{I_k} = \frac{(\bar{r}_k \times \Delta\bar{v}_k)}{k_k^2} \tag{61}$$

where we express the yaw moment of inertia in terms of vehicle *k*'s radius of gyration $I_k = m_k k_k^2$.

*Solving for $\Delta\boldsymbol{v}$ at the CG*

Let's write our accelerometer position in Earth-fixed inertial frame coordinates:

$$\bar{r}_k^A = r_{k,x}^A \hat{x} + r_{k,y}^A \hat{y} \tag{62}$$

We can thus write our cross-product by:

$$\Delta\bar{\omega}_k \times \bar{r}_k^A = \begin{pmatrix} \hat{x} & \hat{y} & \hat{z} \\ 0 & 0 & \Delta\omega_k \\ r_{k,x}^A & r_{k,y}^A & 0 \end{pmatrix}$$

$$= \hat{x}\left(-r_{k,y}^A \cdot \Delta\omega_k\right) - \hat{y}\left(-r_{k,x}^A \cdot \Delta\omega_k\right) = \Delta\omega_k \cdot \left(-r_{k,y}^A \hat{x} + r_{k,x}^A \hat{y}\right) \tag{63}$$

With this, (55) becomes:

$$\begin{pmatrix} \Delta v_{k,x}^A \\ \Delta v_{k,y}^A \end{pmatrix} = \begin{pmatrix} \Delta v_{k,x}^{CG} - \Delta\omega_k \cdot r_{k,y}^A \\ \Delta v_{k,y}^{CG} + \Delta\omega_k \cdot r_{k,x}^A \end{pmatrix} \tag{64}$$

Let's now express the impulse centroid position in the Earth-fixed inertial frame:

$$\bar{r}_k = r_{k,x}\hat{x} + r_{k,y}\hat{y} \tag{65}$$

With this, we can evaluate the cross-product in (61):

$$\bar{r}_k \times \Delta\bar{v}_k^{CG} = \begin{pmatrix} \hat{x} & \hat{y} & \hat{z} \\ r_{k,x} & r_{k,y} & 0 \\ \Delta v_{k,x}^{CG} & \Delta v_{k,y}^{CG} & 0 \end{pmatrix}$$

$$= \hat{z}\left(r_{k,x} \cdot \Delta v_{k,y}^{CG} - r_{k,y} \cdot \Delta v_{k,x}^{CG}\right) \tag{66}$$

We can now rewrite (61) by:

$$\Delta\bar{\omega}_k = \frac{\left(r_{k,x} \cdot \Delta v_{k,y}^{CG} - r_{k,y} \cdot \Delta v_{k,x}^{CG}\right)}{k_k^2}\hat{z} \tag{67}$$

Using (67), (64) becomes:

$$\begin{pmatrix} \Delta v_{k,x}^A \\ \Delta v_{k,y}^A \end{pmatrix} = \begin{pmatrix} \Delta v_{k,x}^{CG} - \frac{\left(r_{k,x} \cdot \Delta v_{k,y}^{CG} - r_{k,y} \cdot \Delta v_{k,x}^{CG}\right)}{k_k^2} \cdot r_{k,y}^A \\ \Delta v_{k,y}^{CG} + \frac{\left(r_{k,x} \cdot \Delta v_{k,y}^{CG} - r_{k,y} \cdot \Delta v_{k,x}^{CG}\right)}{k_k^2} \cdot r_{k,x}^A \end{pmatrix} \tag{68}$$

or

$$\begin{pmatrix} \Delta v_{k,x}^A \\ \Delta v_{k,y}^A \end{pmatrix} = \begin{pmatrix} \Delta v_{k,x}^{CG} \cdot \left(1 + \frac{r_{k,y} \cdot r_{k,y}^A}{k_k^2}\right) + \Delta v_{k,y}^{CG} \cdot \left(\frac{-r_{k,x} \cdot r_{k,y}^A}{k_k^2}\right) \\ \Delta v_{k,x}^{CG} \cdot \left(\frac{-r_{k,y} \cdot r_{k,x}^A}{k_k^2}\right) + \Delta v_{k,y}^{CG} \cdot \left(1 + \frac{r_{k,x} \cdot r_{k,x}^A}{k_k^2}\right) \end{pmatrix} \tag{69}$$

(69) can be re-expressed as an equation that takes vector $\Delta\bar{v}_k^{CG}$ and rotates it to obtain $\Delta\bar{v}_k^A$:

$$\begin{pmatrix} \Delta v_{k,x}^A \\ \Delta v_{k,y}^A \end{pmatrix} = \begin{pmatrix} \left(1 + \frac{r_{k,y} \cdot r_{k,y}^A}{k_k^2}\right) & \left(\frac{-r_{k,x} \cdot r_{k,y}^A}{k_k^2}\right) \\ \left(\frac{-r_{k,y} \cdot r_{k,x}^A}{k_k^2}\right) & \left(1 + \frac{r_{k,x} \cdot r_{k,x}^A}{k_k^2}\right) \end{pmatrix} \begin{pmatrix} \Delta v_{k,x}^{CG} \\ \Delta v_{k,y}^{CG} \end{pmatrix} \tag{70}$$

Where our rotation matrix, **R**, is given by:

$$\mathbf{R} = \begin{pmatrix} \left(1 + \frac{r_{k,y} \cdot r_{k,y}^A}{k_k^2}\right) & \left(\frac{-r_{k,x} \cdot r_{k,y}^A}{k_k^2}\right) \\ \left(\frac{-r_{k,y} \cdot r_{k,x}^A}{k_k^2}\right) & \left(1 + \frac{r_{k,x} \cdot r_{k,x}^A}{k_k^2}\right) \end{pmatrix} \tag{71}$$

Rewriting (70), we have:

$$\Delta\bar{v}_k^A = \mathbf{R} \cdot \Delta\bar{v}_k^{CG} \tag{72}$$

So long as **R** is not singular, we can find its inverse, $\mathbf{R^{-1}}$, such that we can obtain $\Delta\bar{v}_k^{CG}$ by:

$$\Delta\bar{v}_k^{CG} = \mathbf{R^{-1}} \cdot \Delta\bar{v}_k^A \tag{73}$$

Next, we define four new variables, $a$, $b$, $c$, and $d$ given by:

$$a = 1 + \frac{r_{k,y} \cdot r_{k,y}^A}{k_k^2}$$
$$b = \frac{-r_{k,x} \cdot r_{k,y}^A}{k_k^2}$$
$$c = \frac{-r_{k,y} \cdot r_{k,x}^A}{k_k^2} \tag{74}$$
$$d = 1 + \frac{r_{k,x} \cdot r_{k,x}^A}{k_k^2}$$

With (74), we can rewrite our rotation matrix:

$$\mathbf{R} = \begin{pmatrix} a & b \\ c & d \end{pmatrix} \tag{75}$$

With this, (72) becomes:

$$\Delta v_{k,x}^A = a \cdot \Delta v_{k,x}^{CG} + b \cdot \Delta v_{k,y}^{CG} \tag{76}$$
$$\Delta v_{k,y}^A = c \cdot \Delta v_{k,x}^{CG} + d \cdot \Delta v_{k,y}^{CG} \tag{77}$$

The inverse, $\mathbf{R^{-1}}$, is given by:

$$\mathbf{R^{-1}} = \frac{\mathbf{1}}{|\mathbf{R}|}\begin{pmatrix} \boldsymbol{d} & -\boldsymbol{b} \\ -\boldsymbol{c} & \boldsymbol{a} \end{pmatrix} \tag{78}$$

Where the determinate, $|\mathbf{R}|$, is:

$$|\mathbf{R}| = ad - bc = 1 + \frac{r_{k,y} \cdot r_{k,y}^A}{k_k^2} + \frac{r_{k,x} \cdot r_{k,x}^A}{k_k^2} \tag{79}$$

With (78), (73) now becomes:

$$\begin{pmatrix} \Delta v_{k,x}^{CG} \\ \Delta v_{k,y}^{CG} \end{pmatrix} = \frac{1}{|\boldsymbol{R}|}\begin{pmatrix} d & -b \\ -c & a \end{pmatrix} \cdot \begin{pmatrix} \Delta v_{k,x}^A \\ \Delta v_{k,y}^A \end{pmatrix} = \frac{1}{|\boldsymbol{R}|} \cdot \begin{pmatrix} d \cdot \Delta v_{k,x}^A - b \cdot \Delta v_{k,y}^A \\ -c \cdot \Delta v_{k,x}^A + a \cdot \Delta v_{k,y}^A \end{pmatrix} \tag{80}$$

Using (80), we obtain our final form for $\Delta\bar{v}_k^{CG}$:

$$\Delta v_{k,x}^{CG} = \frac{1}{|\boldsymbol{R}|} \cdot \left(d \cdot \Delta v_{k,x}^A - b \cdot \Delta v_{k,y}^A\right) \tag{81}$$
$$\Delta v_{k,y}^{CG} = \frac{1}{|\boldsymbol{R}|} \cdot \left(-c \cdot \Delta v_{k,x}^A + a \cdot \Delta v_{k,y}^A\right) \tag{82}$$

Note, because $r_{k,y}, r_{k,y}^A, r_{k,x}, r_{k,x}^A$ are all signed values, it is possible to obtain some combinations of these values which makes **R** singular ($|\mathbf{R}| = 0$). From basic linear algebra, we know there is a unique solution for $\Delta\bar{v}_k^{CG}$ if and only if **R** is non-singular. We will explore the implications of this further below.

*PDOF*

With (57), we know by obtaining an estimate of $\Delta v_k^{CG}$, we also obtain an estimate of $\bar{J}_k$ – that is, the principal direction of force:

$$J_{k,x} = \frac{m_k}{|\boldsymbol{R}|} \cdot \left(d \cdot \Delta v_{k,x}^A - b \cdot \Delta v_{k,y}^A\right)$$
$$J_{k,y} = \frac{m_k}{|\boldsymbol{R}|} \cdot \left(-c \cdot \Delta v_{k,x}^A + a \cdot \Delta v_{k,y}^A\right) \tag{83}$$

where the direction of the impulse is given by:

$$\hat{J}_k = \frac{J_{k,x}\hat{x} + J_{k,y}\hat{y}}{\sqrt{J_{k,x}^2 + J_{k,y}^2}} \tag{84}$$

*The Lever-Arm*

The lever-arm, $h_k$, is given by the component of the vector $\bar{r}_k$ perpendicular to the impulse direction $\hat{J}_k$, and is given by:

$$h_k = \left|\bar{r}_k \times \hat{J}_k\right| = \left|\bar{r}_k \times \Delta\hat{v}_k\right| \tag{85}$$

Using (66), this becomes:

$$h_k = \frac{\left|r_{k,x} \cdot \Delta v_{k,y}^{CG} - r_{k,y} \cdot \Delta v_{k,x}^{CG}\right|}{\left|\Delta\bar{v}_k^{CG}\right|} \tag{86}$$

*The Closing-Velocity at the Point-of-Contact*

From (55), we know the velocity change at point *P*, fixed within the vehicle *k* frame, can be written as:

$$\Delta\bar{v}_k^P = \Delta\bar{v}_k^{CG} + \left(\Delta\bar{\omega}_k \times \bar{r}_k^P\right) \tag{87}$$

Using (61), we can write this as:

$$\Delta\bar{v}_k^P = \Delta\bar{v}_k^{CG} + \frac{1}{k_k^2}\left(\left(\bar{r}_k^P \times \Delta\bar{v}_k^{CG}\right) \times \bar{r}_k^P\right) \tag{88}$$

Taking the dot-product of $\frac{1}{k_k^2}\left(\left(\bar{r}_k^P \times \Delta\bar{v}_k^{CG}\right) \times \bar{r}_k^P\right)$ with $\Delta\hat{v}_k^{CG}$, and using the scalar triple product, and (85), we have:

$$\left(\left(\bar{r}_k^P \times \Delta\bar{v}_k^{CG}\right) \times \bar{r}_k^P\right) \cdot \Delta\hat{v}_k^{CG} = \left(\bar{r}_k^P \times \Delta\hat{v}_k^{CG}\right) \cdot \left(\bar{r}_k^P \times \Delta\bar{v}_k^{CG}\right) = \left|\Delta\bar{v}_k^{CG}\right| \cdot \left|\bar{r}_k^P \times \Delta\hat{v}_k^{CG}\right|^2 = \left|\Delta\bar{v}_k^{CG}\right| \cdot h_k^2 \tag{89}$$

Thus, taking the dot-product of (88) with $\Delta\hat{v}_k^{CG}$, we have:

$$\Delta\bar{v}_k^P \cdot \Delta\hat{v}_k^{CG} = \left|\Delta\bar{v}_k^{CG}\right| + \left|\Delta\bar{v}_k^{CG}\right| \cdot \frac{h_k^2}{k_k^2} = \left|\Delta\bar{v}_k^{CG}\right| \cdot \left(1 + \frac{h_k^2}{k_k^2}\right) \quad (90)$$

With (90), we now have a way to express the change-in-velocity component along the impulse direction, at the point of contact.

Let's now look at the difference in value for two vehicles:

$$(\Delta\bar{v}_1^P - \Delta\bar{v}_2^P) \cdot \Delta\hat{v}_1^{CG} = \left|\Delta\bar{v}_1^{CG}\right| \cdot \left(1 + \frac{h_1^2}{k_1^2}\right) + \left|\Delta\bar{v}_2^{CG}\right| \cdot \left(1 + \frac{h_2^2}{k_2^2}\right) \quad (91)$$

Note $\Delta\bar{v}_2^P \cdot \Delta\hat{v}_1^{CG}$ is negative since $\Delta\bar{v}_2^P$ is exactly antiparallel with $\Delta\hat{v}_1^{CG}$ from Newton's 3rd Law; therefore, we know:

$$(\Delta\bar{v}_1^P - \Delta\bar{v}_2^P) \cdot \Delta\hat{v}_1^{CG} \geq 0$$

Let's define a new parameter:

$$\gamma_k = \frac{k_k^2}{k_k^2 + h_k^2} \quad (92)$$

(91) can therefore be rewritten:

$$(\Delta\bar{v}_1^P - \Delta\bar{v}_2^P) \cdot \Delta\hat{v}_1^{CG} = \frac{\left|\Delta\bar{v}_1^{CG}\right|}{\gamma_1} + \frac{\left|\Delta\bar{v}_2^{CG}\right|}{\gamma_2} \quad (93)$$

Let's now write out the difference:

$$(\Delta\bar{v}_1^P - \Delta\bar{v}_2^P) \cdot \Delta\hat{v}_1^{CG} = \left(\left(\bar{v}_{1,f}^P - \bar{v}_{1,i}^P\right) - \left(\bar{v}_{2,f}^P - \bar{v}_{2,i}^P\right)\right) \cdot \Delta\hat{v}_1^{CG} = \left(\left(\bar{v}_{1,f}^P - \bar{v}_{2,f}^P\right) - \left(\bar{v}_{1,i}^P - \bar{v}_{2,i}^P\right)\right) \cdot \Delta\hat{v}_1^{CG} \quad (94)$$

The difference

$$\bar{v}_{Rel,i}^P = \bar{v}_{1,i}^P - \bar{v}_{2,i}^P \quad (95)$$

is simply the initial relative velocity of vehicle 1 with respect to vehicle 2 at the moment just prior to impact (the "closing-velocity").

The difference

$$\bar{v}_{Rel,f}^P = \bar{v}_{1,f}^P - \bar{v}_{2,f}^P \quad (96)$$

is simply the final relative velocity of vehicle 1 with respect to vehicle 2 at the moment just after impact (the "separation-velocity").

With the relative velocities defined, we can rewrite (94):

$$(\Delta\bar{v}_1^P - \Delta\bar{v}_2^P) \cdot \Delta\hat{v}_1^{CG} = \left(\bar{v}_{Rel,f}^P - \bar{v}_{Rel,i}^P\right) \cdot \Delta\hat{v}_1^{CG} \quad (97)$$

Let's define the coefficient-of-restitution by[3]:

$$\varepsilon = -\frac{\bar{v}_{Rel,f}^P \cdot \Delta\hat{v}_1^{CG}}{\bar{v}_{Rel,i}^P \cdot \Delta\hat{v}_1^{CG}} \quad (98)$$

Thus, the differences in final and initial relative velocity can be re-expressed by:

$$\left(\bar{v}_{Rel,f}^P - \bar{v}_{Rel,i}^P\right) \cdot \Delta\hat{v}_1^{CG} = \bar{v}_{Rel,f}^P \cdot \Delta\hat{v}_1^{CG} - \bar{v}_{Rel,i}^P \cdot \Delta\hat{v}_1^{CG} = -\varepsilon \cdot \bar{v}_{Rel,i}^P \cdot \Delta\hat{v}_1^{CG} - \bar{v}_{Rel,i}^P \cdot \Delta\hat{v}_1^{CG} = -(1+\varepsilon) \cdot \left(\bar{v}_{Rel,i}^P \cdot \Delta\hat{v}_1^{CG}\right) \quad (99)$$

With (93), (97), and (99), we finally have:

$$-(1+\varepsilon) \cdot \left(\bar{v}_{Rel,i}^P \cdot \Delta\hat{v}_1^{CG}\right) = \frac{\left|\Delta\bar{v}_1^{CG}\right|}{\gamma_1} + \frac{\left|\Delta\bar{v}_2^{CG}\right|}{\gamma_2} \quad (100)$$

Note, because we defined the closing velocity vector as $\bar{v}_{1,i}^P - \bar{v}_{2,i}^P$, we expect $\bar{v}_{Rel,i}^P \cdot \Delta\hat{v}_1^{CG} < 0$, and therefore $-(1+\varepsilon) \cdot \left(\bar{v}_{Rel,i}^P \cdot \Delta\hat{v}_1^{CG}\right) > 0$.

We can thus solve for the magnitude of the closing-velocity vector component parallel with the PDOF axis by[4]:

$$\left|\bar{v}_{Rel,i}^P \cdot \Delta\hat{v}_1^{CG}\right| = \frac{1}{1+\varepsilon} \cdot \left(\frac{\left|\Delta\bar{v}_1^{CG}\right|}{\gamma_1} + \frac{\left|\Delta\bar{v}_2^{CG}\right|}{\gamma_2}\right) \quad (101)$$

*With Knowledge of Only One $\Delta\boldsymbol{v}$*

From Newton's 3rd Law, we know (again, neglecting any external forces):

$$m_1 \cdot \Delta\bar{v}_1^{CG} = -m_2 \cdot \Delta\bar{v}_2^{CG} \quad (102)$$

Therefore, (91) can be written:

$$(\Delta\bar{v}_1^P - \Delta\bar{v}_2^P) \cdot \Delta\hat{v}_k^{CG} = \left|\Delta\bar{v}_1^{CG}\right| \cdot \left(1 + \frac{h_1^2}{k_1^2}\right) + \frac{m_1}{m_2}\left|\Delta\bar{v}_1^{CG}\right| \cdot \left(1 + \frac{h_2^2}{k_2^2}\right) = \left|\Delta\bar{v}_1^{CG}\right| \cdot \left[\left(1 + \frac{h_1^2}{k_1^2}\right) + \frac{m_1}{m_2} \cdot \left(1 + \frac{h_2^2}{k_2^2}\right)\right] = \left|\Delta\bar{v}_1^{CG}\right| \cdot \left[\frac{1}{\gamma_1} + \frac{m_1}{m_2} \cdot \frac{1}{\gamma_2}\right] = m_1 \cdot \left|\Delta\bar{v}_1^{CG}\right| \cdot \left[\frac{1}{\gamma_1 m_1} + \frac{1}{\gamma_2 m_2}\right] \quad (103)$$

Using (97) and (99), this becomes:

$$-(1+\varepsilon) \cdot \left(\bar{v}_{Rel,i}^P \cdot \Delta\hat{v}_1^{CG}\right) = m_1 \cdot \left|\Delta\bar{v}_1^{CG}\right| \cdot \left[\frac{1}{\gamma_1 m_1} + \frac{1}{\gamma_2 m_2}\right] \quad (104)$$

We can thus solve for the magnitude of the closing-velocity vector component parallel with the PDOF axis by:

$$\left|\bar{v}_{Rel,i}^P \cdot \Delta\hat{v}_1^{CG}\right| = \frac{1}{1+\varepsilon} \cdot \left(m_1 \cdot \left|\Delta\bar{v}_1^{CG}\right| \cdot \left[\frac{1}{\gamma_1 m_1} + \frac{1}{\gamma_2 m_2}\right]\right) \quad (105)$$

With our mathematical formalism on firm footing, we now demonstrate the method using staged collision data.

**Demonstration of Method**

Above, we derived the corrections needed to transform $\Delta\bar{v}$ estimates based on data from accelerometers positioned away from the CG, to the equivalent values at the CG. Ideally, this method is tested using staged collisions where the test vehicles are instrumented with perfectly accurate accelerometers distributed at various locations within the test vehicles. In what follows below, we present the results of applying our transformation to EDR-based $\Delta\bar{v}$ estimates from four EDRs distributed throughout a test vehicle subjected to a staged collision event. Though the soundness of the transformation method is demonstrated, using EDR data for this purpose comes with its own challenges related potential errors in the EDR-based $\Delta\bar{v}$ values themselves. The tangential but important issue of accounting for potential EDR-based $\Delta\bar{v}$ errors when such transformations are applied is also discussed below.

**2018 IPTM Crash Test 3**

*Experimental Set-up*

A crash test was performed on May 21, 2018 in Orlando, Florida. The crash test was crash test number 3 from IPTM's Symposium on Traffic Safety. The crash configuration was of the T-bone type. The bullet vehicle struck the target vehicle behind the rear axle.

The target vehicle was a 1998 Chevrolet Malibu LS 4-door bearing VIN 1G1NE52M3W6XXXXXX (see Figure 4). Note, we will use the index "1", "vehicle 1", "Chevy", and "target vehicle" interchangeably below. The Chevy was stationary at impact. Its weight was obtained with Rebco 1200-pound scales. The weight on the front axle was 1806 pounds and the weight on the rear axle was 1026 pounds, for a total weight of 2832 pounds.

The bullet vehicle was a 2002 Buick LeSabre Custom 4-door bearing VIN 1G4HP54K72UXXXXXX (see Figure 5). Note, we will use the index "2", "vehicle 2", Buick, and "bullet vehicle" interchangeably below. The bullet was driven into impact at 27 MPH by a volunteer driver. The impact speed was obtained with a VBox Sport. The Vbox Sport measures speed with a 20 Hz GPS engine. The Buick and its driver were weighed with Rebco 1200-pound scales. The weight on the front axle was 2334 pounds and the weight on the rear axle was 1421 pounds for a total weight for the vehicle and driver of 3755 pounds. The impact configuration is shown in Figure 6.

[3] Note, in this model, where $\Delta\hat{v}_1^{CG}$ is approximately aligned with the direction of crush, we define restitution as the point of contact negative ratio of separation-velocity to closing-velocity vector components directed along the $\Delta\hat{v}_1^{CG}$ axis. In cases where frictional effects are non-negligible, one may wish to use the more generally applicable formalism, where $\Delta\hat{v}_1^{CG}$ is represented in a normal "crush axis" and tangent "friction axis" basis, and where restitution is defined as the point of contact negative ratio of separation-velocity to closing-velocity vector components directed along the normal axis (see for example, R. Brach's planar impact mechanics). This topic will be considered in a future article.

[4] See Appendix 1 for discussion on solving for pre-impact ground speeds.

The Chevy was instrumented with 2 laboratory-grade +/-250G accelerometers and 1 laboratory-grade +/-600 deg/sec rate gyro. The rate gyro was a Summit Instruments model 31206B and the two accelerometers were Measurement Specialties model 34208A (see Figures 7 and 8). One accelerometer was mounted at the CG and the rate gyro was mounted just behind the accelerometer mounted at the CG. The second accelerometer was mounted on the firewall, inside the engine compartment. The measured locations of the accelerometers are documented in Table 1. The position of the rate gyro was not documented because angular rate is constant within a rigid body.

The data acquisition equipment used for the laboratory-grade equipment were two Vericom Computers VC4000DAQs. The VC4000s were set to sample data at 1000 Hz. One of the VC4000s was used to record longitudinal and lateral acceleration data at the CG as well as the yaw rate. The other VC4000 recorded longitudinal and lateral acceleration at the firewall. Acceleration along the z-axis was not recorded. All data was stored as voltage, which was later post-processed in ROOT [4] into accelerations and yaw rate.

Seven "ride-along" EDRs were installed in the Chevy (see Figure 7). A ride-along EDR is an airbag control module that is attached to the structure of a vehicle for capturing the crash pulse. The EDRs used in this crash test were GM sensing and diagnostic modules (SDM) that were used in the 2005 to 2009 Chevrolet Trailblazer and GMC Envoys (Bosch cable 3293). Ride-along EDRs are not connected to the vehicle's CAN bus, so no pre-crash data may be obtained. The ride-along EDRs are powered by a small external battery back. In the subject collision, four of the seven ride-along EDRs recorded an event. Two of the four ride-along EDRs were installed in the trunk of the Chevy, close to the impulse-centroid. One EDR was installed on the center tunnel just behind the center-of-gravity, and two were installed on the front passenger floor pan, to the right and ahead of the center-of-gravity. The two ride-along EDRs in the trunk were installed with their longitudinal axes aligned with the negative *y*-axis of the Chevy. The two ride-along EDRs in the passenger compartment were installed with their longitudinal axes aligned with the positive *y*-axis of the Chevy. The measured locations of the ride-along EDRs are documented in Table 1.

The crash test was documented with several video cameras including one high-speed camera running at 240 fps and one unmanned aerial system (UAS). Photographs were taken of both vehicles before and after the crash test. The scene was photographed after the test and it was also documented with a Riegl 3-D laser scanner. The final rest positions of both vehicles were documented with hand measurements, as well as the Riegl scanner and the UAS.

*Accelerometer Measurements*

Table 1 shows the location and the cumulative $\Delta v$ from the instruments and the ride-along EDRs. Locations are given with respect to the $CG_{xy}$ using SAE conventions. The instrumentation-grade accelerometers were post-processed in ROOT from voltage to acceleration and then integrated to get cumulative $\Delta v$. Figure 9 shows the longitudinal, lateral acceleration, and yaw rate graphs from the laboratory-grade instruments. The black lines show the acceleration values with a 60 CFC Butterworth filter applied to the acceleration data. Figure 10 shows the corresponding longitudinal and lateral $\Delta v$ graphs, as well as the change-in-yaw, which were obtained by numerically integrating the accelerometer and rate gyro data. From this data, we estimate that the cumulative local $\Delta\bar{v}^{\text{CG}}_{\text{Chevy}} = (1.04\text{ mph}, -6.30\text{ mph})$. Correcting for vehicle coordinate axis rotation (see "$\Delta\boldsymbol{v}$ in Continuous Time" section above) using yaw versus time, the global frame values are $\Delta\bar{v}^{\text{CG}}_{\text{Chevy}} = (1.31\text{ mph}, -6.14\text{ mph})$. Note, the analyst will typically need global frame $\Delta\bar{v}^{\text{CG}}$, and therefore global frame $\Delta\bar{v}^{\text{A}}$, for various calculations; however, because yaw rate data often isn't available in EDR data, the analyst will not be able to precisely calculate global frame $\Delta\bar{v}^{\text{A}}$ since the exact yaw versus time behavior is unknown. In the results that follow below, the true $\Delta\bar{v}^{\text{CG}}_{\text{Chevy}}$ values from our accelerometer at the CG are obtained by applying rotational corrections to the components of $\bar{a}^{CG}(t)$. We make no attempts to apply similar corrections to $\Delta\bar{v}^{\text{EDR}}_{\text{Chevy}}$ in order to approximate the real-world scenario encountered by many analysts.

Note, the 250 G accelerometers mounted at the firewall did not record a crash pulse.

*Rate Gyro Measurement*

The rate gyro showed an average peak rate of 204.5 deg/sec and a rotation of 12.3° during the ~100 ms crash pulse. The integrated total rotation from impact to final rest was 134.6° (see Figure 11). Hand measurements determined that the total rotation was 135° and that the Chevy's CG translated 15.3 feet to final rest.

*EDR Measurements*

Figure 12 depicts the location of the four EDRs used in this analysis.

EDR A was placed in the trunk, toward the rear, at position (-7.8 ft, -1.5 ft) in the Chevy's reference frame. The data obtained from EDR A is shown in Figure 13. From this data, we estimate $\Delta\bar{v}^{\text{EDRA}}_{\text{Chevy}} = (5.09\text{ mph}, -24.19\text{ mph})$.

EDR B was placed in the trunk, toward the front, at position (-6.98 ft, -1.5 ft) in the Chevy's reference frame. The data obtained from EDR B is shown in Figure 14. From this data, we estimate $\Delta\bar{v}^{\text{EDRB}}_{\text{Chevy}} = (5.09\text{ mph}, -21.01\text{ mph})$.

EDR C was placed in the occupant cabin, behind the center-of-gravity, at position (-0.67 ft, 0.0 ft) in the Chevy's reference frame. The data obtained from EDR C is shown in Figure 15. From this data, we estimate $\Delta\bar{v}^{\text{EDRC}}_{\text{Chevy}} = (0.64\text{ mph}, -6.37\text{ mph})$.

EDR D was placed in the occupant cabin, in front of the center-of-gravity, at position (0.3 ft, 0.73 ft) in the Chevy's reference frame. The data obtained from EDR D is shown in Figure 16. From this data, we estimate $\Delta\bar{v}^{\text{EDRD}}_{\text{Chevy}} = (0.0\text{ mph}, -5.73\text{ mph})$.

*Crush Damage*

The damage profile of the Chevy was documented by hand measurements as well as a Carlson total station. Two separate sets of measurements were taken by hand. One set of measurements included the induced damage, and one set of measurements included only the contact damage. The hand measurements are shown in Table 2. In the analysis that follows, the impulse centroid was taken at the point of maximum crush on the direct contact damage only crush profile of the Chevy.

*$\Delta v_x$ and $\Delta v_y$ Estimates*

Using equations (81) and (82), the transformed $\Delta\bar{v}^{CG}_{Chevy}$ values were obtained. Because all parameters were not well controlled in the experiment, we used a Monte Carlo analysis script written for ROOT to obtain best-estimates, as well as upper and lower limits for our $\Delta\bar{v}^{CG}_{Chevy}$ values. The inputs used in the Monte Carlo script are shown in Table 3. In our first round of results, we assumed no uncertainty on our EDR-based estimates of $\Delta v_x^A$ and $\Delta v_y^A$. In the sections that follow, we explore the issue of EDR inaccuracies. Uniform probability distributions were used for all inputs. The best-estimate $\Delta\bar{v}^{CG}_{Chevy}$ and closing-speed were obtained by using the best-estimate input values. The minimum and maximum values were obtained directly by finding the endpoints of the resulting output distributions. The accelerometer-based $\Delta\bar{v}$ at the CG is shown in Table 4. The EDR-based results are shown in Table 5. The differences between best-estimate EDR-based results and results estimated from the accelerometer are shown in Table 6. The average of the differences shown in this table for both components is 0.25 mph.

The results from Tables 4, 5, and 6 are shown in Figure 17. The red line illustrates the accelerometer-based $\Delta v_x$ and $\Delta v_y$ estimates. The gray boxes illustrate the uncorrected EDR-based $\Delta v_x$ and $\Delta v_y$ values. The black dots represent the best-estimate EDR-based $\Delta v_x$ and $\Delta v_y$ values at the CG. The upper and lower bound corrected EDR values are illustrated by the black lines. These results illustrate that the correction method properly brackets the accelerometer-based $\Delta v_x$ and $\Delta v_y$ estimates. Note, the shaded region about the red line indicates the minimum and maximum accelerometer-based $\Delta v_x$ and $\Delta v_y$. The upper and lower bound estimates are based on randomly sampling the pre-impact acceleration bias of the accelerometer, which was estimated by examining data during a 10 second window before impact.

*Closing-speed Estimates*

Table 7 shows the EDR-based closing speed estimates for each EDR, along with the corresponding uncertainties obtained using our Monte Carlo script. Table 8 shows the difference between the best-estimate EDR-based value closing-speed and the true value. The average of the differences is approximately -2.7 mph.

Figure 18 illustrates the EDR-based closing-speed estimates versus source of EDR data. The black dots represent the best-estimate. The lines represent the upper and lower-bound estimates. We see in comparison to the true closing-speed, our EDR-based estimates properly bracket the true results.

An example end-to-end calculation is provided in the Appendix 2.

*Uncertainty Due to EDR Inaccuracy and the $\Delta v$ Corridor*

In the above presented results, we assumed no uncertainty on EDR-based estimates of $\Delta v_x^A$ and $\Delta v_y^A$; however, inaccuracies on these input values can have important consequences for one's minimum and maximum uncertainty range on $\Delta v_x^{CG}$ and $\Delta v_y^{CG}$ and closing-speed. This is explored below.

The accuracy of $\Delta v$ estimates from EDRs has been the subject of numerous studies [5,6,7,8,9,10,11]. Indeed, the authors of reference [12] delineate a helpful checklist of potential error sources which can cause inaccuracies in EDR-reported speed-change data. The reader is strongly encouraged to review this reference. We briefly summarize those error sources here:

1. Internal acceleration thresholds: Algorithm enable acceleration trigger threshold. Trigger threshold typically in range of 1g to 2g.
2. Short EDR time-window: EDR time-window for recording $\Delta v$ too short to capture full acceleration pulse. Can be ruled out by examining if $\Delta v$ reaches maximum value and possibly decreases prior to end of window.
3. Long EDR time-window: Recording window may be too long, which may cause post-impact ground-contact tire forces to contribute to $\Delta v$ over-estimates. This can be ruled out by examining the EDR data for a local $\Delta v$ maximum, possibly followed by decrease, then an upward drift.
4. "Clipping": The true acceleration at the EDR may exceed the EDR accelerometer's minimum or maximum limit. This can cause a truncation of the true peak acceleration and therefore a corresponding underestimate of $\Delta v$ . This may be discerned by looking for a flattened portion of the EDR's acceleration curve if possible.
5. Off-axis: The EDR may be away from the vehicle CG. This is the subject of this paper.
6. Vehicle crush: The EDR is located in the direct region of crush. This can cause an overestimate of $\Delta v$ since the crushed region will undergo accelerations exceeding that of the vehicle CG. This can also result in underestimates if the EDR orientation changes during the collision. For example, material crushing could cause the EDR's local *x*-axis to rotate into the vehicle's *x*-*y* plane thereby causing the EDR to lose sensitivity to longitudinal vehicle acceleration.
7. EDR power loss: The EDR may lose power before completely recording $\Delta v$. This will result in an underestimate of $\Delta v$. Depending on model year, the EDR report may indicate if an event's recording is complete.

Because our formalism to obtain $\Delta v_x^{CG}$ and $\Delta v_y^{CG}$ relies on $\Delta v_x^A$ and $\Delta v_y^A$ as inputs, EDR errors introduced by sources such as those listed above will naturally propagate to our estimates of $\Delta v_x^{CG}$ and $\Delta v_y^{CG}$, as well as to closing-speed. In some cases, the errors propagated to the final $\Delta v_x^{CG}$ and $\Delta v_y^{CG}$ estimates can be quite large due to delicate numerical cancelations between $\Delta v_{k,x}^A$ and $\Delta v_{k,y}^A$ and the inverse proportionality to $|\mathbf{R}|$. This important issue is explored in more detail below.

In order to account for potential sources of error such as those enumerated above, we examined EDR-based versus accelerometer-based $\Delta v$ values for GM models ranging from 2002 to 2009 production years from references [5,6,7,8,9,10,11]. Using this data, a representative "$\Delta v$ corridor" was constructed as a function of EDR $\Delta v$ (Figure 19 and Figure 20). Ideally, such a corridor is defined based on bench test experiments where the orientation of a given EDR can be precisely controlled, and the input true $\Delta\vec{v}$ at the accelerometer is known to a high degree of accuracy. Unfortunately, such tests are rare, and therefore our corridor relies on data mostly from staged collisions. While staged collision data is quite useful for the researcher, using EDR data from staged collisions makes it difficult to disentangle $\Delta v$ inaccuracies due to physical effects such as vehicle rotation during impact versus inaccuracies due to underlying algorithm design and accelerometer performance characteristics. Therefore, the corridor depicted in Figure 19 is meant to represent a worst-case potential minimum/maximum range for true $\Delta v$ as a function of our subject EDR $\Delta v$. The corridor is defined as follows. For component *j*, we have for the lower-bound of the corridor:

$$\Delta v_j^{True,Low} = \mathrm{sign}(\Delta v_j^{EDR}) \cdot (1 - 10\%)\left|\Delta v_j^{EDR}\right|$$

For the upper-bound, we have the piecewise continuous function:

$$\Delta v_j^{True,High} = \begin{cases} \mathrm{sign}(\Delta v_j^{EDR}) \cdot (4.4\text{ kph}), \text{for } \left|\Delta v_j^{EDR}\right| \le 1.5\text{ kph} \\ \mathrm{sign}(\Delta v_j^{EDR}) \cdot (\left|\Delta v_j^{EDR}\right| + 2.9\text{ kph}), \quad \text{for } 1.5\text{ kph} < \left|\Delta v_j^{EDR}\right| \le \frac{2.9\text{ kph}}{17\%} \\ \mathrm{sign}(\Delta v_j^{EDR}) \cdot (1 + 17\%) \cdot \left|\Delta v_j^{EDR}\right|, \text{for } \left|\Delta v_j^{EDR}\right| > \frac{2.9\text{ kph}}{17\%} \end{cases}$$

Thus, for high $\Delta v$, the oft quoted $\pm 10\%$ uncertainty on $\Delta v$, typically attributed to finite accelerometer accuracy [13], becomes a $+17\%$ upper-bound uncertainty and $-10\%$ lower-bound uncertainty. Thus, our uncertainty on $\Delta v$ is asymmetric.

For low $\Delta v$, the behavior is more complex due to both threshold effects and offset effects (see Figure 20). The threshold effects cause $\Delta v^{EDR} = 0$ for $\Delta v^{True} < \Delta v^{Threshold}$, whereas for $\Delta v^{True} \ge \Delta v^{Threshold}$, we have $\Delta v^{EDR} = \Delta v^{True} - \Delta v^{Offset}$. This behavior can be attributed to error type (1) and is explored in great depth in reference [6]. For our $\Delta v$ corridor, we have $\Delta v^{Threshold} = 4.4$ kph and $\Delta v^{Offset} = 2.9$ kph. A non-zero $\Delta v^{Threshold}$ implies, without any other knowledge, that $\Delta v = 0$ may actually imply $\Delta v = \Delta v^{Threshold}$ in the worst-case. This has important consequences that are explored further below.

With our $\Delta v$ corridor defined, we can now better understand our uncertainty ranges for $\Delta v_x^{CG}$ and $\Delta v_y^{CG}$. Note, the corridor described above was defined based on tests related to longitudinal $\Delta v$. A recent bench test study involving 2012 GM EDRs indicates that uncertainty in lateral $\Delta v$ component may be symmetric about 0 and less than 10% [18]; however, such an experiment is likely less sensitive to clipping effects since the acceleration pulse used by the test apparatus is well controlled without large fluctuations. Another recent study on lateral $\Delta v$ EDR accuracy using vehicles from 2010 – 2012 model years (including GMs) that were subjected to side-impact tests as a part of the National Highway Safety Administration Side-Impact New Car Assessment Program showed EDRs tended to underestimate lateral $\Delta v$ [19].

Figure 21 and Figure 22 illustrate $\Delta v_x^{CG}$, $\Delta v_y^{CG}$, and closing-speed applying the same corridor for both lateral and longitudinal components of the EDR $\Delta v$ values. For all three values, the uncertainty windows have widened due to accounting for worst-case potential EDR-based $\Delta v$ inaccuracies. How the analyst can try to reduce the effect of such inaccuracies is discussed below.

**Mitigation of Uncertainties**

*Reduction of Input Uncertainties*

The usefulness of speed and change-in-speed estimates can directly depend on the uncertainty of those estimates. Though the uncertainties illustrated in Figure 21 and Figure 22 appear formidable, they are primarily driven by only a few factors. For the $\Delta v_x$ and $\Delta v_y$ estimates, the most important contributor to uncertainty is the a priori unknown EDR accuracy. This source of uncertainty can be reduced if one has data from tests conducted on the EDR of same year, make, and model as the subject vehicle. Ideally, the tests would be conducted over a wide spectrum of $\Delta v$ values.

The second largest contributor to uncertainty is related to the physical location of the accelerometer onboard the EDR circuit board. This can be easily addressed by removing the EDR housing and visually locating and measuring the accelerometer with respect to the EDR's geometrical center.

Finally, the coefficient of restitution contributes to large uncertainties in closing-speed. Though a large range (from 5% to 25%) was used in this analysis, based on data from staged collisions of similar severity [15,16,17], using reasonable exemplar vehicles could help narrow this range of restitution values.

*Use of Additional Constraints*

Additional evidence collected during scene and vehicle inspections such as crush damage, departure angles, and post-impact trajectory lengths can be used to provide additional constraints on both closing-speed and $\Delta\vec{v}^{CG}$. 3D computer simulation applications can be used to quickly simulate post-impact trajectories over scene data. Such simulations can provide further constraints. This is explored in the next section.

**Full Virtual CRASH 4 Simulation**

We performed a simulation of the 2018 IPTM Crash Test 3 using Virtual CRASH 4 [20]. Virtual CRASH 4 is a software application for accident reconstruction which includes the ability to simulate motor vehicle collisions using an impulse-momentum based model.

Starting with point cloud data created with the Pix4D application [21] using drone photographs taken after the crash test, the data was automatically aligned using the output .tfw file. A 3D surface mesh was created in Virtual CRASH 4 on top of which the simulated vehicles were placed (Figure 23). The goal of the analysis was to determine if, primarily using knowledge of the post-impact motion of the vehicles and the crush damage on the vehicles, we could use the simulation engine to find estimates for the Buick's pre-impact ground speed.

*Focus on Post-Impact Motion*

In our subject crash, simply focusing on the post-impact trajectory using the Virtual CRASH simulation model can provide useful constraints on the $|\Delta\bar{v}_1^{CG}|$,$|\Delta\bar{\omega}_1|$, and $|\Delta\bar{\omega}_2|$. Using the real-time feedback given by the Virtual CRASH simulation engine, it is easy to probe, as initial conditions to the simulation, various combinations of $\Delta\bar{v}_1^{CG}$ and $\Delta\bar{\omega}_1$ that simultaneously satisfy (67) and the correct post-impact motion for the Chevy as loose conditions. Using such an approach, we can place rough upper and lower bounds on $|\Delta\bar{v}_1^{CG}|$ and $|\Delta\bar{\omega}_1|$, which can be included into our Monte Carlo script as selection cuts to eliminate Monte Carlo trials that exceed those bounds. Using this same simulation approach, we can also place constraints on $|\Delta\bar{\omega}_2|$ by searching for upper bound values beyond which the Buick's post-impact heading is directed too far from its documented area of rest. The results of applying these simulation-based selection cuts to our Monte Carlo analysis are illustrated in Figure 24 and Figure 25. The reduction in uncertainty for $\Delta\bar{v}_1^{CG}$ and closing-speed is evident.

*Simulation Optimization of Full Event*

In addition to using simulation for post-impact motion studies, we also simulated the full collision event. Using the post-impact motion path of the Chevy and point of maximum crush as primary constraints, we obtained a pre-crash ground speed of 28.9 mph for the Buick, in good agreement with the known pre-impact speed. The resulting simulated change-in-velocity for the Chevy is $\bar{v}_{Chevy}^{CG} = (0.74\text{ mph}, 5.65\text{ mph})$. This is also in good agreement with the values obtained from the accelerometer-based estimate. The simulated motion sequence can be seen in Figure 26.

As demonstrated above, the uncertainties on $\Delta v$ corrections presented above can be sensitive to both EDR position and EDR inaccuracies. In the section below we explore these dependencies in more detail.

**Implications of a Singular R**

*Point of Zero Motion*

Let's examine the case where the impulse centroid and accelerometer can be at any arbitrary position within the Chevy (vehicle 1). The condition $|\mathbf{R}| = 0$ implies:

$$1 + \frac{r_{1,y} \cdot r_{1,y}^A}{k_1^2} + \frac{r_{1,x} \cdot r_{1,x}^A}{k_1^2} = 0 \qquad (106)$$

Thus, there is an imaginary line (the $R_0$ line) along which we cannot solve for a unique $\Delta\bar{v}_1^{CG}$ given the non-homogenous condition $\Delta\bar{v}_1^A \neq \bar{0}$. We can describe the $R_0$ line as a function of $r_{1,x}^A$:

$$r_{1,y}^A = -\frac{k_1^2}{r_{1,y}} - \left(\frac{r_{1,x}}{r_{1,y}}\right) \cdot r_{1,x}^A \qquad (107)$$

Along the $R_0$ line, we have for *a*:

$$\begin{aligned} a &= 1 + \frac{r_{1,y} \cdot r_{1,y}^A}{k_1^2} \\ &= 1 + \frac{r_{1,y}}{k_1^2} \cdot \left(-\frac{k_1^2}{r_{1,y}} - \left(\frac{r_{1,x}}{r_{1,y}}\right) \cdot r_{1,x}^A\right) \\ &= \frac{r_{1,x} \cdot r_{1,x}^A}{k_1^2} = 1 - d \end{aligned} \qquad (108)$$

for *b*:

$$\begin{aligned} b &= -\frac{r_{1,x} \cdot r_{1,y}^A}{k_1^2} \\ &= -\frac{r_{1,x}}{k_1^2} \cdot \left(-\frac{k_1^2}{r_{1,y}} - \left(\frac{r_{1,x}}{r_{1,y}}\right) \cdot r_{1,x}^A\right) \\ &= \frac{r_{1,x}}{r_{1,y}} \cdot d \end{aligned} \qquad (109)$$

and for *c*:

$$\begin{aligned} c &= -\frac{r_{1,y} \cdot r_{1,x}^A}{k_1^2} \\ &= \frac{r_{1,y}}{r_{1,x}} \cdot (1 - d) \end{aligned} \qquad (110)$$

Therefore, for an accelerometer on the $R_0$ line, we have:

$$\begin{aligned} \Delta v_{1,x}^A &= (1 - d) \cdot \Delta v_{1,x}^{CG} + \frac{r_{1,x}}{r_{1,y}} \cdot d \cdot \Delta v_{1,y}^{CG} \\ \Delta v_{1,y}^A &= \frac{r_{1,y}}{r_{1,x}} \cdot (1 - d) \cdot \Delta v_{1,x}^{CG} + d \cdot \Delta v_{1,y}^{CG} \end{aligned} \qquad (111)$$

which implies the components of $\Delta\bar{v}_1^A$ must be related by:

$$\frac{\Delta v_{1,y}^A}{\Delta v_{1,x}^A} = \frac{r_{1,y}}{r_{1,x}} \qquad (112)$$

This interesting result indicates that if an accelerometer happens to sit on the $R_0$ line, $\Delta\bar{v}_1^A$ will point either parallel or anti-parallel to the vector pointing from the center-of-gravity to the impulse-centroid. Comparing this to the slope of the $R_0$ line (equation (107)), the vector $\Delta\bar{v}_1^A$ must be perpendicular to the $R_0$ line.

With the condition $|\mathbf{R}| = 0$, for the homogeneous condition, $\Delta\bar{v}_1^A = \bar{0}$, we know we cannot solve for a unique $\Delta\bar{v}_1^{CG}$. We can, however, solve for the point along the $R_0$ line (the $R_0$ point), which remains in its pre-impact velocity state immediately after impact, by solving:

$$\begin{pmatrix} 0 \\ 0 \end{pmatrix} = \begin{pmatrix} \Delta v_{1,x}^{CG} - \Delta\omega_1 \cdot r_{1,y}^A \\ \Delta v_{1,y}^{CG} + \Delta\omega_1 \cdot r_{1,x}^A \end{pmatrix} \qquad (113)$$

for $r_{1,x}^A$ and $r_{1,y}^A$. This gives:

$$r_{1,x}^A = -\frac{\Delta v_{1,y}^{CG}}{\Delta\omega_1} = \frac{-\Delta v_{1,y}^{CG} \cdot k_1^2}{r_{1,x} \cdot \Delta v_{1,y}^{CG} - r_{1,y} \cdot \Delta v_{1,x}^{CG}} \qquad (114)$$

$$r_{1,y}^A = \frac{\Delta v_{1,x}^{CG}}{\Delta\omega_1} = \frac{\Delta v_{1,x}^{CG} \cdot k_1^2}{r_{1,x} \cdot \Delta v_{1,y}^{CG} - r_{1,y} \cdot \Delta v_{1,x}^{CG}}$$

which implies:

$$-\frac{\Delta v_{1,y}^{CG}}{\Delta v_{1,x}^{CG}} = \frac{r_{1,x}^A}{r_{1,y}^A} \qquad (115)$$

Therefore the $R_0$ point must sit on a line going through the center-of-gravity that is also perpendicular to $\Delta\bar{v}_1^{CG}$.

Solving for $r_{1,x}^A$ in (115), then substituting into (107), we have:

$$r_{1,y}^A = -\frac{k_1^2}{r_{1,y}} - \left(\frac{r_{1,x}}{r_{1,y}}\right) \cdot \left(-\frac{\Delta v_{1,y}^{CG}}{\Delta v_{1,x}^{CG}} \cdot r_{1,y}^A\right) \qquad (116)$$

Solving the above for $r_{1,y}^A$ gives:

$$r_{1,y}^A = \frac{-\Delta v_{1,x}^{CG} \cdot k_1^2}{r_{1,x} \cdot \Delta v_{1,y}^{CG} - r_{1,y} \cdot \Delta v_{1,x}^{CG}} \qquad (117)$$

Thus, comparing (117) to (114) confirms that the $R_0$ point must be a point on the $R_0$ line. Much like the ground contact point of an ideal (rigid) rolling wheel that is in motion with no slip, the $R_0$ point can be thought of as the post-impact instantaneous center of rotation for all points within the vehicle. We can confirm this by defining a radial vector, $\bar{L}_1$, originating from the $R_0$ point and pointing to some point in the vehicle frame $P$ ($\bar{r}_1^P$). This vector is given by:

$$\bar{L}_1 = \left(r_{1,x}^P + \frac{\Delta v_{1,y}^{CG}}{\Delta\omega_1}\right)\hat{x} + \left(r_{1,y}^P - \frac{\Delta v_{1,x}^{CG}}{\Delta\omega_1}\right)\hat{y} \qquad (118)$$

The change in velocity at this point is given by:

$$\begin{aligned} \Delta\bar{v}_1^P = &\left(\Delta v_{1,x}^{CG} - \Delta\omega_1 \cdot r_{1,y}^P\right)\hat{x} \\ &+ \left(\Delta v_{1,y}^{CG} + \Delta\omega_1 \cdot r_{1,x}^P\right)\hat{y} \end{aligned} \qquad (119)$$

Using (118) and (119), it is easy to show that $\bar{L}_1 \cdot \Delta\bar{v}_1^P = 0$ and $\bar{L}_1 \times \Delta\bar{\omega} = \Delta\bar{v}_1^P$ for any point $P$. This implies that the $R_0$ point acts as the effective instantaneous center of rotation in the Earth frame for any point within vehicle 1.

*Uncertainty Near the $R_0$ Line*

Recall (81) and (82) tell us how to relate the true change-in-velocity at the accelerometer position to the true change-in-velocity at the CG. Let's assume we have perfect knowledge of our geometry parameters *a*, *b*, *c*, and *d* with no uncertainties. What effect does EDR measurement uncertainty have on our estimates of change-in-velocity at the center-of-gravity? Suppose our accelerometer-based change-in-velocity estimates at *A* differ from the true values by a simple scale factor (remember, we do not have access to the "true" value, but only the experimentally determined estimate). To simplify the analysis, let's suppose the scale factor is the same for both components. That is:

$$\begin{aligned} \Delta v_{1,x}^{A\,Est} &= \beta \cdot \Delta v_{1,x}^{A\,True} \\ \Delta v_{1,y}^{A\,Est} &= \beta \cdot \Delta v_{1,y}^{A\,True} \end{aligned} \qquad (120)$$

For example, we may have $\beta = 1/(1 + 17\%)$ as shown in the above discussion on the $\Delta v$ corridor.

With this, our change-in-velocity estimates at the CG become:

$$\Delta v_{k,x}^{CG\,Est} = \frac{1}{|\boldsymbol{R}|} \cdot \left(d \cdot \beta \cdot \Delta v_{1,x}^{A\,True} - b \cdot \beta \cdot \Delta v_{1,y}^{A\,True}\right)$$
$$\Delta v_{k,y}^{CG\,Est} = \frac{1}{|\boldsymbol{R}|} \cdot \left(-c \cdot \beta \cdot \Delta v_{1,x}^{A\,True} + a \cdot \beta \cdot \Delta v_{1,y}^{A\,True}\right) \tag{121}$$

or:

$$\Delta v_{1,x}^{CG\,Est} = \beta \cdot \Delta v_{1,x}^{CG\,True}$$
$$\Delta v_{1,y}^{CG\,Est} = \beta \cdot \Delta v_{1,y}^{CG\,True} \tag{122}$$

Thus implying, if our change-in-velocity estimates obtained from our accelerometer differ from the true value by a simple scale factor, we can expect our change-in-velocity estimates at the center-of-gravity to differ by the same scale factor with respect to the true values.

Now suppose instead, our accelerometer-based change-in-velocity values at $A$ differ from the true values by (see reference [6] for examples):

$$\Delta v_{1,x}^{A\,Est} = \beta \cdot \Delta v_{1,x}^{A\,True} + \delta$$
$$\Delta v_{1,y}^{A\,Est} = \beta \cdot \Delta v_{1,y}^{A\,True} + \delta \tag{123}$$

where $\delta$ may be due to an acceleration threshold effect.

In this case, we have:

$$\begin{aligned}\Delta v_{1,x}^{CG\,Est} &= \frac{1}{|\boldsymbol{R}|} \cdot \left(d \cdot \Delta v_{1,x}^{A\,Est} - b \cdot \Delta v_{1,y}^{A\,Est}\right) \\ &= \frac{1}{|\boldsymbol{R}|} \cdot \left(d \cdot \beta \cdot \Delta v_{1,x}^{A\,True} - b \cdot \beta \cdot \Delta v_{1,y}^{A\,True}\right) \\ &+ \frac{1}{|\boldsymbol{R}|} \cdot (d \cdot \delta - b \cdot \delta) \\ &= \beta \cdot \Delta v_{1,x}^{CG\,True} + \delta \cdot \frac{1}{|\boldsymbol{R}|} \cdot (d - b)\end{aligned} \tag{124}$$

and

$$\begin{aligned}&\Delta v_{1,y}^{CG\,Est} \\ &= \frac{1}{|\boldsymbol{R}|} \cdot \left(-c \cdot \Delta v_{1,x}^{A\,Est} + a \cdot \Delta v_{1,y}^{A\,Est}\right) \\ &= \frac{1}{|\boldsymbol{R}|} \cdot \left(-c \cdot \beta \cdot \Delta v_{1,x}^{A\,True} + a \cdot \beta \cdot \Delta v_{1,y}^{A\,True}\right) \\ &+ \frac{1}{|\boldsymbol{R}|} \cdot (-c \cdot \delta + a \cdot \delta) \\ &= \beta \cdot \Delta v_{1,y}^{CG\,True} + \delta \cdot \frac{1}{|\boldsymbol{R}|} \cdot (a - c)\end{aligned} \tag{125}$$

Thus, here we see the $\Delta v_{1,x}^{CG\,Est}$ and $\Delta v_{1,y}^{CG\,Est}$ values will differ from the true values by terms dependent on $1/|\boldsymbol{R}|$. This can have important consequences for one's uncertainty analysis. That is, if our accelerometer is moved closer to the $R_0$ line, we can expect our uncertainty on $\Delta\bar{v}_1^{CG}$ to increase if $\Delta\bar{v}_1^{A}$ differs from the true value by a linear constant. This means, when conducting an error analysis, we should expect to see larger error bars for EDRs near the $R_0$ line. This is indeed what we see in our data. For our subject crash, we estimate the $R_0$ line to intersect with the center-line of the Chevy about 9 inches behind the front axle. In our results presented above, it is observed that the uncertainty range in $\Delta\bar{v}_{Chevy}^{CG}$ increases as the EDR position gets closer to the front axle. Since closing-speed is directly proportional to the magnitude of $\Delta\bar{v}_{Chevy}^{CG}$, this same pattern is observed in our closing-speed results.

**Threshold Effects**

We can gain more insight into the effects of the $R_0$ line and $R_0$ point using a forward evaluation calculation C++ script written for ROOT. In this script, we tested all possible EDR (*x*,*y*) positions within the Chevy. With the script, we can assume a known $\Delta\bar{v}_1^{CG\,True}$, as well as impulse centroid location. With the known (true) impulse as an input, we can then use equations (76) and (77) to simulate the exact $\Delta v_{1,x}^{A\,True}$ and $\Delta v_{1,y}^{A\,True}$ values expected at a given point within the vehicle. Starting with these expected true $\Delta v_{1,x}^{A\,True}$ and $\Delta v_{1,y}^{A\,True}$, we can probe the effects introduced by EDR measurement error by modifying them such that $\Delta v_{1,x}^{A\,True} \to \Delta v_{1,x}^{A\,Est}$ and $\Delta v_{1,y}^{A\,True} \to \Delta v_{1,y}^{A\,Est}$, where our modifications are based on behavior observed in staged collision data (see Figure 19). The $\Delta v_{1,x}^{A\,Est}$ and $\Delta v_{1,y}^{A\,Est}$ values can then be used to calculate $\Delta\bar{v}_1^{CG\,Est}$. This is done for each point within the vehicle. Using this framework, we can also study the effect of adding corrections back to $\Delta v_{1,x}^{A\,Est}$ and $\Delta v_{1,y}^{A\,Est}$ to account for thresholding effects and offsets. After adding corrections, we can then apply equations (81) and (82) to obtain our estimates for $\Delta v_{1,x}^{CG\,Est}$ and $\Delta v_{1,y}^{CG\,Est}$ as a function of EDR position. The results introduced below assume: $\Delta v_{1,x}^{CG\,True} = 1.04$ mph and $\Delta v_{1,y}^{CG\,True} = -6.3$ mph.

Figure 27 shows $\Delta v_{1,x}^{CG\,Est}$ and $\Delta v_{1,y}^{CG\,Est}$ assuming no modifications to $\Delta\bar{v}_1^{A}$ and with no corrections applied. As expected, the $\Delta v_{1,x}^{CG\,Est}$ and $\Delta v_{1,y}^{CG\,Est}$ values are constant and independent of position, except at the $R_0$ point where $\Delta v_{1,x}^{A\,Est} = 0$ and $\Delta v_{1,y}^{A\,Est} = 0$. Note, the impulse centroid ("IC") and impulse unit vector are also depicted in the figure.

In low-speed tests of EDR performance, it has been demonstrated that for a given EDR, there is a minimum $\Delta v_{1,j}^{A\,True}$ value, $\Delta v^{Threshold}$, below which we expect $\Delta v_{1,j}^{A\,Est} \to 0$ for component *j*. $\Delta v^{Threshold}$ is both EDR dependent as well as dependent on peak acceleration and pulse width [4]. The upper-bound $\Delta v$ corridor line shown in Figure 19 is constructed assuming a worst-case scenario $\Delta v^{Threshold} = 4.4$ kph. It is easy to solve for "zero-corridors" within the *x-y* plane where either $\Delta v_{1,x}^{A\,Est}$ or $\Delta v_{1,y}^{A\,Est}$ will always equal 0 (assuming the worst-case scenario in both lateral and longitudinal directions). The $\Delta v_{1,y}^{A\,Est}$ zero-corridor is defined along the *x*-axis by:

$$\frac{-v_{1,y}^{CG\,True} - \Delta v^{Threshold}}{\Delta\omega_1} \leq r_{1,x}^{A} \leq \frac{-\Delta v_{1,y}^{CG\,True} + \Delta v^{Threshold}}{\Delta\omega_1}$$

Within this corridor, $\Delta v_{1,y}^{A\,Est} = 0$ for any $r_{1,y}^{A}$. Similarly, the $\Delta v_{x}^{A\,Est}$ zero-corridor is given by:

$$\frac{v_{1,x}^{CG\,True} - \Delta v^{Threshold}}{\Delta\omega_1} \leq r_{1,y}^{A} \leq \frac{\Delta v_{1,x}^{CG\,True} + \Delta v^{Threshold}}{\Delta\omega_1}$$

Within this corridor, $\Delta v_{1,x}^{A\,Est} = 0$ for any $r_{1,x}^{A}$.

Figure 28 illustrates zero corridors using the upper-bound $\Delta v$ corridor line condition where $\Delta v^{Threshold} = 4.4$ kph.

The intersection of these two zero-corridors defines a "zero box" whose sides are given by $2 \cdot \Delta v^{Threshold}/\Delta\omega_1$. Within this box, we are guaranteed to have both $\Delta v_{1,x}^{A\,Est} = 0$ and $\Delta v_{1,y}^{A\,Est} = 0$ which implies our calculations must yield $\Delta v_{1,x}^{CG\,Est} = 0$ and $\Delta v_{1,y}^{CG\,Est} = 0$ for EDRs within this box.

Figure 29 shows $\Delta v_{x}^{CG\,Est}$ and $\Delta v_{y}^{CG\,Est}$ assuming the $\Delta v_{1,x}^{A\,Est}$ and $\Delta v_{1,y}^{A\,Est}$ values were adjusted to follow the upper bound of the $\Delta v$ corridor ($\Delta v^{Threshold} = 4.4$ kph) and no corrections applied. Because of the threshold, the $\Delta v_{x}^{CG\,Est}$ and $\Delta v_{y}^{CG\,Est}$ values are dependent on position. All points within the white box, including the $R_0$ point, have $\Delta v_{x}^{A\,Est} = 0$ and $\Delta v_{y}^{A\,Est} = 0$; this yields the trivial solution: $\Delta v_{x}^{CG\,Est} = 0$ and $\Delta v_{y}^{CG\,Est} = 0$. For this plot, we assume $\Delta v_{1,x}^{CG\,True} = 1.31$ mph and $\Delta v_{1,y}^{CG\,True} = -6.14$ mph.

Figure 30 shows $\Delta v_{x}^{CG\,Est}$ and $\Delta v_{y}^{CG\,Est}$ assuming the $\Delta v_{1,x}^{A\,Est}$ and $\Delta v_{1,y}^{A\,Est}$ values were adjusted to follow the upper bound of the $\Delta v$ corridor ($\Delta v^{Threshold} = 4.4$ kph) and corrections applied. Because $\Delta v_{x}^{A\,Est} \to 0$ and $\Delta v_{y}^{A\,Est} \to 0$ for values below 4.4 kph, the correction applied in this region takes $0$ kph $\to$ 2.9 kph.

It is evident from Figure 27, Figure 29, and Figure 30 that near the $R_0$ line, $\Delta\bar{v}^{CG\,Est}$ is extremely sensitive to measurement inaccuracies of $\Delta\bar{v}^{A\,Est}$ introduced by the measuring device. Even in cases where the exact correction is known, threshold effects, which cause low $\Delta v_{x}^{A\,Est} \to 0$ and low $\Delta v_{y}^{A\,Est} \to 0$, when corrected for, will still yield problematic regions near the $R_0$ line where $|\Delta\bar{v}^{CG\,Est}|$ can tend toward extremely large values, and thus will greatly increase estimate uncertainties. This is an irreducible effect that the analyst should be aware of.

In a recent study on combined EDR lateral and longitudinal $\Delta v$ accuracy, it was demonstrated that for the 2012 EDRs tested, lateral $\Delta v$ error was symmetric about 0 and less than 10% [18]. Figure 31 illustrates the combined effect of reducing longitudinal $\Delta v$ (using our corridor) while leaving lateral $\Delta v$ unchanged.

Finally, to illustrate the dependence on PDOF, Figure 32 illustrates the changing behavior of $\Delta v_{x}^{CG\,Est}$ and $\Delta v_{y}^{CG\,Est}$ versus various PDOFs assuming $|\Delta\bar{v}| = 5$ mph.

**Conclusions**

Using a 2D rigid-body dynamics approach, we have created a mathematical model which allows one to transform change-in-velocity estimates at any position within a vehicle to the center-of-gravity equivalent value. We have demonstrated the method by reproducing experimentally measured change-in-velocity values from a staged collision. We have also demonstrated the possibility of reconstructing pre-impact ground speeds with the Virtual CRASH simulation.

**About the Authors**

Bob Scurlock, Ph.D., ACTAR, is a Research Associate at the University of Florida, Department of Physics. He is also the owner Scurlock Scientific Services, LLC, an accident reconstruction consulting firm based out of Gainesville, Florida, USA. He is also the CEO of Virtual CRASH, LLC. He can be reached at Bob@ScurlockSci.com.
Andrew Rich, BSME, ACTAR, is the owner of Rich Consulting LLC, an accident reconstruction consulting firm based out of Fairlawn, Ohio, USA. He can be reached at: andy@rich-llc.com.

Kyle Poe is an undergraduate research assistant in Physics at the University of Florida.

**Acknowledgements**

The authors thank the 2018 IPTM crash team for their hard work putting together this crash test. They would also like to thank Jeremy Daily, Ph.D. for the use of his instrumentation.

| Instrument | X (ft) | Y (ft) | $\Delta v_x$ (mph) | $\Delta v_y$ (mph) | Remarks |
|---|---|---|---|---|---|
| 250G Accel. | 0.00 | 0.00 | 1.29 | -3.94 | At $CG_{xy}$ position |
| 250G Accel. | 2.83 | 0.31 | N/A | N/A | In engine compartment |
| EDR A | -7.80 | -1.50 | 5.09 | -24.19 | In trunk near impulse |
| EDR B | -6.98 | -1.50 | 5.09 | -21.01 | In trunk near impulse |
| EDR C | -0.67 | 0.00 | 0.64 | -6.37 | Just behind $CG_{xy}$ |
| EDR D | 0.30 | 0.73 | 0.0 | 5.73 | Passenger-front floor pan |

**Table 1: Location of instruments and cumulative $\Delta v$ values.**

| Measurement | Contact Damage Only (in) | Contact and Induced Damage (in) |
|---|---|---|
| Indentation Length (L) | 23.0 | 38.0 |
| Offset (D) | -91.5 | -84.0 |
| C1 | 13.0 | 13.0 |
| C2 | 16.5 | 16.5 |
| C3 | 15.5 | 13.0 |
| C4 | 14.0 | 12.0 |
| C5 | 12.0 | 3.0 |
| C6 | 8.0 | 0.0 |
| Calculated Damage Centroid Longitudinal Position | -92.4 | -89.2 |
| Calculated Damage Centroid Lateral Position | 27.0 | 27.5 |

**Table 2: Chevy damage profile hand measurements**

| Input Parameter | Best Estimate | Uncertainty | Uncertainty Basis |
|---|---|---|---|
| LeSabre Pre-impact Heading | -90 degrees | ±5 degrees | Video analysis |
| $I_{Chevy}^{Yaw}$ | 1836.7 slug-ft$^2$ | ±4.8% | [14] |
| $I_{Buick}^{Yaw}$ | 2683.3 slug-ft$^2$ | ±4.8% | [14] |
| $(r_{Chevy,x}^{EDRA}, r_{Chevy,y}^{EDRA})$ in Chevy frame | $(-7.8$ ft , $-1.5$ ft) | ($\pm 4$ in , $\pm 2$ in) | Geometrical size of EDR |
| $(r_{Chevy,x}^{EDRB}, r_{Chevy,y}^{EDRB})$ in Chevy frame | $(-6.98$ ft , $-1.5$ ft) | ($\pm 4$ in , $\pm 2$ in) | Geometrical size of EDR |
| $(r_{Chevy,x}^{EDRC}, r_{Chevy,y}^{EDRC})$ in Chevy frame | $(-0.67$ ft , 0 ft) | ($\pm 4$ in , $\pm 2$ in) | Geometrical size of EDR |
| $(r_{Chevy,x}^{EDRD}, r_{Chevy,y}^{EDRD})$ in Chevy frame | (0.3 ft , 0.73 ft) | ($\pm 4$ in , $\pm 2$ in) | Geometrical size of EDR |
| $(r_{Chevy,x}, r_{Chevy,y})$ in Chevy frame | $(-8.2$ ft , 1.46 ft) | ($\pm 4.6$ in , $\pm$ 1 in) | Sampling distance / Measuring uncertainty |
| $(r_{Buick,x}, r_{Buick,y})$ in Buick frame | (7.05 ft , 1.3 ft) | (Range = $-1$ ft to 0 ft , 0 ft) | Measuring uncertainty |
| $\varepsilon$ | 15% | ±10% | [15,16,17] |
| $(\Delta v_{Chevy,x}^{EDRA}, \Delta v_{Chevy,y}^{EDRA})$ in Chevy frame | (5.09 mph , $-24.19$ mph) | See discussion in text. | See discussion in text. |
| $(\Delta v_{Chevy,x}^{EDRB}, \Delta v_{Chevy,y}^{EDRB})$ in Chevy frame | (5.09 mph , $-21.01$ mph) | See discussion in text. | See discussion in text. |
| $(\Delta v_{Chevy,x}^{EDRC}, \Delta v_{Chevy,y}^{EDRC})$ in Chevy frame | (0.64 mph , $-6.37$ mph) | See discussion in text. | See discussion in text. |
| $(\Delta v_{Chevy,x}^{EDRD}, \Delta v_{Chevy,y}^{EDRD})$ in Chevy frame | (0.0 mph , $-5.73$ mph) | See discussion in text. | See discussion in text. |

**Table 3: Inputs to Monte Carlo script.**

| Source | $\Delta v_{Chevy,x}^{CG\ True}$ | $\Delta v_{Chevy,y}^{CG\ True}$ |
|---|---|---|
| Accelerometer (local) | $1.04_{-0.28}^{+0.28}$ mph | $-6.3_{-0.25}^{+0.25}$ mph |
| Accelerometer (global) | $1.31_{-0.27}^{+0.27}$ mph | $-6.14_{-0.27}^{+0.27}$ mph |

**Table 4: EDR-based $\Delta v$ estimates at Chevy CG (Equations 81 and 82).**

| Source | $\Delta v_{Chevy,x}^{CG\ Est}$ | $\Delta v_{Chevy,y}^{CG\ Est}$ |
|---|---|---|
| EDR A | $1.62_{-0.97}^{+0.90}$ mph | $-6.17_{-0.67}^{+0.62}$ mph |
| EDR B | $1.83_{-0.61}^{+0.55}$ mph | $-5.85_{-0.67}^{+0.62}$ mph |
| EDR C | $0.64_{-0.40}^{+0.39}$ mph | $-5.07_{-0.64}^{+0.61}$ mph |
| EDR D | $1.76_{-0.66}^{+1.05}$ mph | $-6.45_{-1.32}^{+0.80}$ mph |

**Table 5: EDR-based local $\Delta v$ estimates at Chevy CG (Equations 81 and 82).**

| Source | $\delta\Delta v_{Chevy,x}^{CG,Best}$ | $\delta\Delta v_{Chevy,y}^{CG,Best}$ |
|---|---|---|
| EDR A | 0.31 mph | 0.03 mph |
| EDR B | 0.52 mph | 0.29 mph |
| EDR C | −0.67 mph | 1.07 mph |
| EDR D | 0.45 mph | −0.31 mph |

**Table 6: Difference between best estimate $\Delta v$ and accelerometer value.**

| Source | Closing-speed |
|---|---|
| EDR A | $25.82^{+7.21}_{-3.57}$ mph |
| EDR B | $24.56^{+6.70}_{-3.89}$ mph |
| EDR C | $21.44^{+8.51}_{-4.94}$ mph |
| EDR D | $27.07^{+15.38}_{-7.48}$ mph |

**Table 7: EDR-based closing-speed estimates.**

| Source | Best Closing-speed Difference |
|---|---|
| EDR A | $-1.18$ mph |
| EDR B | $-2.44$ mph |
| EDR C | $-5.56$ mph |
| EDR D | 0.02 mph |

**Table 8: Difference between best EDR-based closing-speed estimates and true value.**

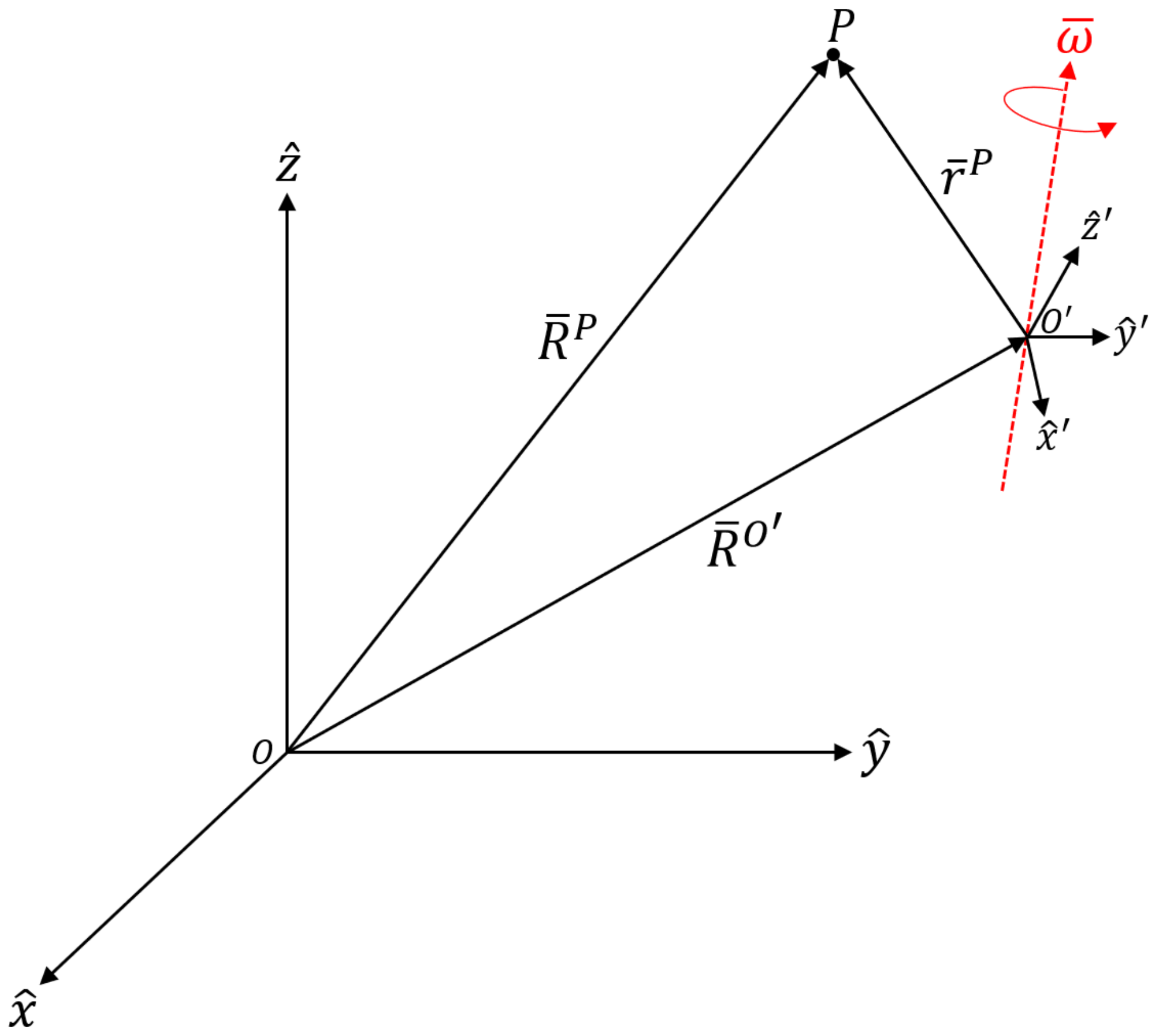

**Figure 1: Illustration of point *P* position vector in inertial and moving frames.**

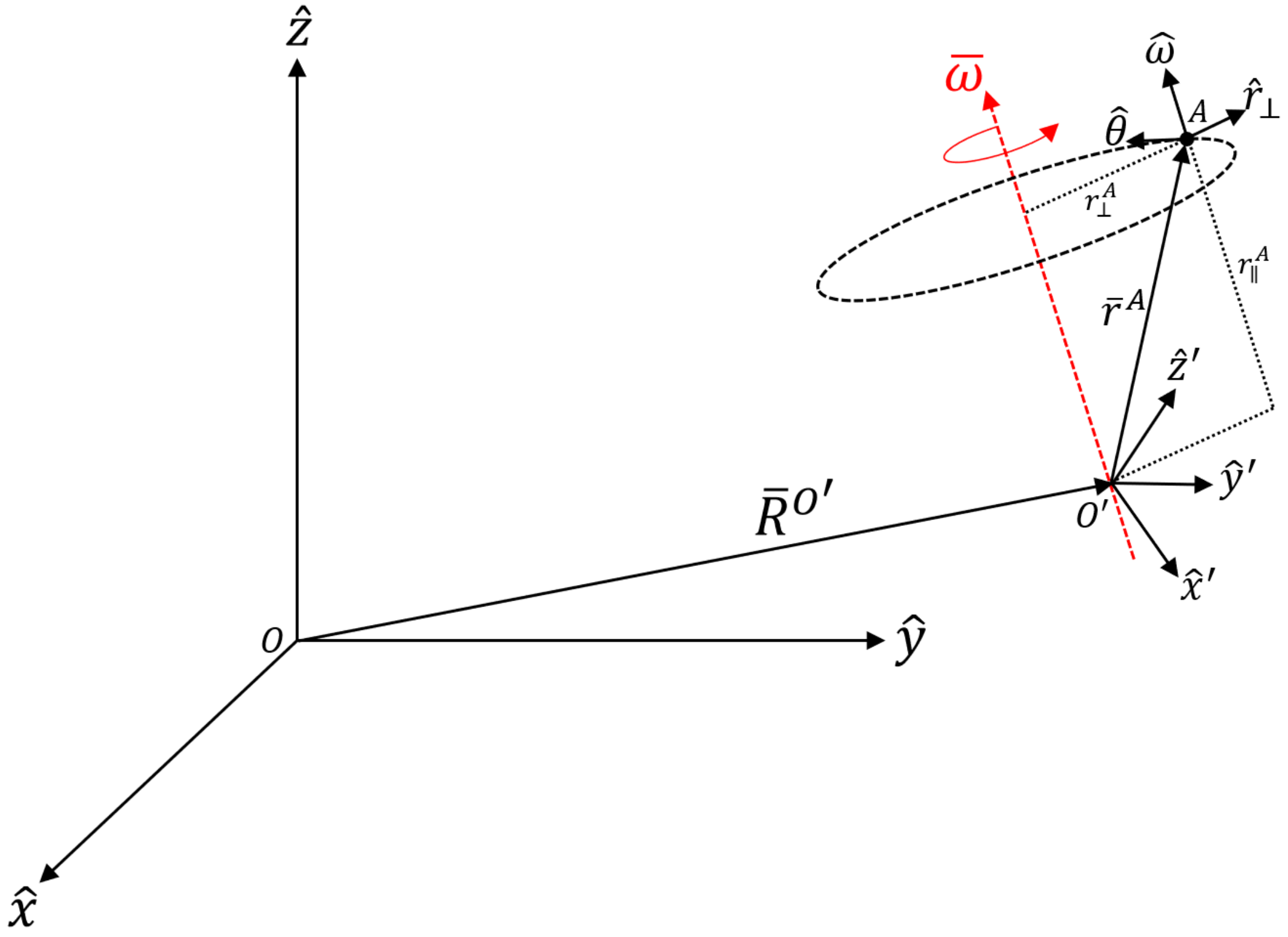

**Figure 2: Illustration of instantaneous cylindrical coordinate unit vectors at accelerometer position *A*.**

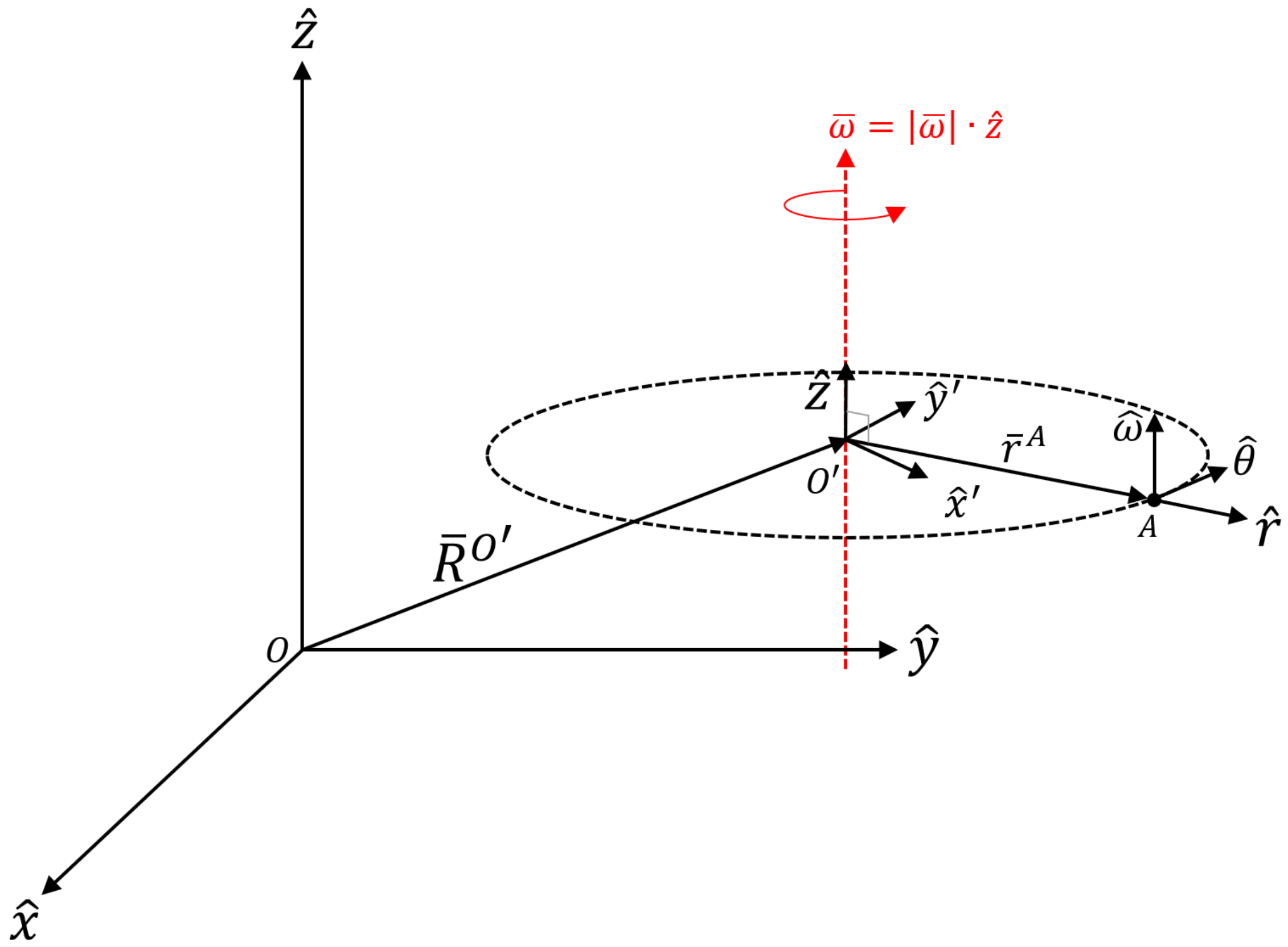

**Figure 3: Illustration of instantaneous cylindrical coordinate unit vectors at accelerometer position *A* in simplified model. Here the angular velocity vector is aligned with the global *z*-axis and $\bar{r}_k^A$ lies in the $\hat{x}' - \hat{y}'$ plane**

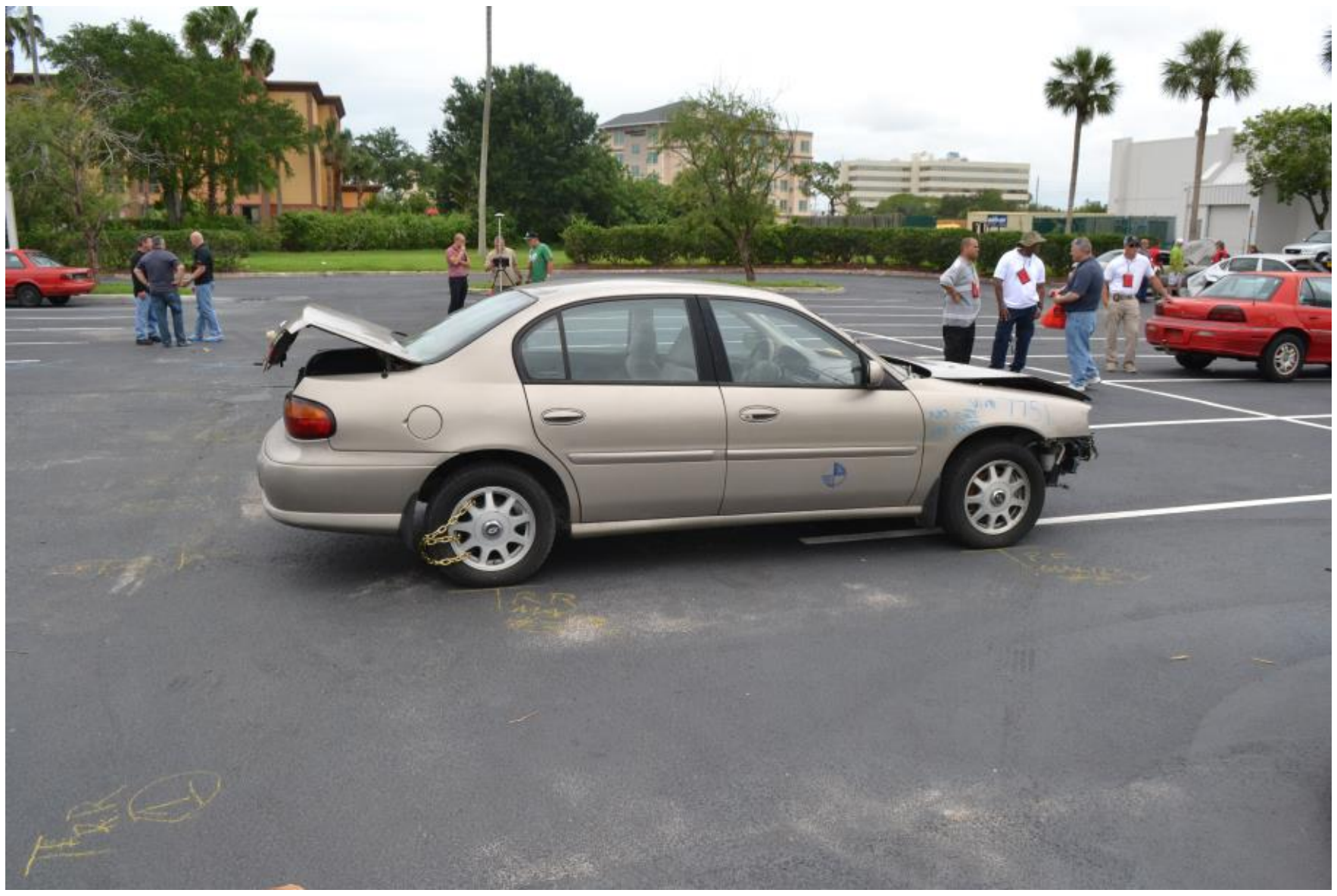

**Figure 4: Photograph of 1998 Chevy Malibu in its pre-impact configuration (target vehicle).**

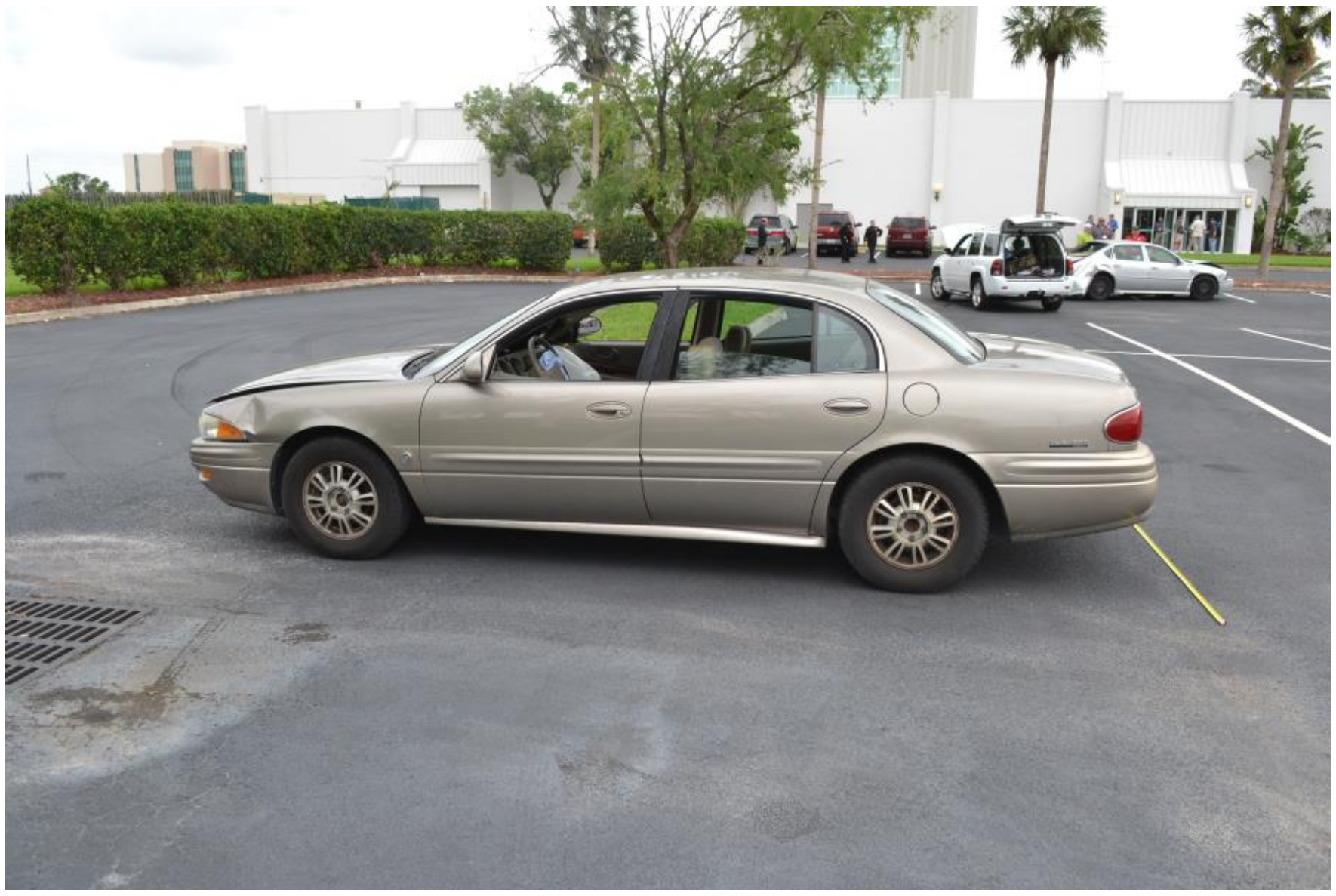

**Figure 5: Photograph of 2002 Buick LeSabre (bullet vehicle).**

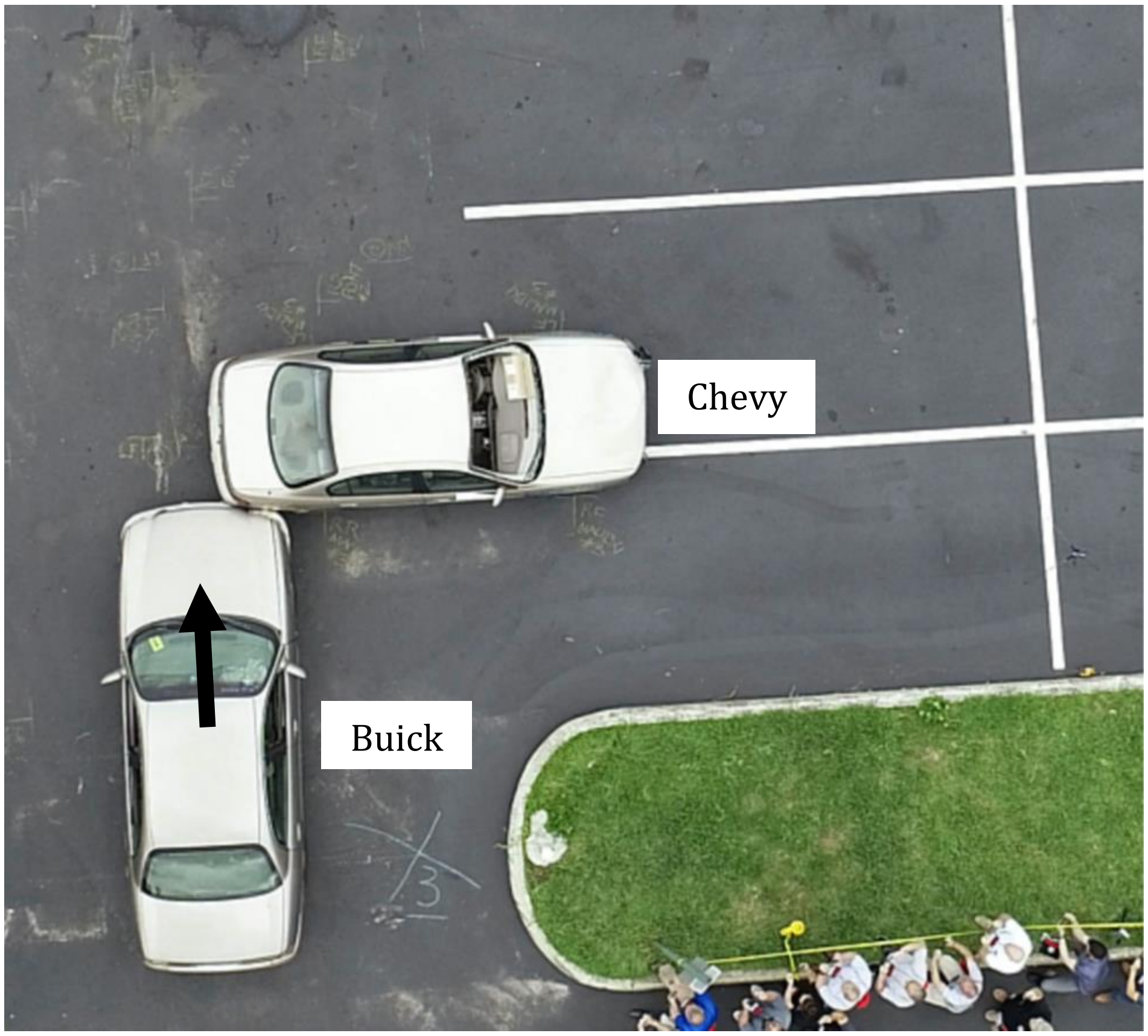

**Figure 6: Impact configuration**

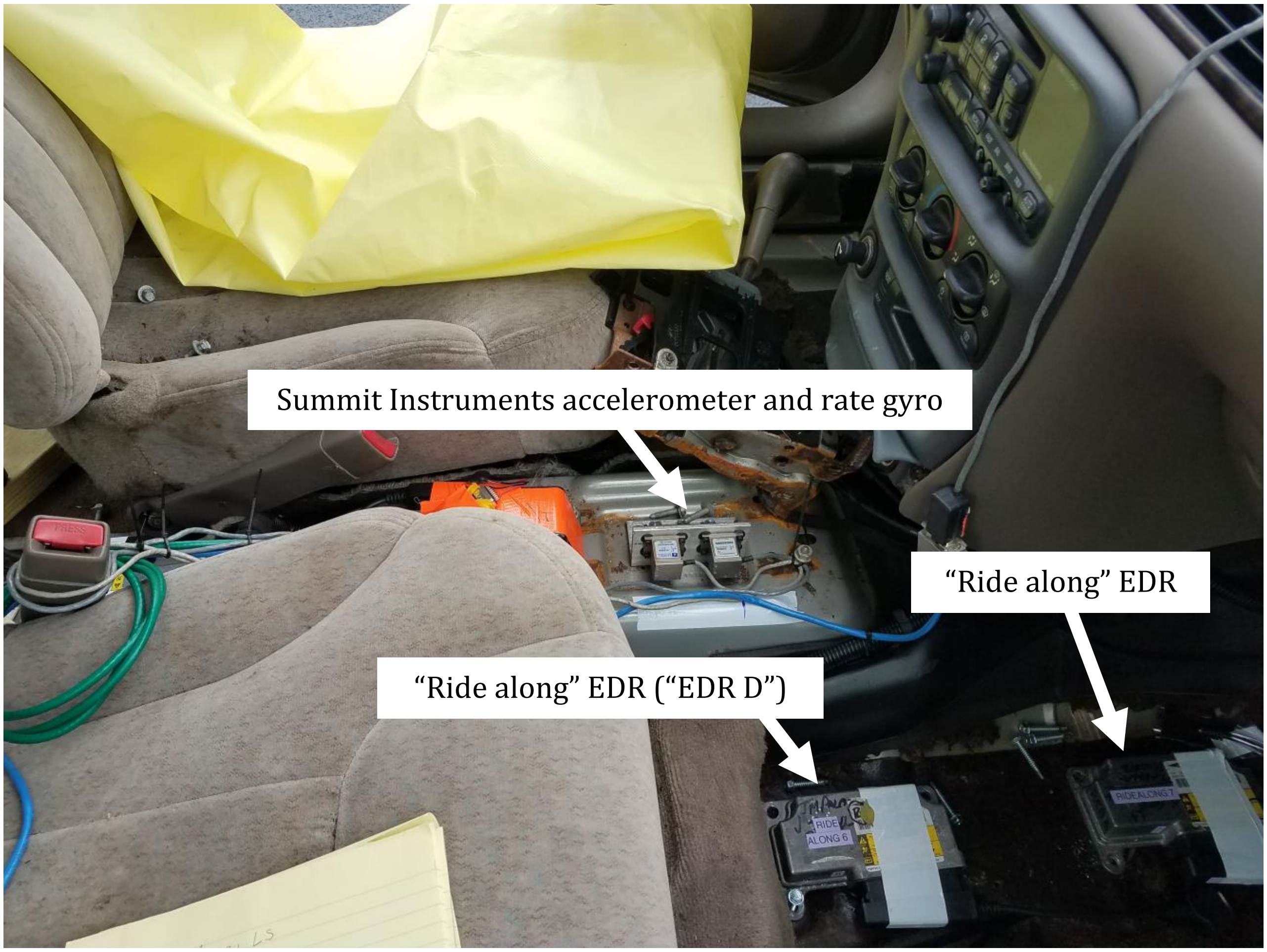

**Figure 7: Photograph showing accelerometer and two "ride along" ACMs inside Chevy Malibu cabin.**

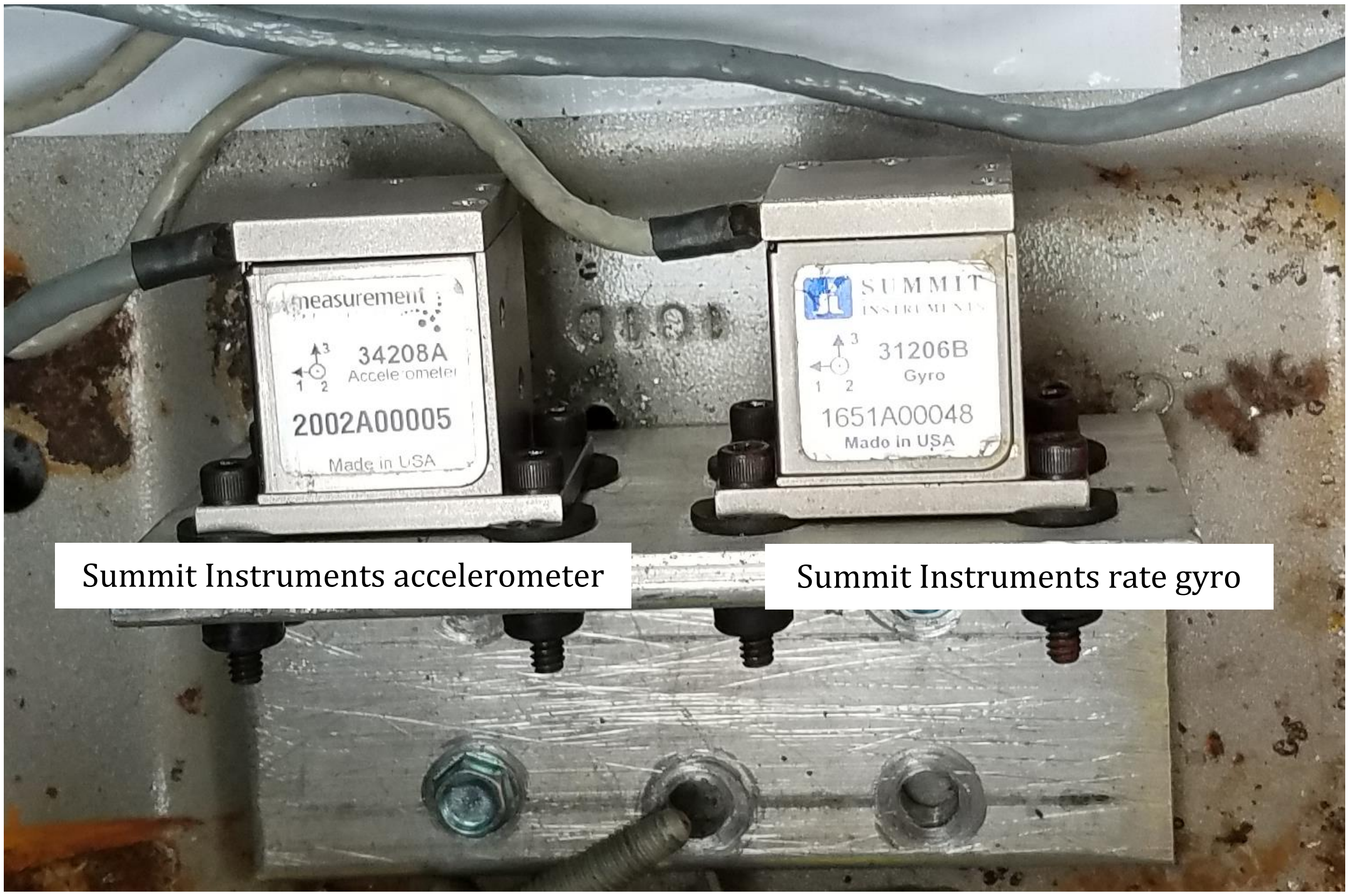

**Figure 8: Close-up view of accelerometer and rate gryo.**

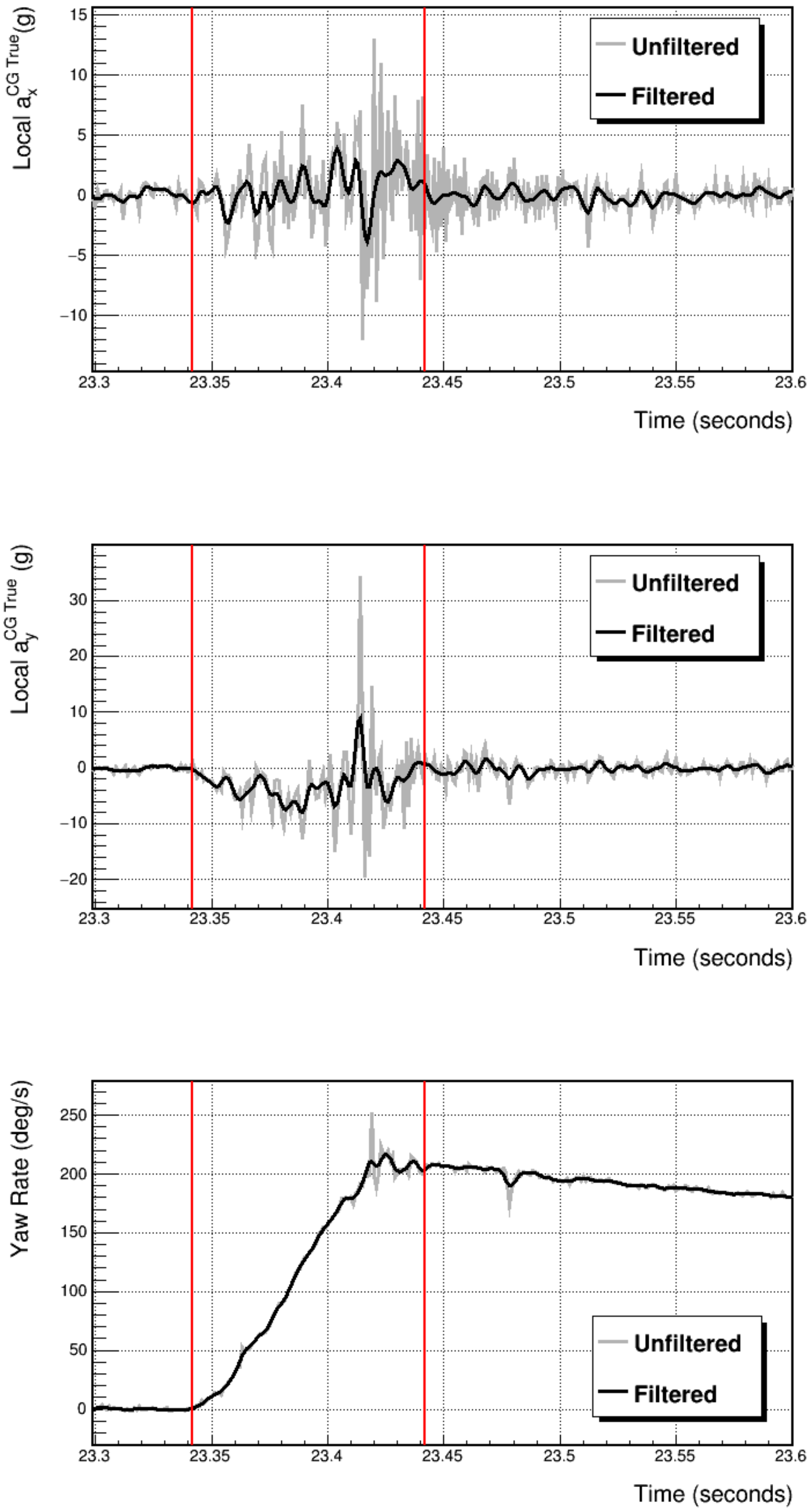

**Figure 9: Longitudinal (top) and lateral (middle) acceleration graphs from accelerometer as well as yaw rate (bottom). Black lines illustrate acceleration with CFC60 Butterworth filter applied, while gray shows unfiltered data.**

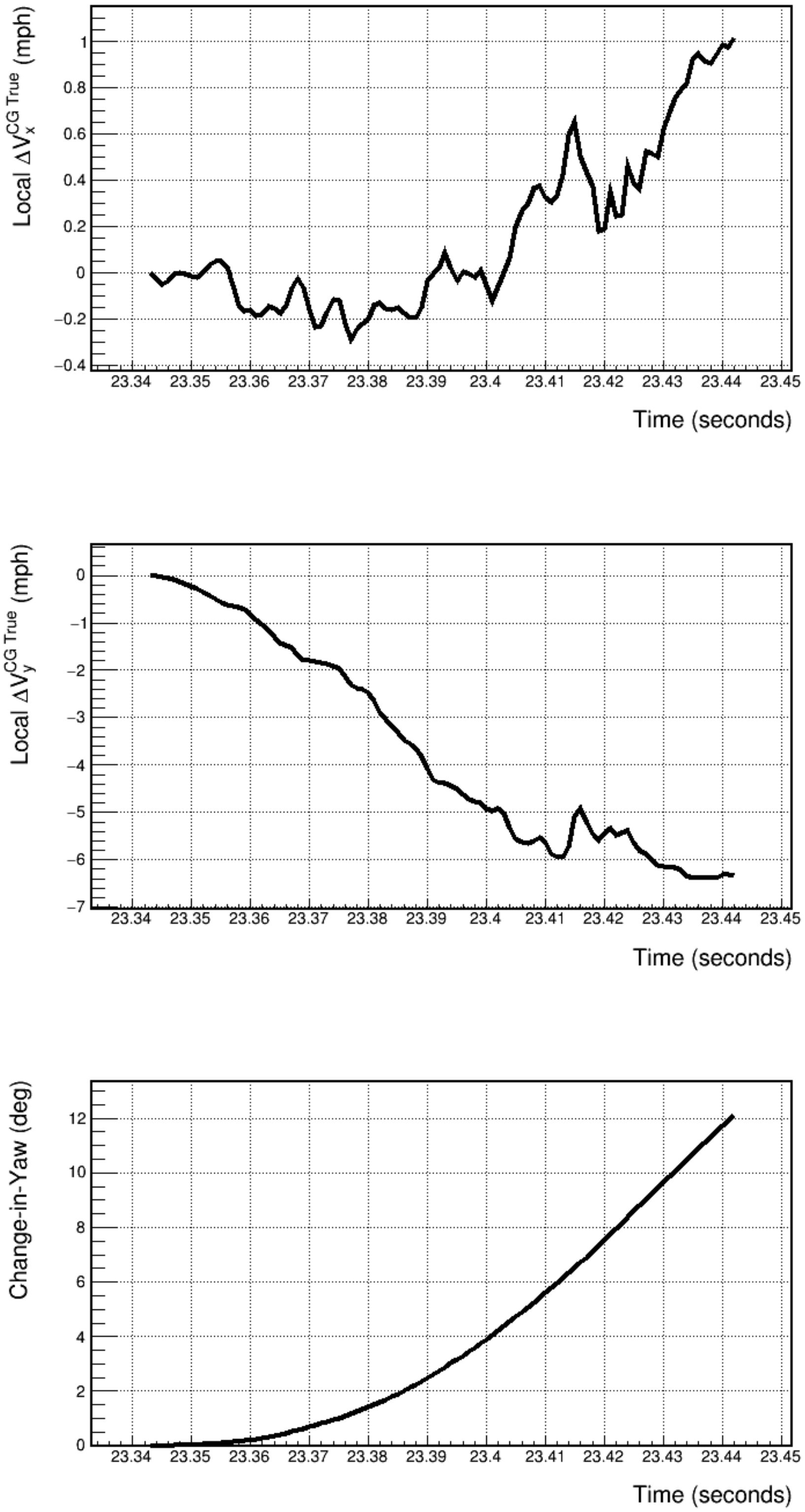

**Figure 10: Cumulative longitudinal (top) and lateral (middle) change-in-velocity graphs, as well as change-in-yaw (bottom).**

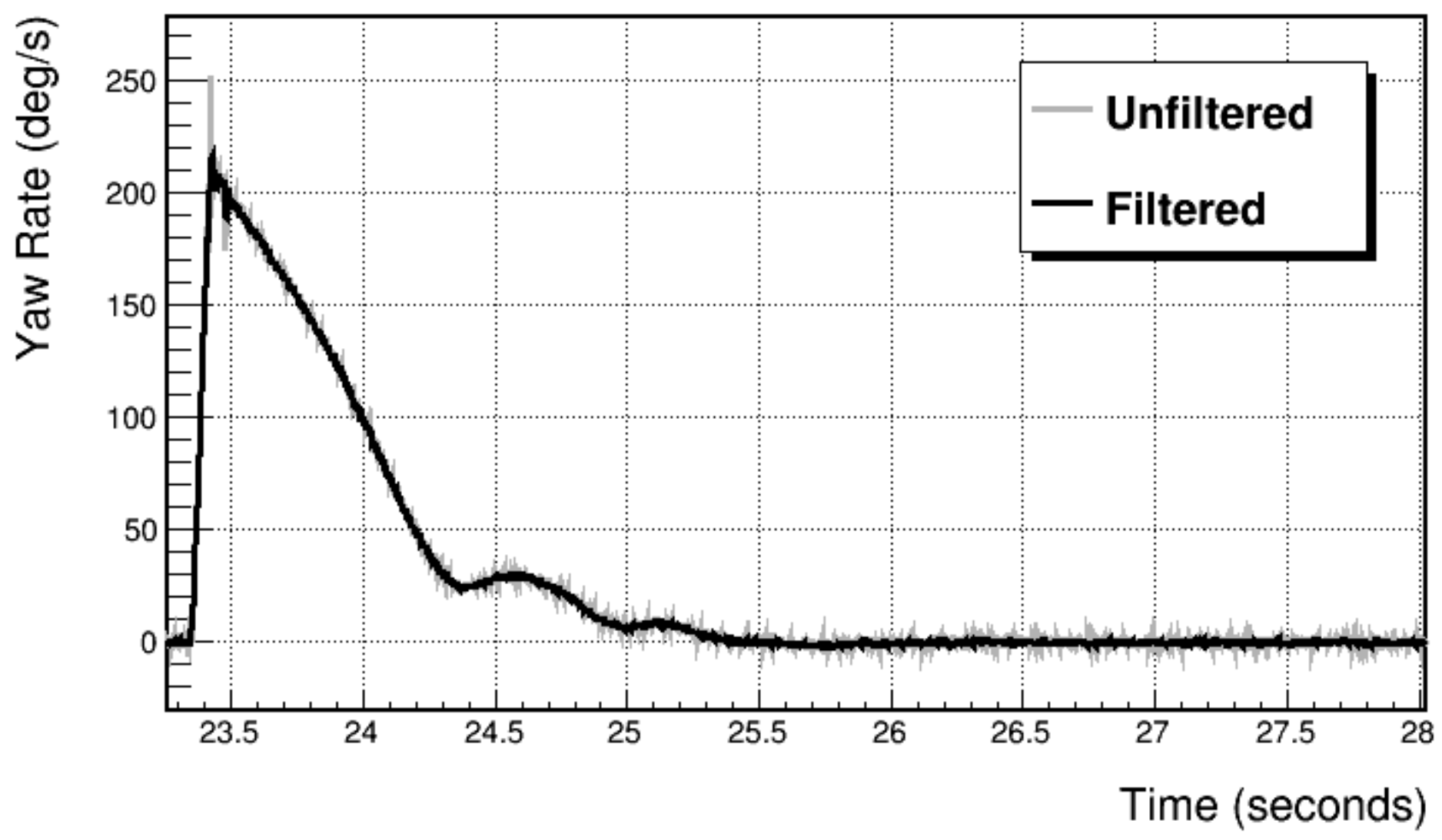

**Figure 11: Yaw rate versus time reported by rate gyro. Gray shows unfiltered data and black shows 60CFC Butterworth filter applied.**

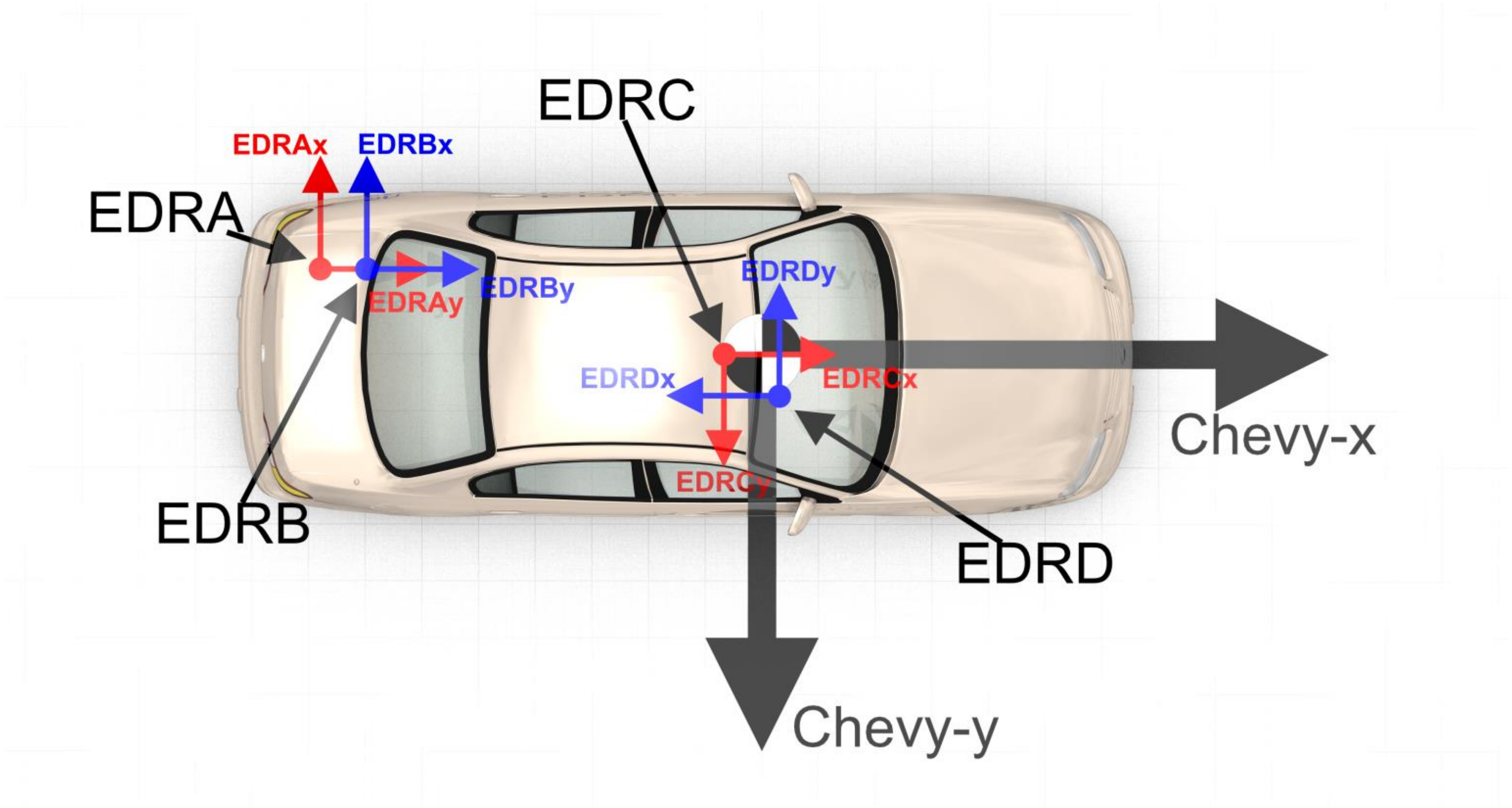

**Figure 12: Positions of EDRs A, B, C, and D.**

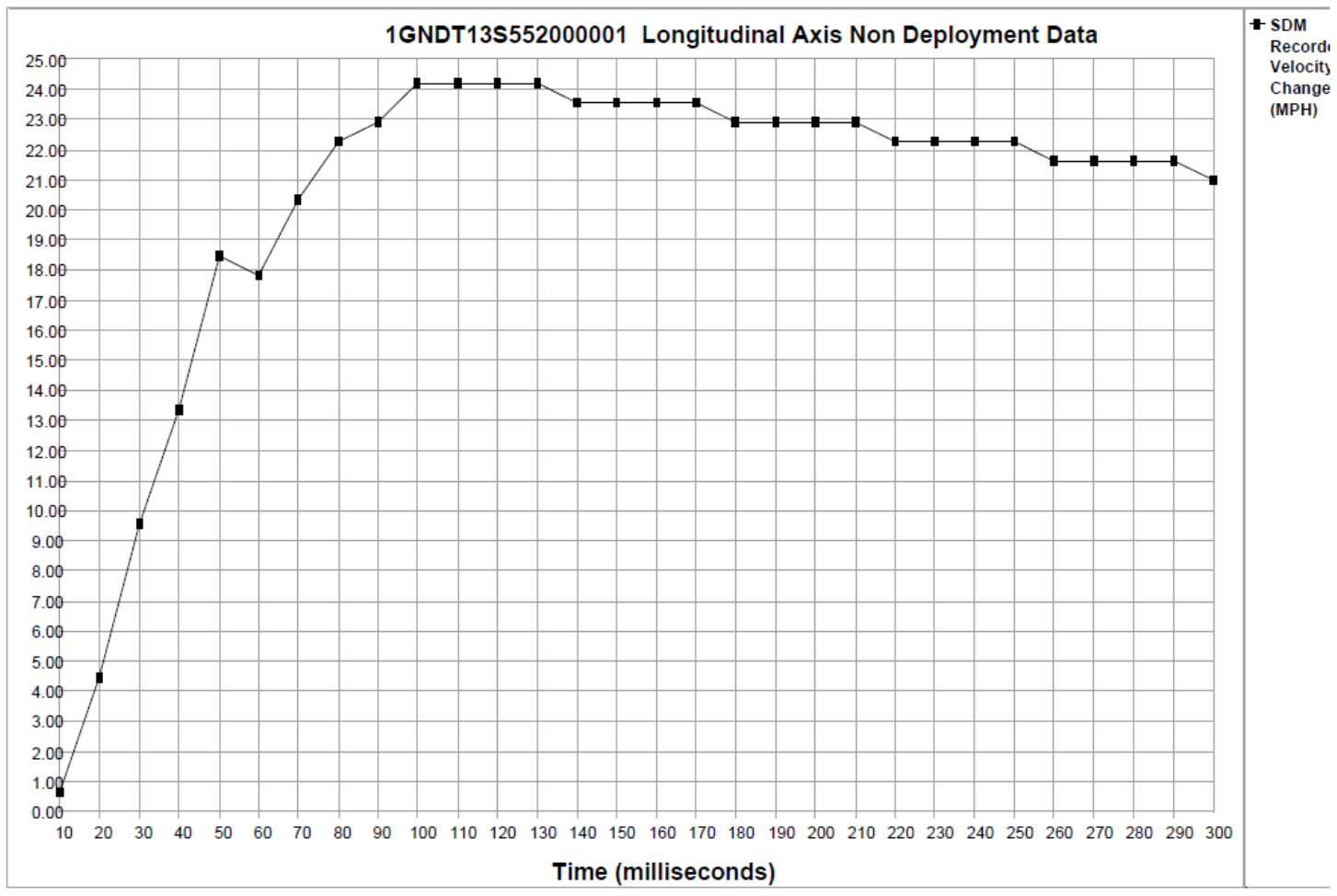

| | | | | | | | | | | | | | | | |
|---|---|---|---|---|---|---|---|---|---|---|---|---|---|---|---|
| Time (milliseconds) | 10 | 20 | 30 | 40 | 50 | 60 | 70 | 80 | 90 | 100 | 110 | 120 | 130 | 140 | 150 |
| SDM Longitudinal Axis Recorded Velocity Change (MPH) | 0.64 | 4.46 | 9.55 | 13.37 | 18.46 | 17.82 | 20.37 | 22.28 | 22.92 | 24.19 | 24.19 | 24.19 | 24.19 | 23.55 | 23.55 |
| Time (milliseconds) | 160 | 170 | 180 | 190 | 200 | 210 | 220 | 230 | 240 | 250 | 260 | 270 | 280 | 290 | 300 |
| SDM Longitudinal Axis Recorded Velocity Change (MPH) | 23.55 | 23.55 | 22.92 | 22.92 | 22.92 | 22.92 | 22.28 | 22.28 | 22.28 | 22.28 | 21.64 | 21.64 | 21.64 | 21.64 | 21.01 |

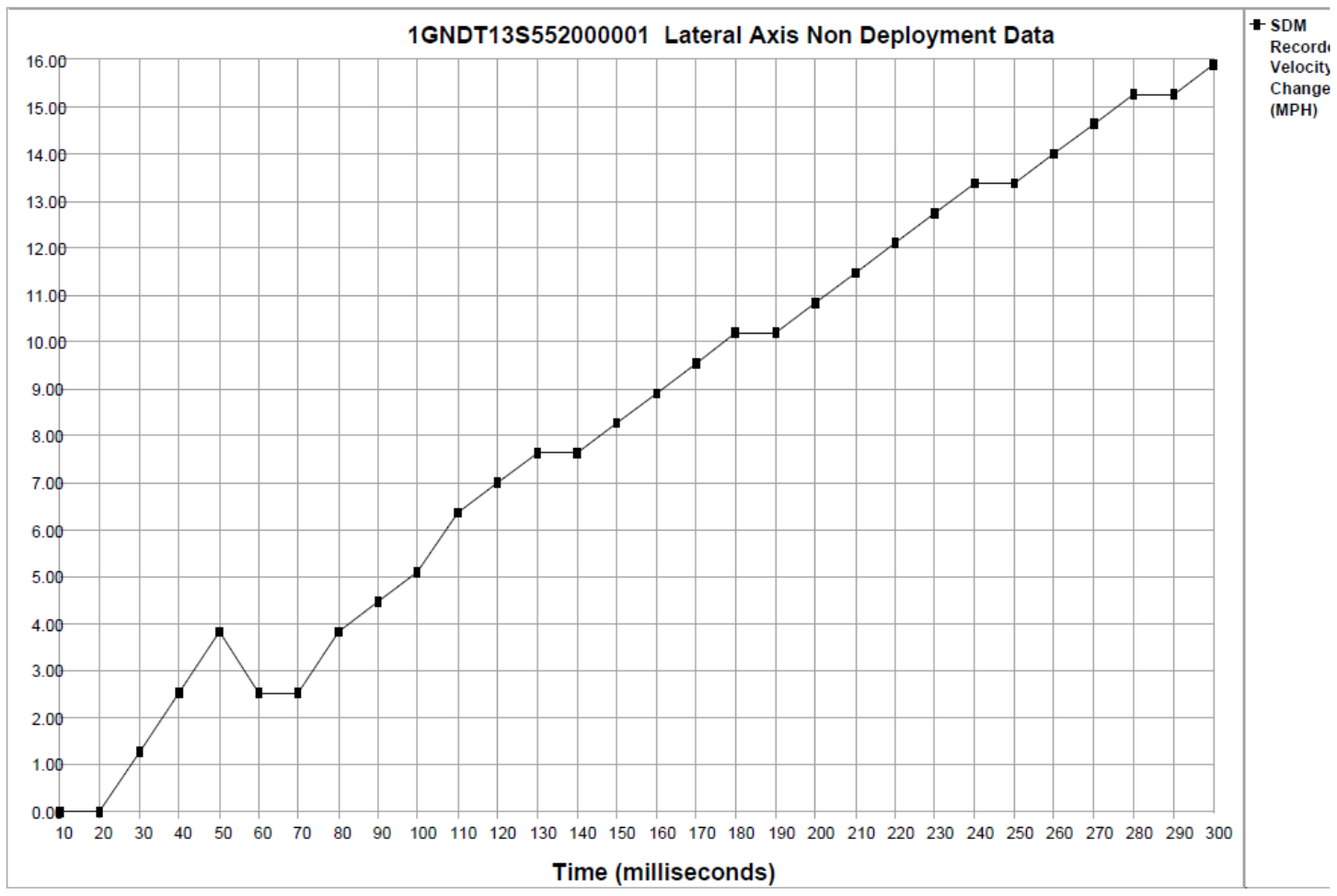

| | | | | | | | | | | | | | | | |
|---|---|---|---|---|---|---|---|---|---|---|---|---|---|---|---|
| Time (milliseconds) | 10 | 20 | 30 | 40 | 50 | 60 | 70 | 80 | 90 | 100 | 110 | 120 | 130 | 140 | 150 |
| SDM Lateral Axis Recorded Velocity Change (MPH) | 0.00 | 0.00 | 1.27 | 2.55 | 3.82 | 2.55 | 2.55 | 3.82 | 4.46 | 5.09 | 6.37 | 7.00 | 7.64 | 7.64 | 8.28 |
| Time (milliseconds) | 160 | 170 | 180 | 190 | 200 | 210 | 220 | 230 | 240 | 250 | 260 | 270 | 280 | 290 | 300 |
| SDM Lateral Axis Recorded Velocity Change (MPH) | 8.91 | 9.55 | 10.19 | 10.19 | 10.82 | 11.46 | 12.10 | 12.73 | 13.37 | 13.37 | 14.01 | 14.64 | 15.28 | 15.28 | 15.91 |

**Figure 13: Data from EDR A obtained using CDR kit. Upper graph and table show longitudinal change-in-velocity (in EDR A's frame). The bottom graph and table show lateral change-in-velocity (in EDR A's frame).**

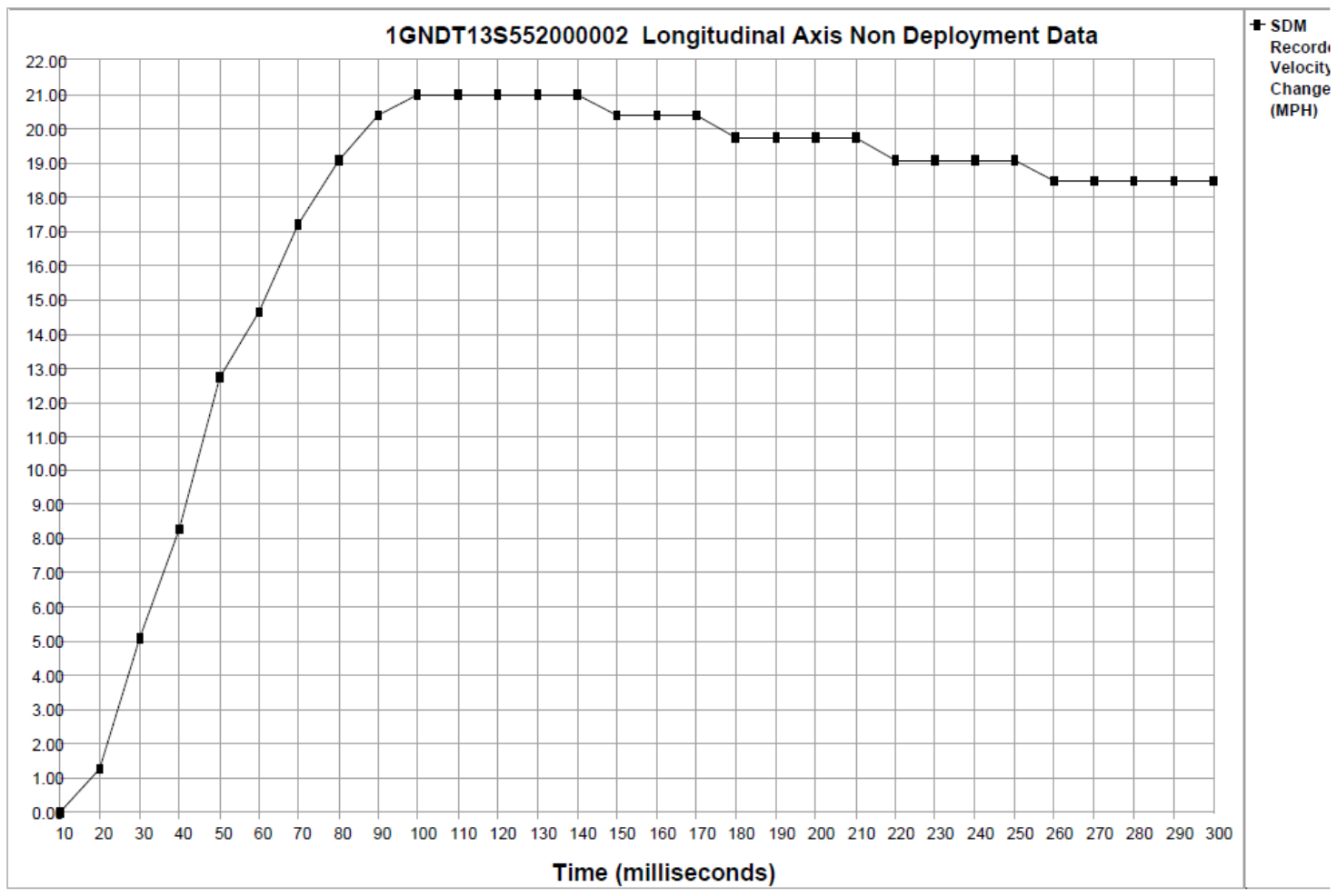

| | | | | | | | | | | | | | | | |
|---|---|---|---|---|---|---|---|---|---|---|---|---|---|---|---|
| Time (milliseconds) | 10 | 20 | 30 | 40 | 50 | 60 | 70 | 80 | 90 | 100 | 110 | 120 | 130 | 140 | 150 |
| SDM Longitudinal Axis Recorded Velocity Change (MPH) | 0.00 | 1.27 | 5.09 | 8.28 | 12.73 | 14.64 | 17.19 | 19.10 | 20.37 | 21.01 | 21.01 | 21.01 | 21.01 | 21.01 | 20.37 |
| Time (milliseconds) | 160 | 170 | 180 | 190 | 200 | 210 | 220 | 230 | 240 | 250 | 260 | 270 | 280 | 290 | 300 |
| SDM Longitudinal Axis Recorded Velocity Change (MPH) | 20.37 | 20.37 | 19.73 | 19.73 | 19.73 | 19.73 | 19.10 | 19.10 | 19.10 | 19.10 | 18.46 | 18.46 | 18.46 | 18.46 | 18.46 |

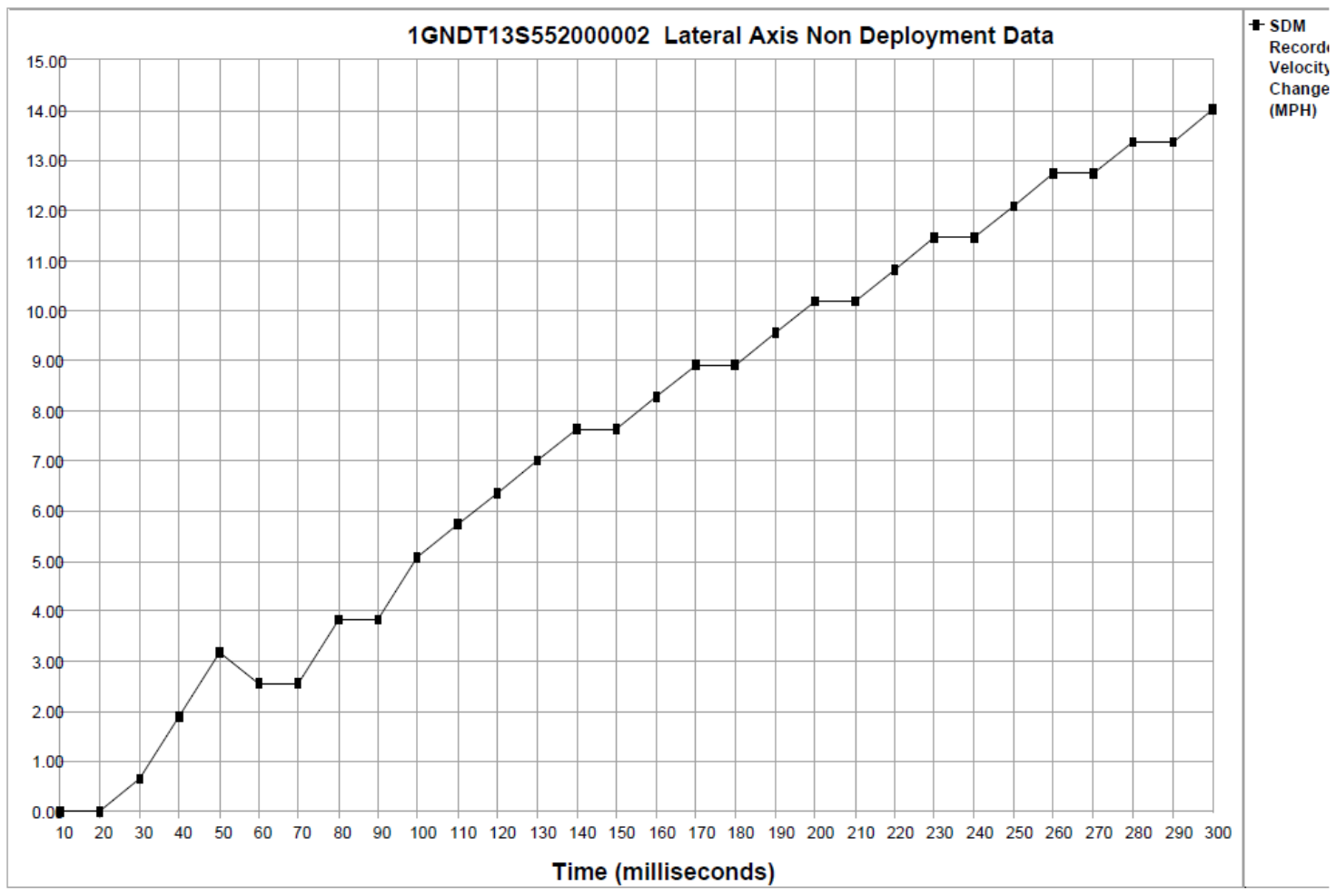

| | | | | | | | | | | | | | | | |
|---|---|---|---|---|---|---|---|---|---|---|---|---|---|---|---|
| Time (milliseconds) | 10 | 20 | 30 | 40 | 50 | 60 | 70 | 80 | 90 | 100 | 110 | 120 | 130 | 140 | 150 |
| SDM Lateral Axis Recorded Velocity Change (MPH) | 0.00 | 0.00 | 0.64 | 1.91 | 3.18 | 2.55 | 2.55 | 3.82 | 3.82 | 5.09 | 5.73 | 6.37 | 7.00 | 7.64 | 7.64 |
| Time (milliseconds) | 160 | 170 | 180 | 190 | 200 | 210 | 220 | 230 | 240 | 250 | 260 | 270 | 280 | 290 | 300 |
| SDM Lateral Axis Recorded Velocity Change (MPH) | 8.28 | 8.91 | 8.91 | 9.55 | 10.19 | 10.19 | 10.82 | 11.46 | 11.46 | 12.10 | 12.73 | 12.73 | 13.37 | 13.37 | 14.01 |

**Figure 14: Data from EDR B obtained using CDR kit. Upper graph and table show longitudinal change-in-velocity (in EDR B's frame). The bottom graph and table show lateral change-in-velocity (in EDR B's frame).**

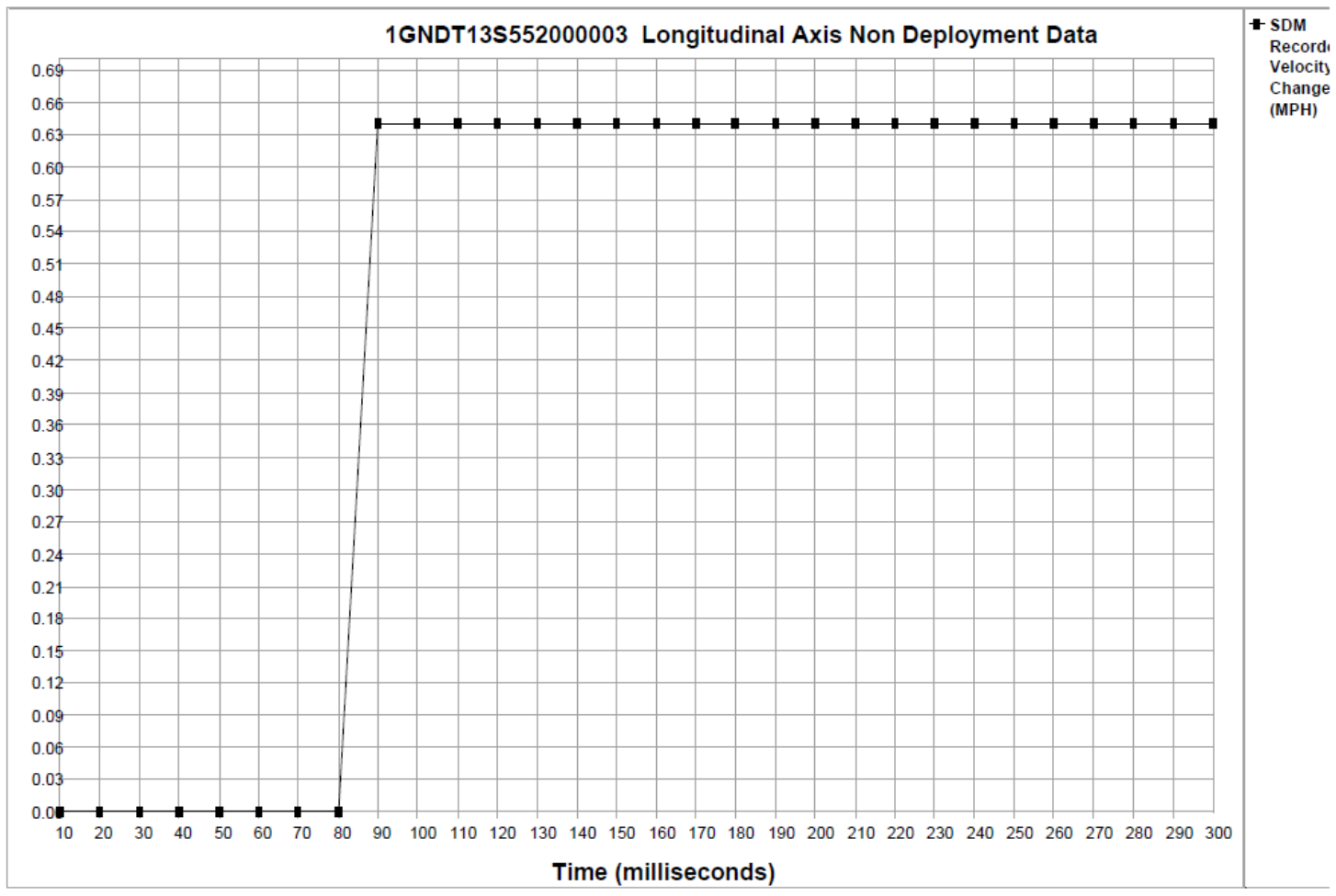

| Time (milliseconds) | 10 | 20 | 30 | 40 | 50 | 60 | 70 | 80 | 90 | 100 | 110 | 120 | 130 | 140 | 150 |
|---|---|---|---|---|---|---|---|---|---|---|---|---|---|---|---|
| SDM Longitudinal Axis Recorded Velocity Change (MPH) | 0.00 | 0.00 | 0.00 | 0.00 | 0.00 | 0.00 | 0.00 | 0.00 | 0.64 | 0.64 | 0.64 | 0.64 | 0.64 | 0.64 | 0.64 |
| Time (milliseconds) | 160 | 170 | 180 | 190 | 200 | 210 | 220 | 230 | 240 | 250 | 260 | 270 | 280 | 290 | 300 |
| SDM Longitudinal Axis Recorded Velocity Change (MPH) | 0.64 | 0.64 | 0.64 | 0.64 | 0.64 | 0.64 | 0.64 | 0.64 | 0.64 | 0.64 | 0.64 | 0.64 | 0.64 | 0.64 | 0.64 |

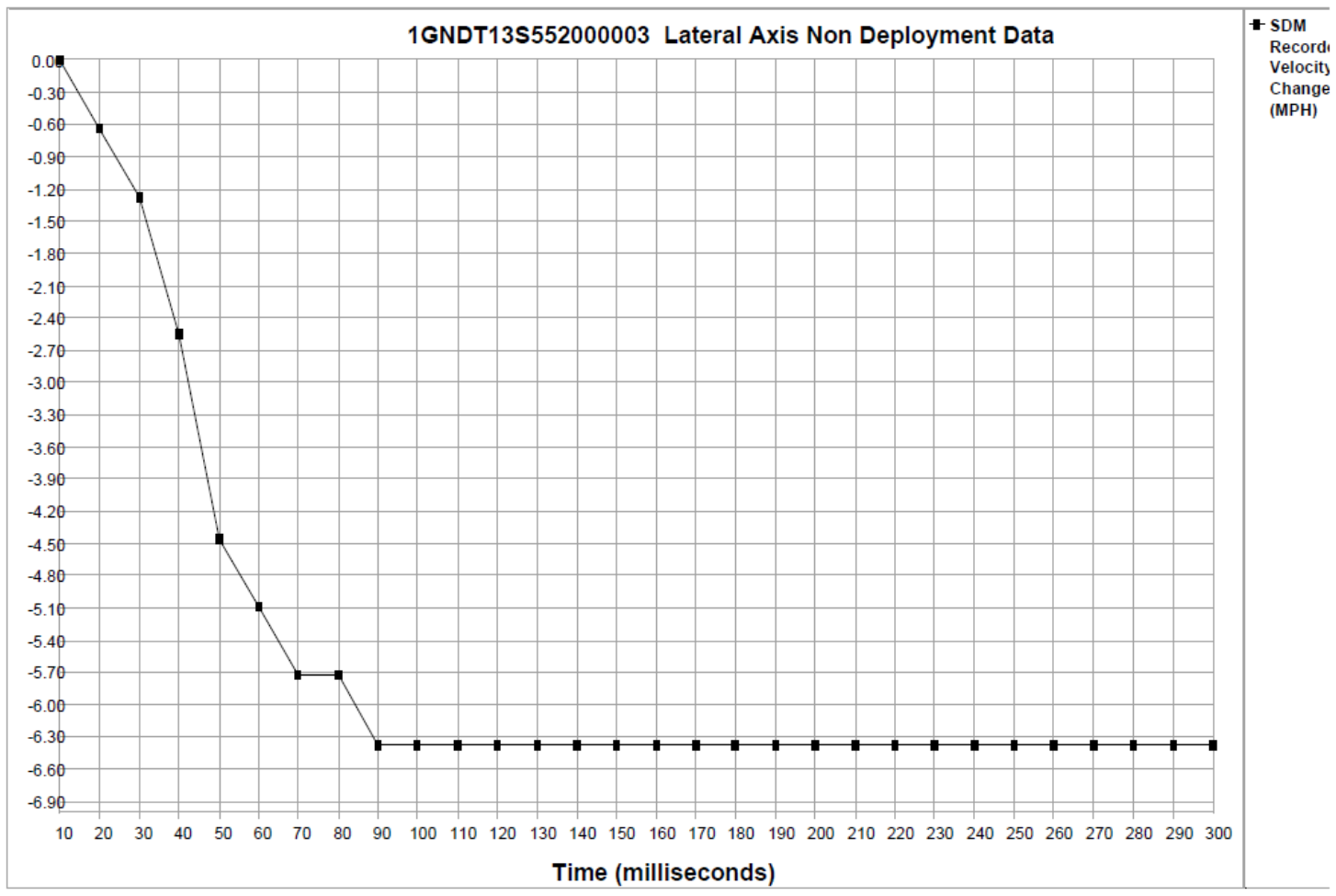

| Time (milliseconds) | 10 | 20 | 30 | 40 | 50 | 60 | 70 | 80 | 90 | 100 | 110 | 120 | 130 | 140 | 150 |
|---|---|---|---|---|---|---|---|---|---|---|---|---|---|---|---|
| SDM Lateral Axis Recorded Velocity Change (MPH) | 0.00 | -0.64 | -1.27 | -2.55 | -4.46 | -5.09 | -5.73 | -5.73 | -6.37 | -6.37 | -6.37 | -6.37 | -6.37 | -6.37 | -6.37 |
| Time (milliseconds) | 160 | 170 | 180 | 190 | 200 | 210 | 220 | 230 | 240 | 250 | 260 | 270 | 280 | 290 | 300 |
| SDM Lateral Axis Recorded Velocity Change (MPH) | -6.37 | -6.37 | -6.37 | -6.37 | -6.37 | -6.37 | -6.37 | -6.37 | -6.37 | -6.37 | -6.37 | -6.37 | -6.37 | -6.37 | -6.37 |

**Figure 15: Data from EDR C obtained using CDR kit. Upper graph and table show longitudinal change-in-velocity (in EDR C's frame). The bottom graph and table show lateral change-in-velocity (in EDR C's frame).**

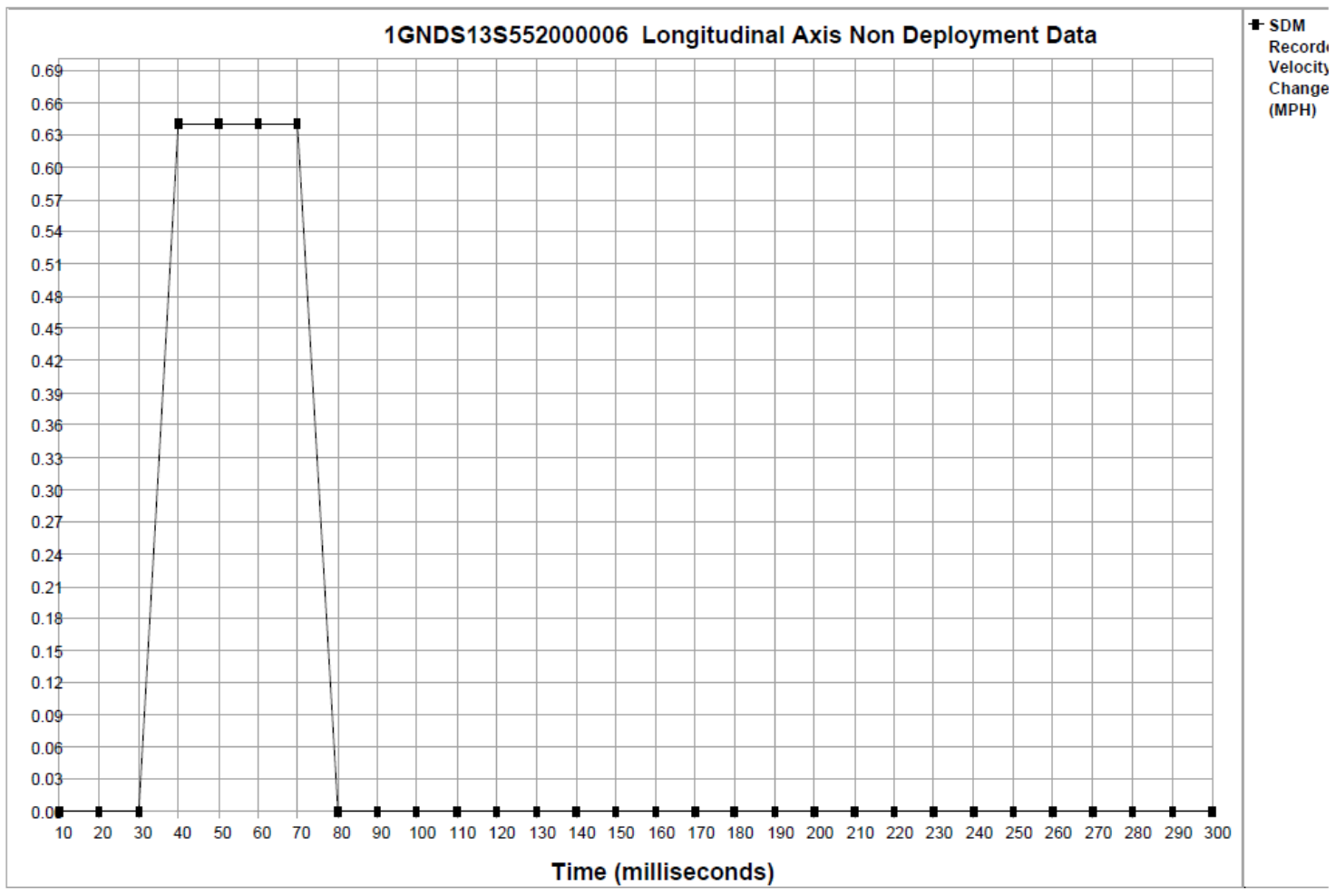

| | | | | | | | | | | | | | | | |
|---|---|---|---|---|---|---|---|---|---|---|---|---|---|---|---|
| Time (milliseconds) | 10 | 20 | 30 | 40 | 50 | 60 | 70 | 80 | 90 | 100 | 110 | 120 | 130 | 140 | 150 |
| SDM Longitudinal Axis Recorded Velocity Change (MPH) | 0.00 | 0.00 | 0.00 | 0.64 | 0.64 | 0.64 | 0.64 | 0.00 | 0.00 | 0.00 | 0.00 | 0.00 | 0.00 | 0.00 | 0.00 |
| Time (milliseconds) | 160 | 170 | 180 | 190 | 200 | 210 | 220 | 230 | 240 | 250 | 260 | 270 | 280 | 290 | 300 |
| SDM Longitudinal Axis Recorded Velocity Change (MPH) | 0.00 | 0.00 | 0.00 | 0.00 | 0.00 | 0.00 | 0.00 | 0.00 | 0.00 | 0.00 | 0.00 | 0.00 | 0.00 | 0.00 | 0.00 |

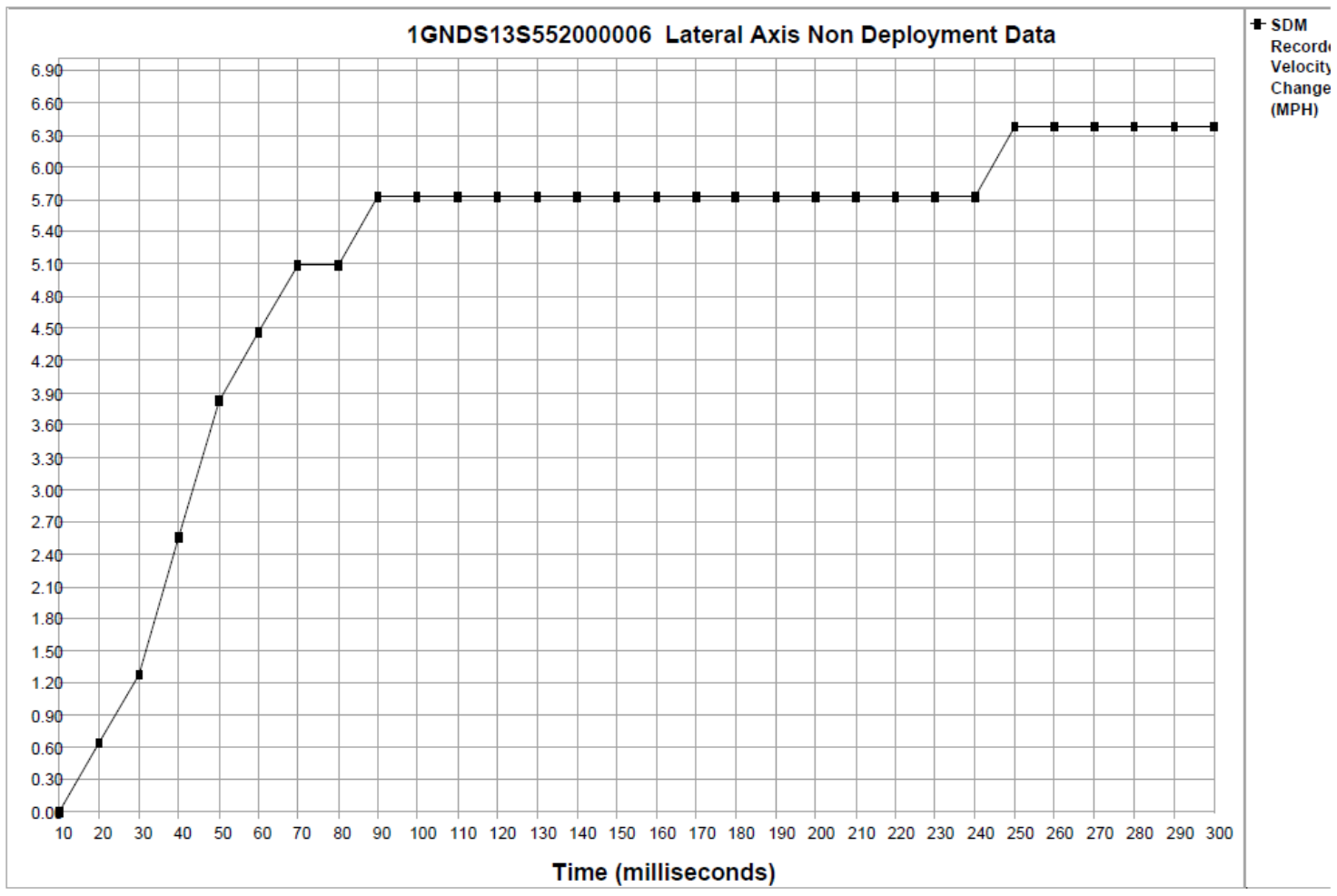

| | | | | | | | | | | | | | | | |
|---|---|---|---|---|---|---|---|---|---|---|---|---|---|---|---|
| Time (milliseconds) | 10 | 20 | 30 | 40 | 50 | 60 | 70 | 80 | 90 | 100 | 110 | 120 | 130 | 140 | 150 |
| SDM Lateral Axis Recorded Velocity Change (MPH) | 0.00 | 0.64 | 1.27 | 2.55 | 3.82 | 4.46 | 5.09 | 5.09 | 5.73 | 5.73 | 5.73 | 5.73 | 5.73 | 5.73 | 5.73 |
| Time (milliseconds) | 160 | 170 | 180 | 190 | 200 | 210 | 220 | 230 | 240 | 250 | 260 | 270 | 280 | 290 | 300 |
| SDM Lateral Axis Recorded Velocity Change (MPH) | 5.73 | 5.73 | 5.73 | 5.73 | 5.73 | 5.73 | 5.73 | 5.73 | 5.73 | 6.37 | 6.37 | 6.37 | 6.37 | 6.37 | 6.37 |

**Figure 16: Data from EDR D obtained using CDR kit. Upper graph and table show longitudinal change-in-velocity (in EDR D's frame). The bottom graph and table show lateral change-in-velocity (in EDR D's frame).**

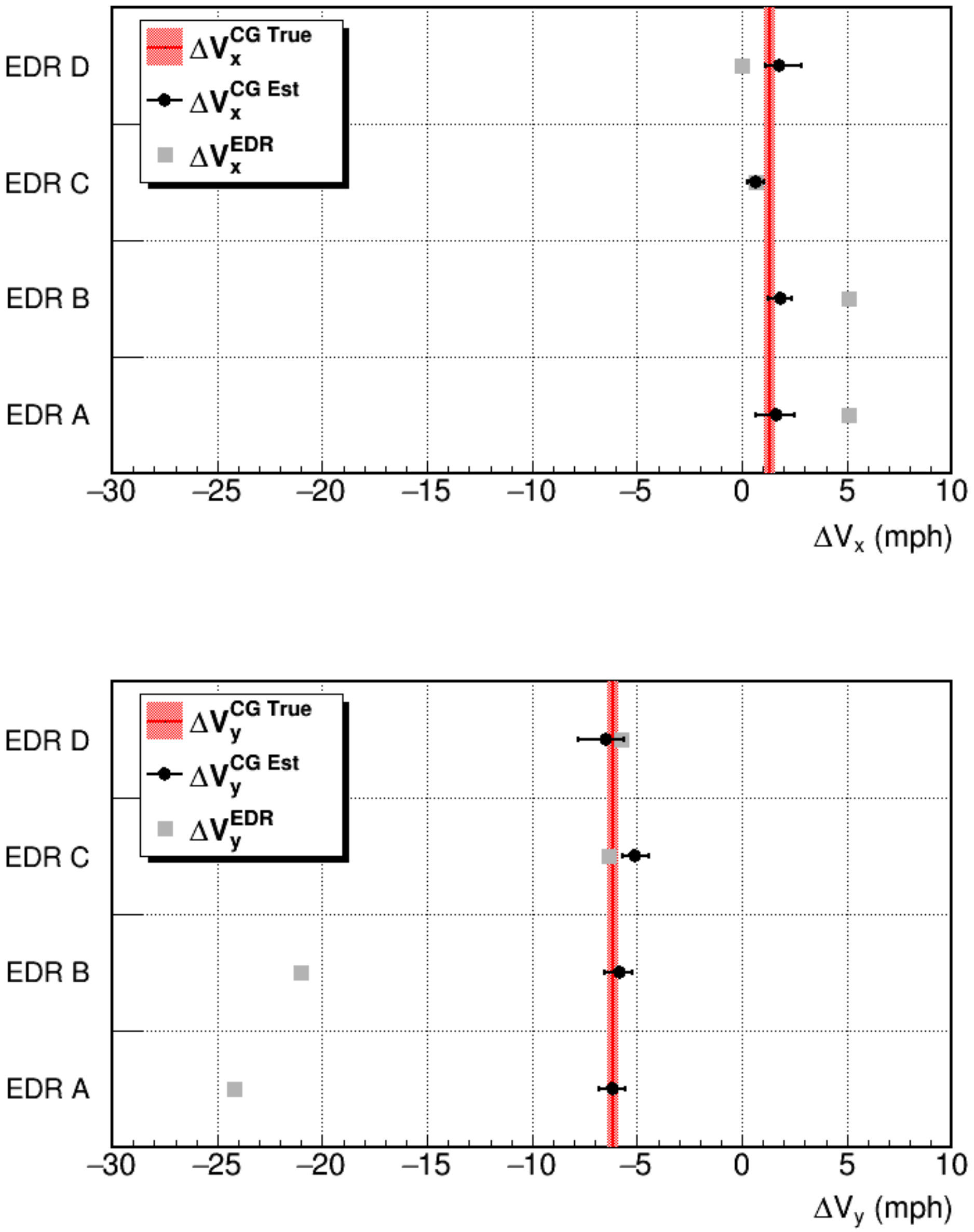

**Figure 17: Longitudinal (top) and lateral (bottom) $\Delta v$ displayed for each EDR. In red we show the accelerometer measured $\Delta v$ values. Gray dots represent the EDR $\Delta v$ values without correction. Black dots represent the best-estimate $\Delta v$ values based on correcting EDR data, along with minimum and maximum estimates. Here no uncertainties for EDR $\Delta v$ input values are accounted for.**

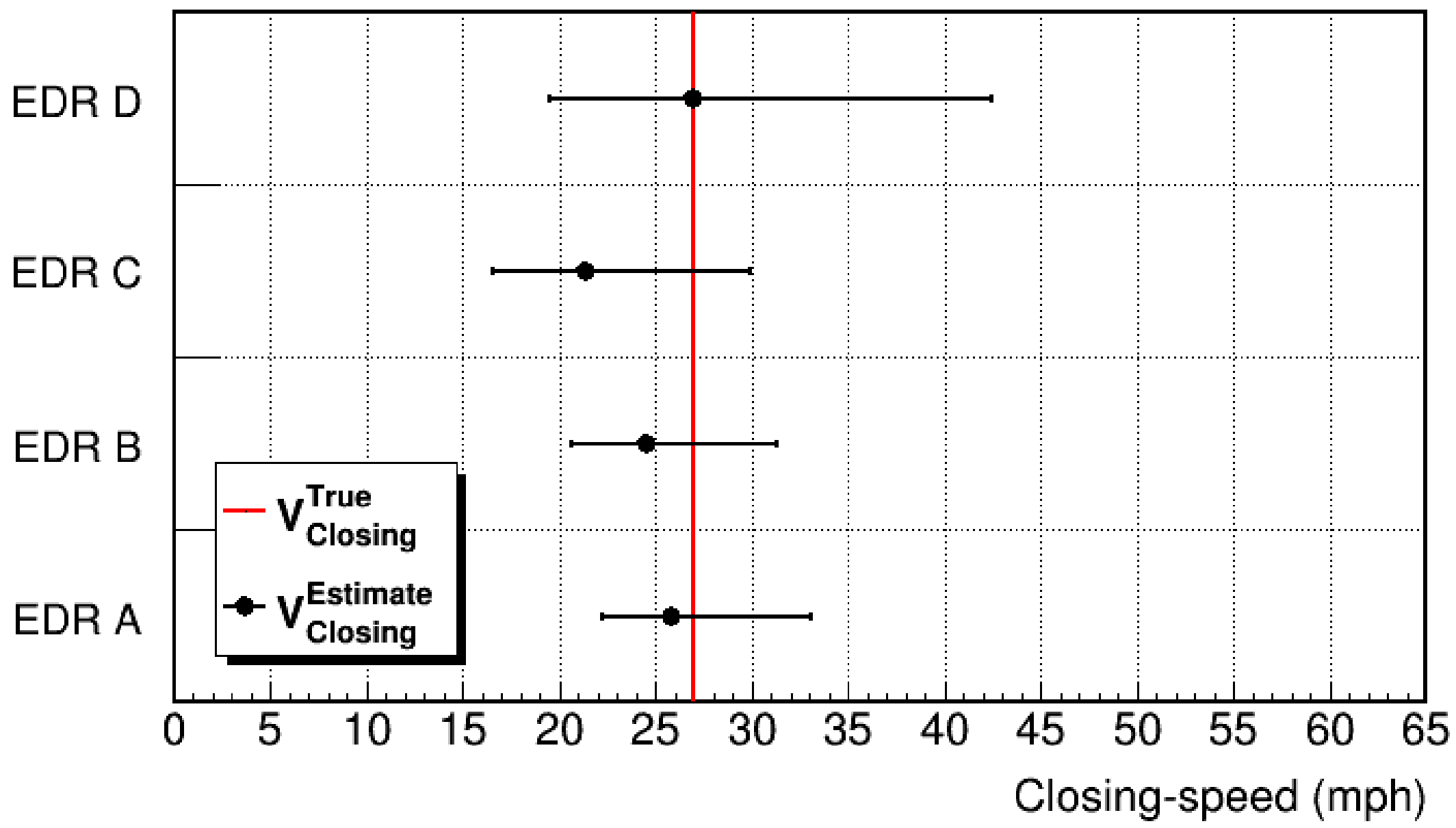

**Figure 18: In red we show the true closing-speed for the test. Black dots represent the best-estimate closing-speed values based on correcting EDR data, along with minimum and maximum estimates. Here no uncertainties for EDR $\Delta v$ input values are accounted for.**

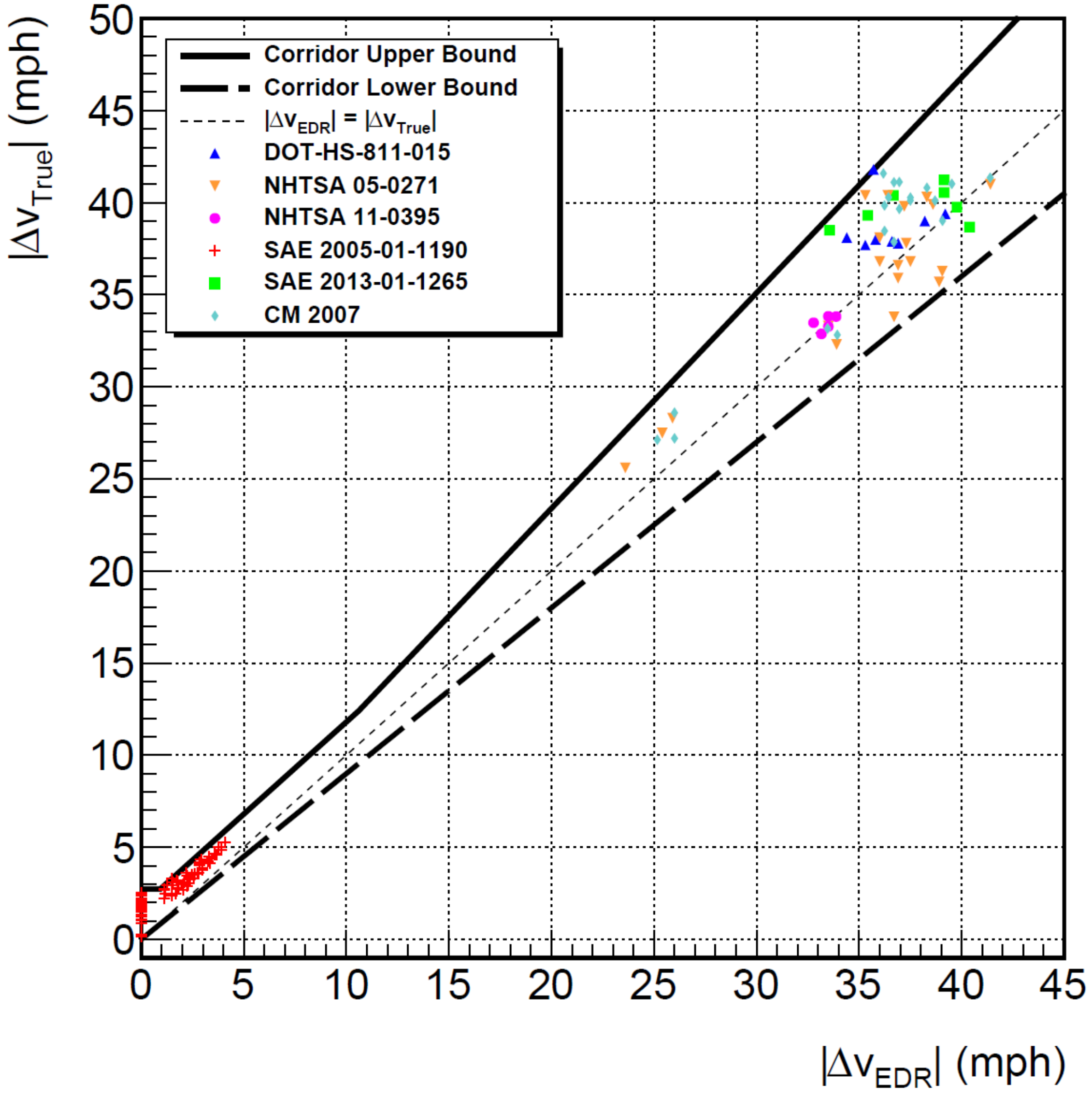

**Figure 19: $\Delta v$ corridor defining the upper and lower true $\Delta v$ versus EDR $\Delta v$.**

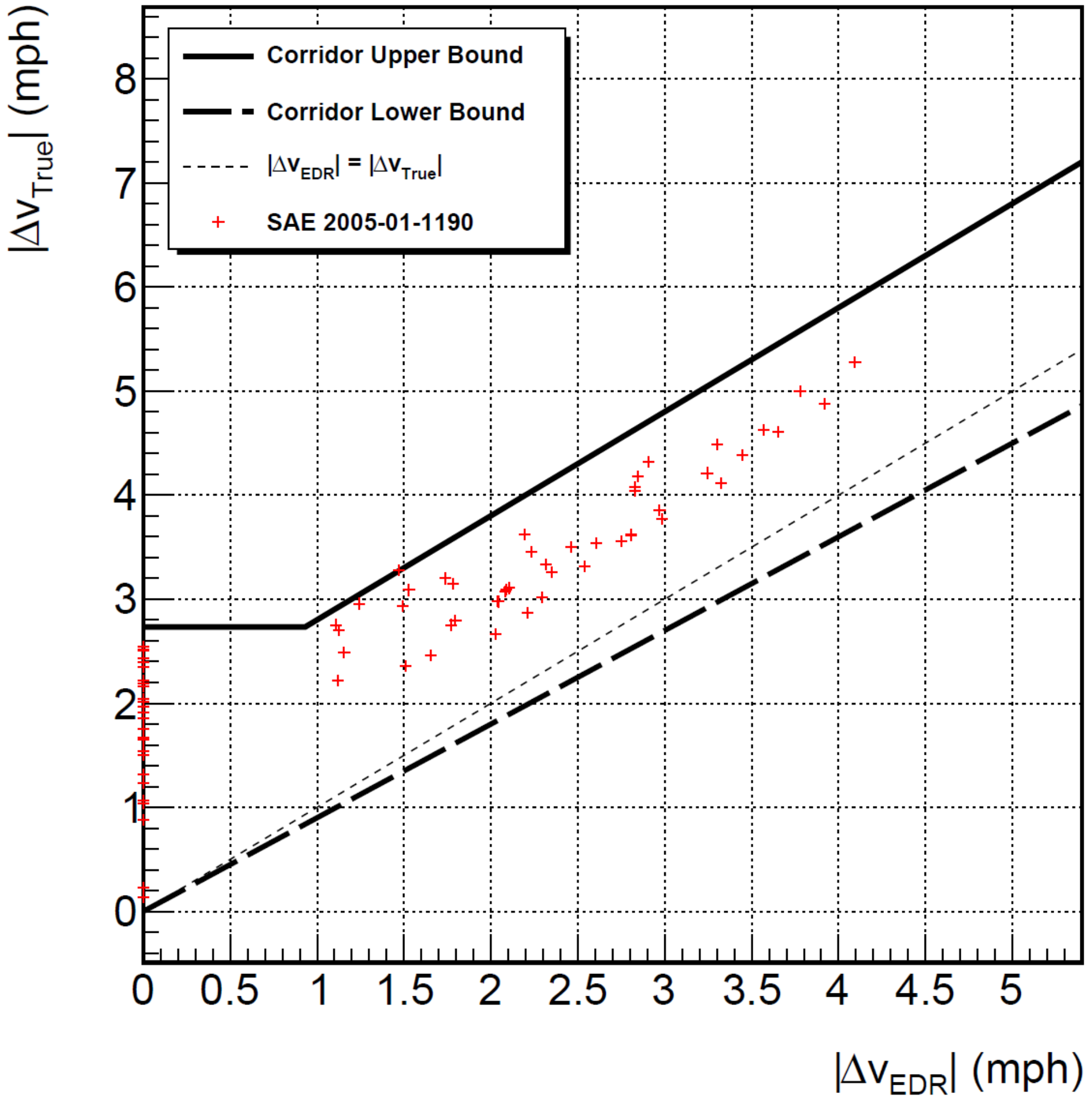

**Figure 20: $\Delta v$ corridor defining the upper and lower true $\Delta v$ versus EDR $\Delta v$. Here we focus on low EDR $\Delta v$ values.**

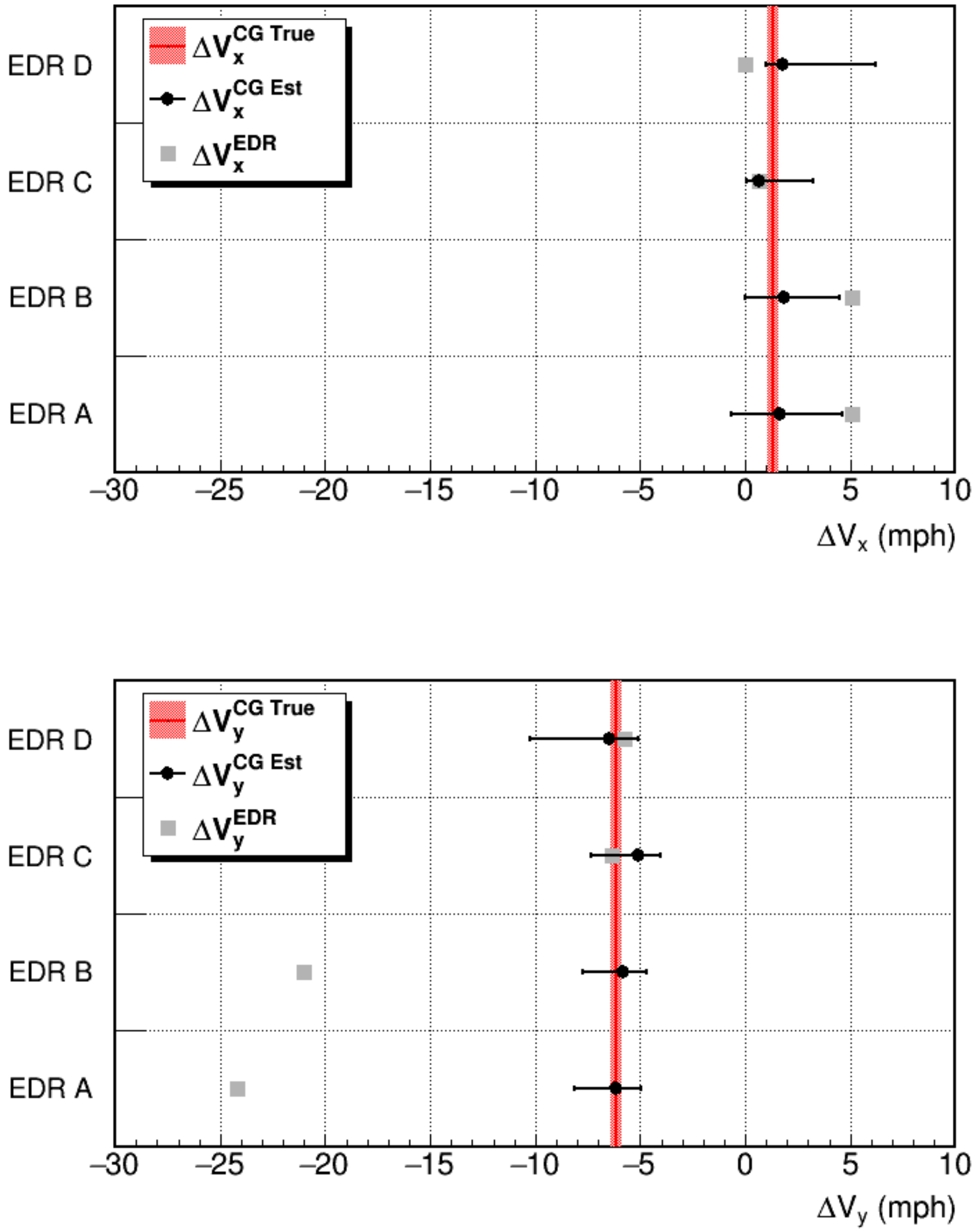

**Figure 21: Longitudinal (top) and lateral (bottom) $\Delta v$ displayed for each EDR. In red we show the accelerometer measured $\Delta v$ values. Gray dots represent the EDR $\Delta v$ values without correction. Black dots represent the best-estimate $\Delta v$ values based on correcting EDR data, along with minimum and maximum estimates. Here we apply the same $\Delta v$ corridor to both lateral and longitudinal components.**

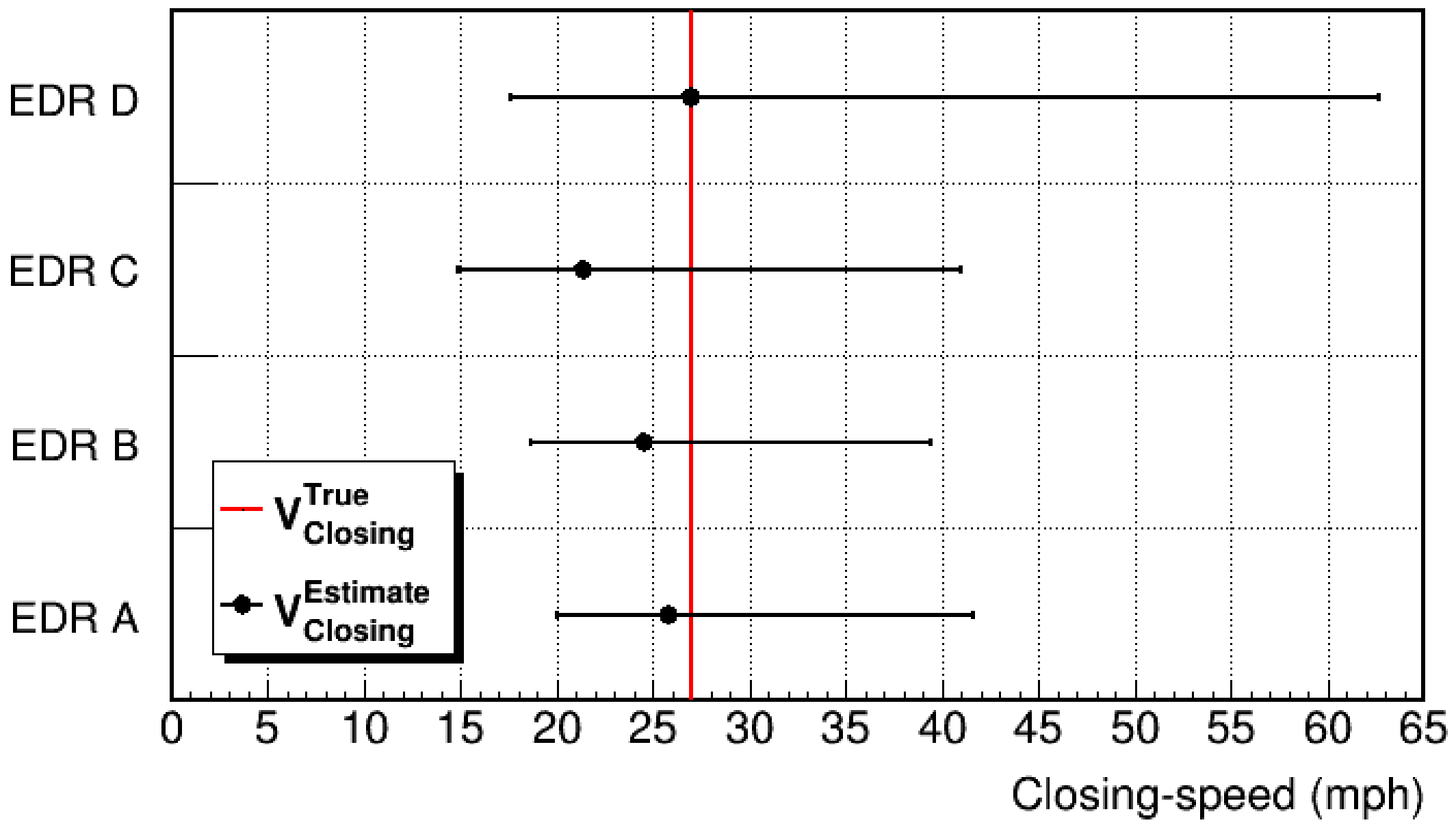

**Figure 22: In red we show the true closing-speed for the test. Black dots represent the best-estimate closing-speed values based on correcting EDR data, along with minimum and maximum estimates. Here no uncertainties for EDR $\Delta v$ input values are accounted for. Here we apply the same $\Delta v$ corridor to both lateral and longitudinal components.**

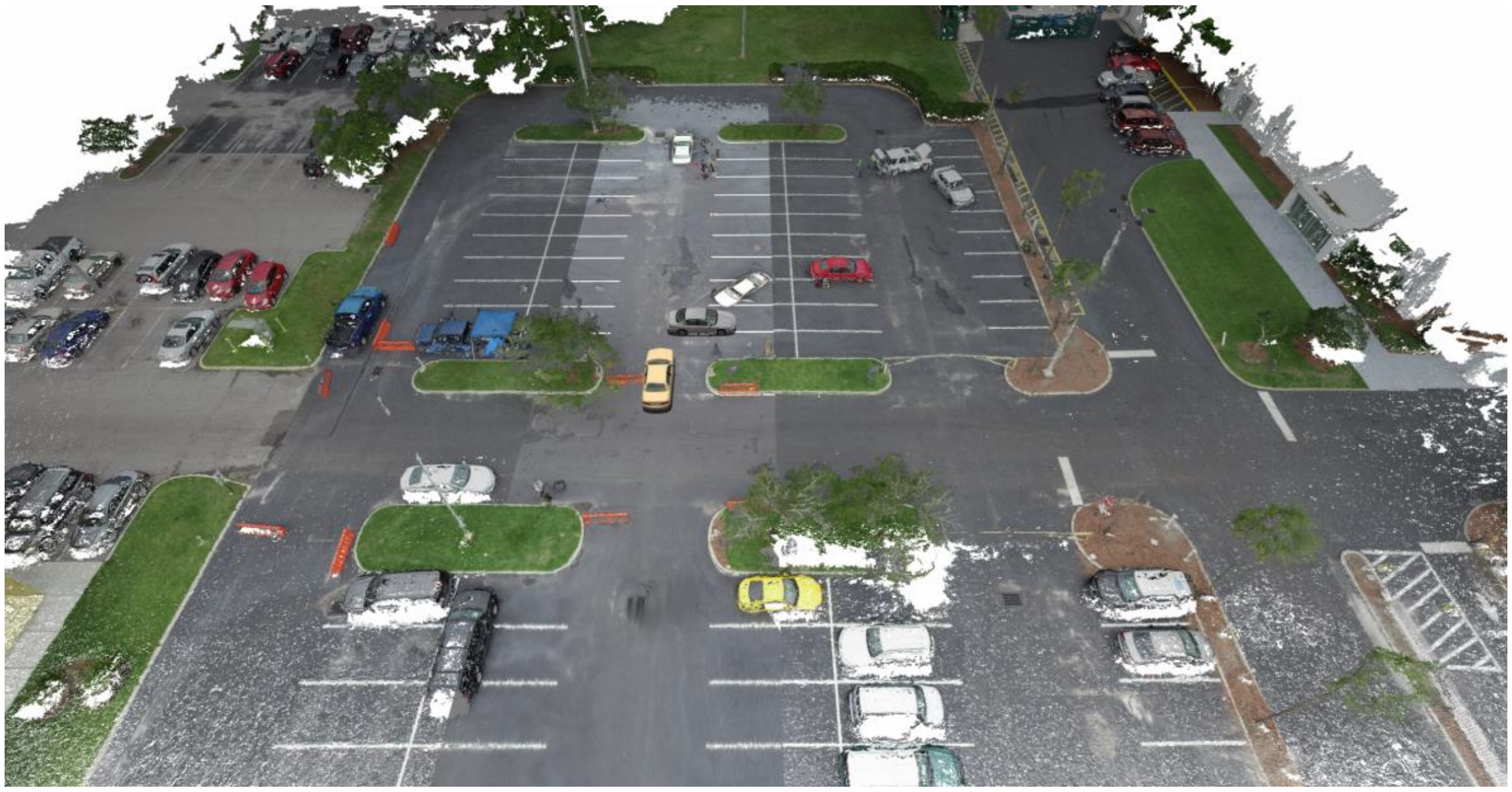

**Figure 23: 3D simulation environment created in Virtual CRASH 4 using point cloud data.**

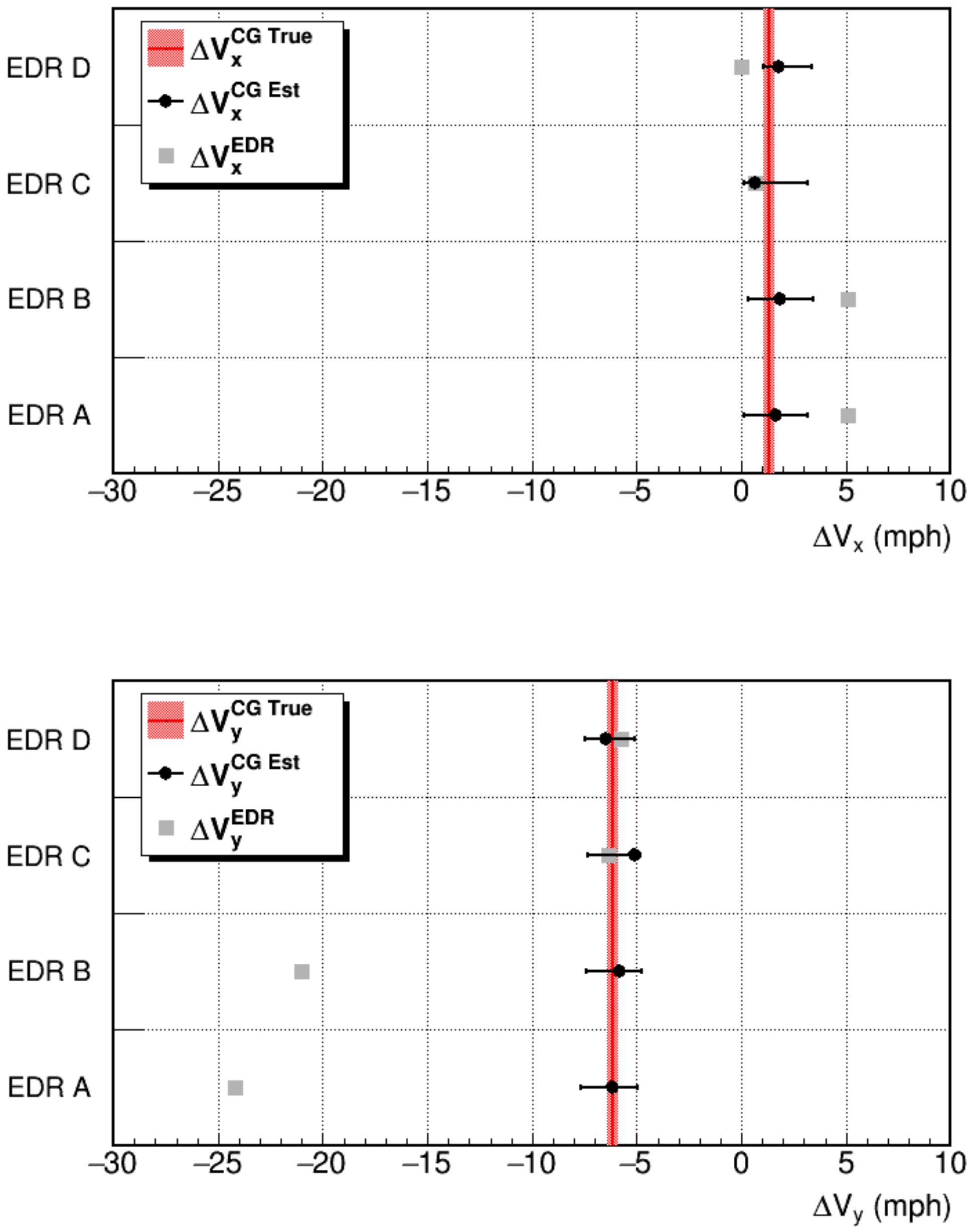

**Figure 24: Longitudinal (top) and lateral (bottom) $\Delta v$ displayed for each EDR. In red we show the accelerometer measured $\Delta v$ values. Gray dots represent the EDR $\Delta v$ values without correction. Black dots represent the best-estimate $\Delta v$ values based on correcting EDR data, along with minimum and maximum estimates. Here we apply the same $\Delta v$ corridor to both lateral and longitudinal components. Monte Carlo selection cuts based on post-impact motion studies conducted with Virtual CRASH 4 are used to reduce the uncertainty range.**

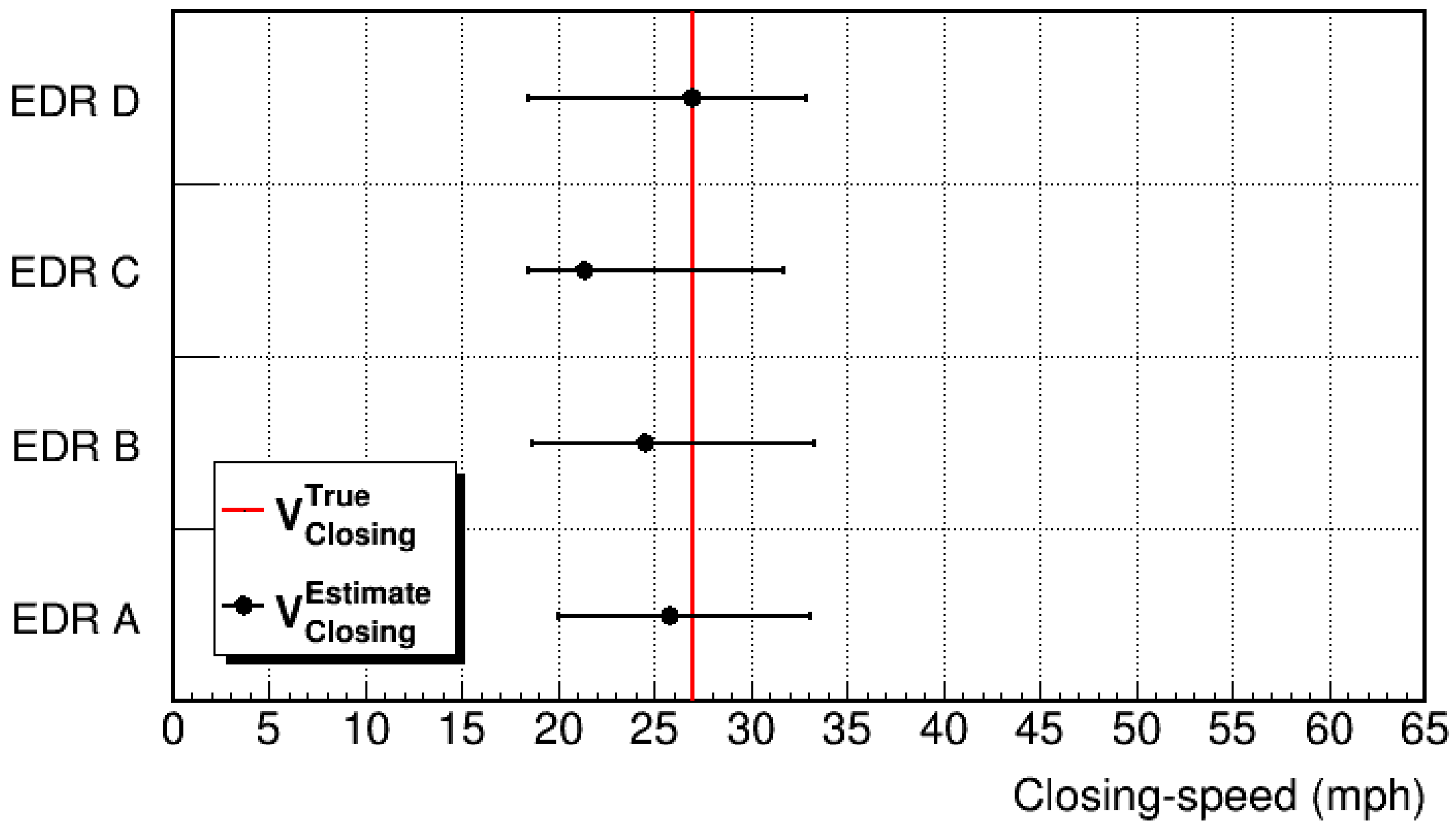

**Figure 25: In red we show the true closing-speed for the test. Black dots represent the best-estimate closing-speed values based on correcting EDR data, along with minimum and maximum estimates. Here no uncertainties for EDR $\Delta v$ input values are accounted for. Here we apply the same $\Delta v$ corridor to both lateral and longitudinal components. Monte Carlo selection cuts based on post-impact motion studies conducted with Virtual CRASH 4 are used to reduce the uncertainty range.**

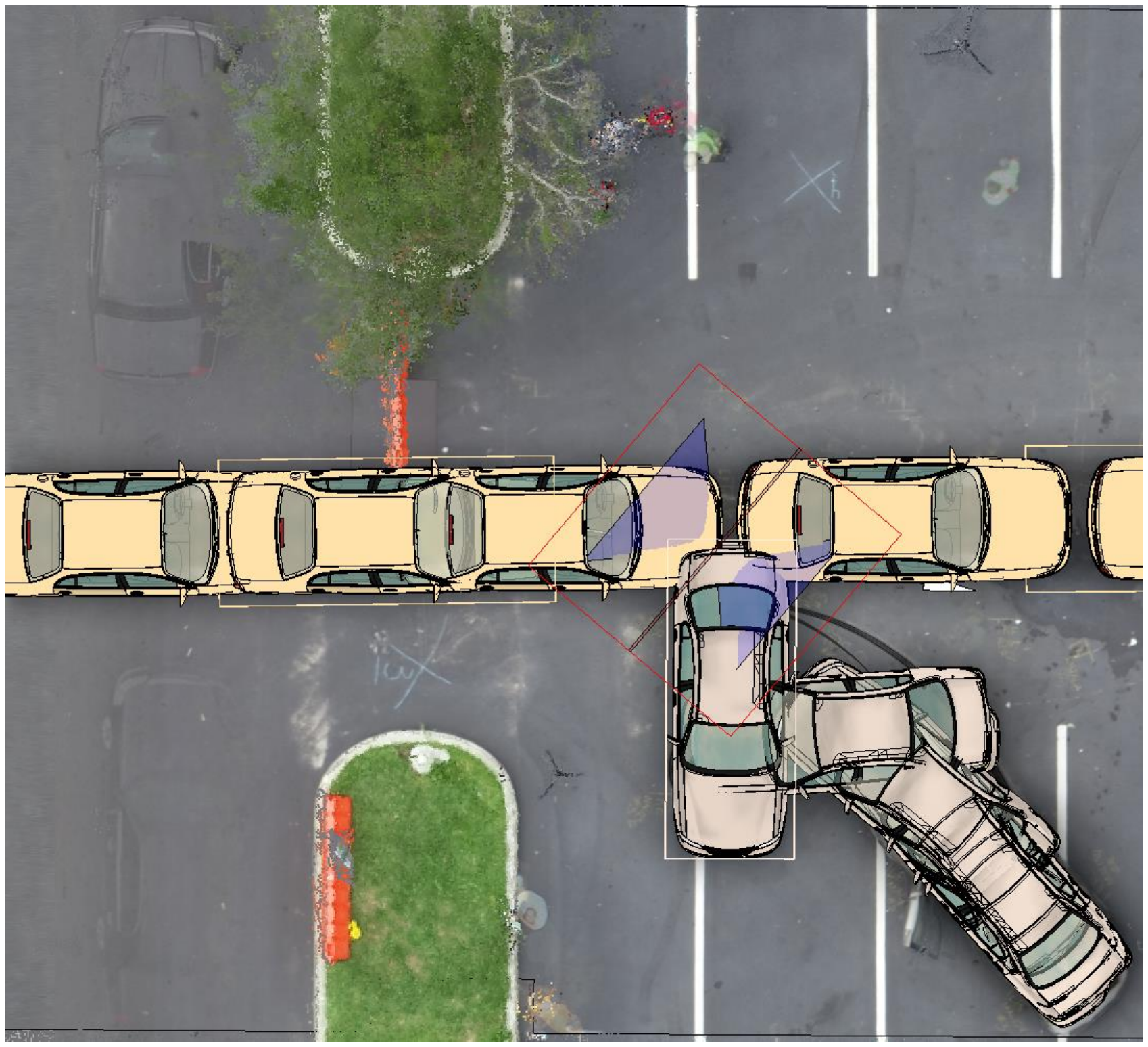

**Figure 26: Diagram showing Virtual CRASH 4 simulation sequence.**

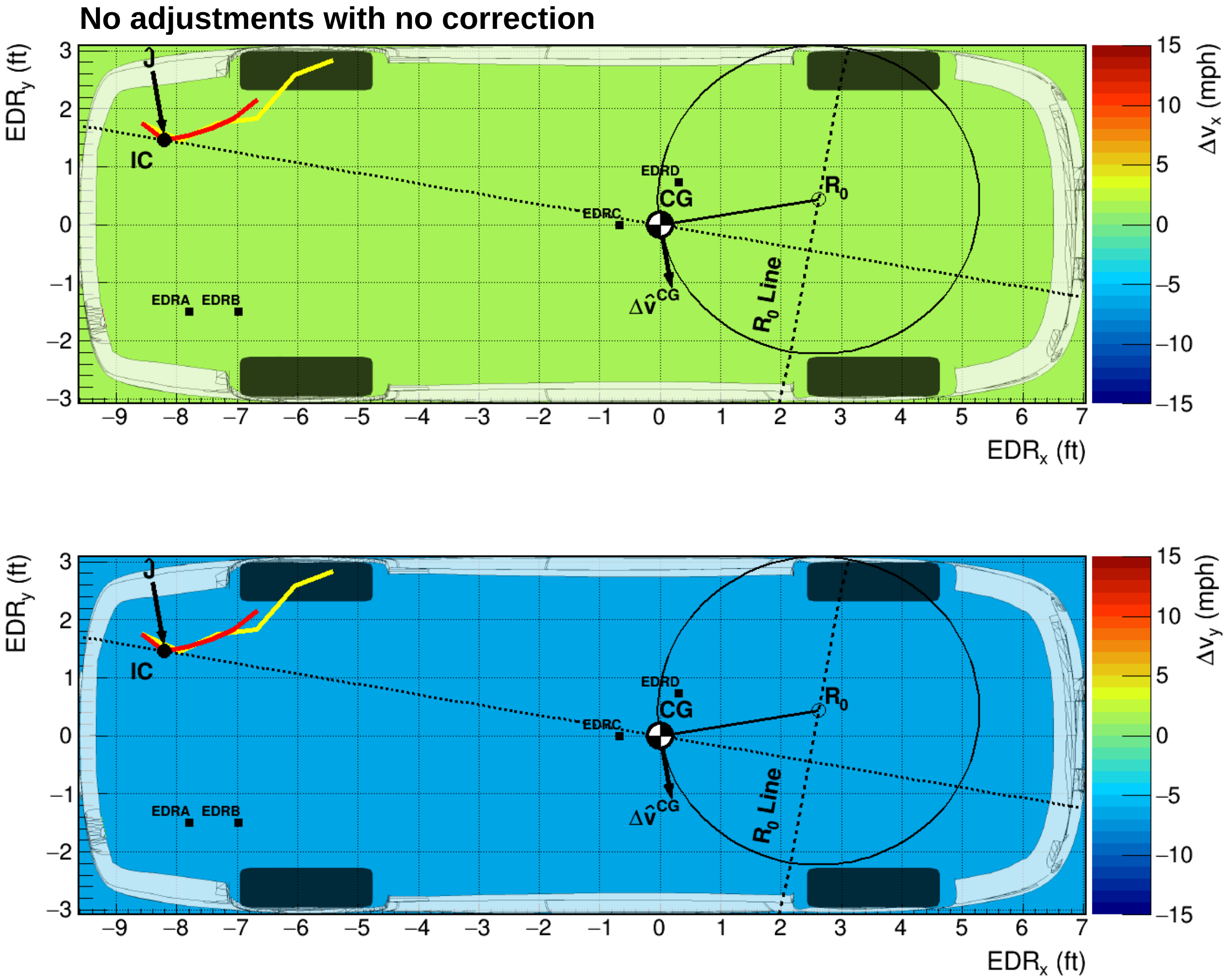

**Figure 27: (Top) Estimated $|\Delta v_x|$ at the center-of gravity as a function of EDR position, assuming no thresholds or corrections are applied. (Bottom) Estimated $|\Delta v_y|$ at the center-of gravity as a function of EDR position, assuming no thresholds or corrections are applied. Note, view is from below vehicle looking up.**

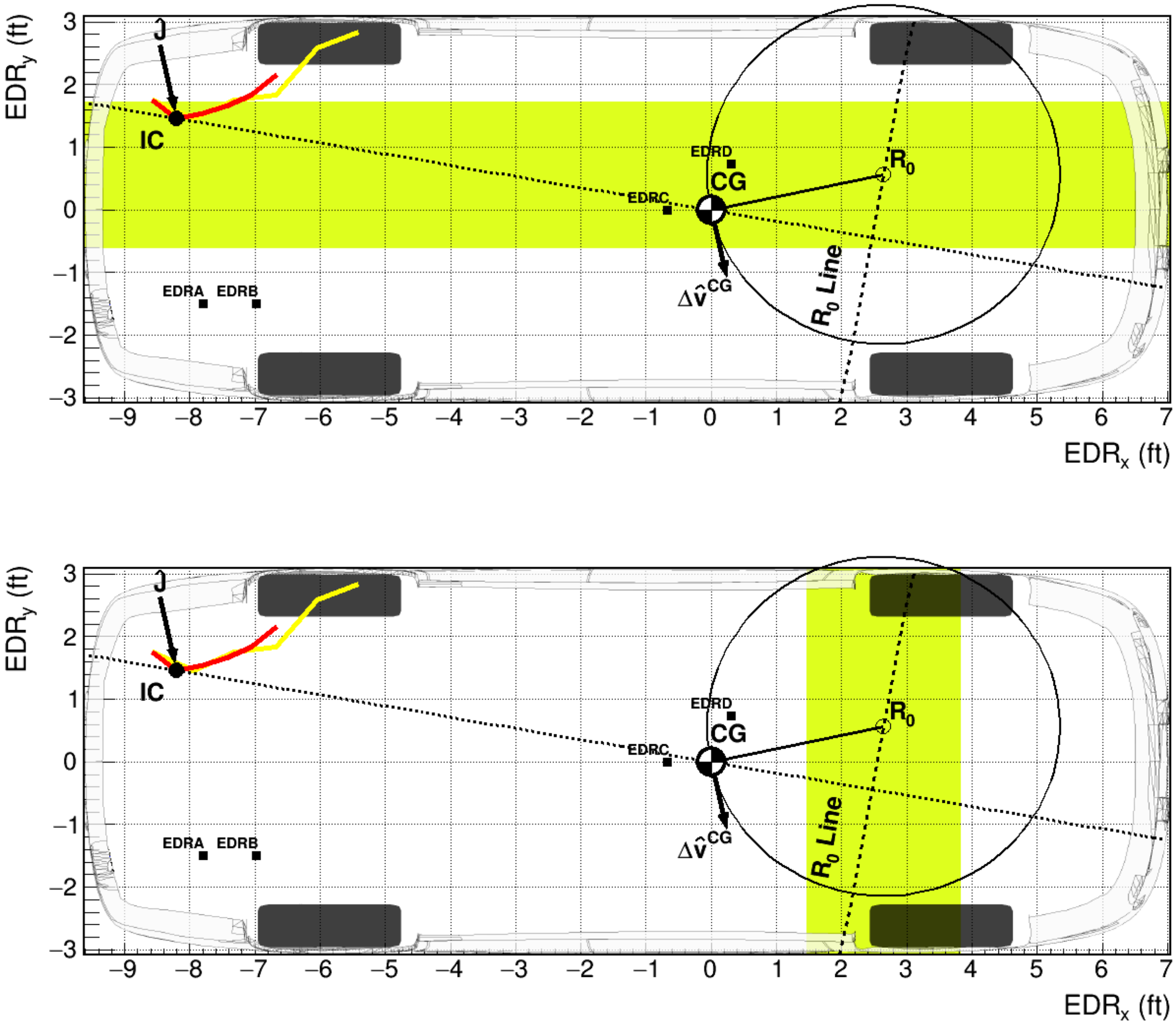

**Figure 28: Zero-corridors shown in yellow highlighted areas for $\Delta v_{1,x}^{A\,Est}$ (top) and $\Delta v_{1,y}^{A\,Est}$ (bottom) assuming $\Delta v^{Threshold} = 4.4$ kph. Note, view is from below vehicle looking up.**

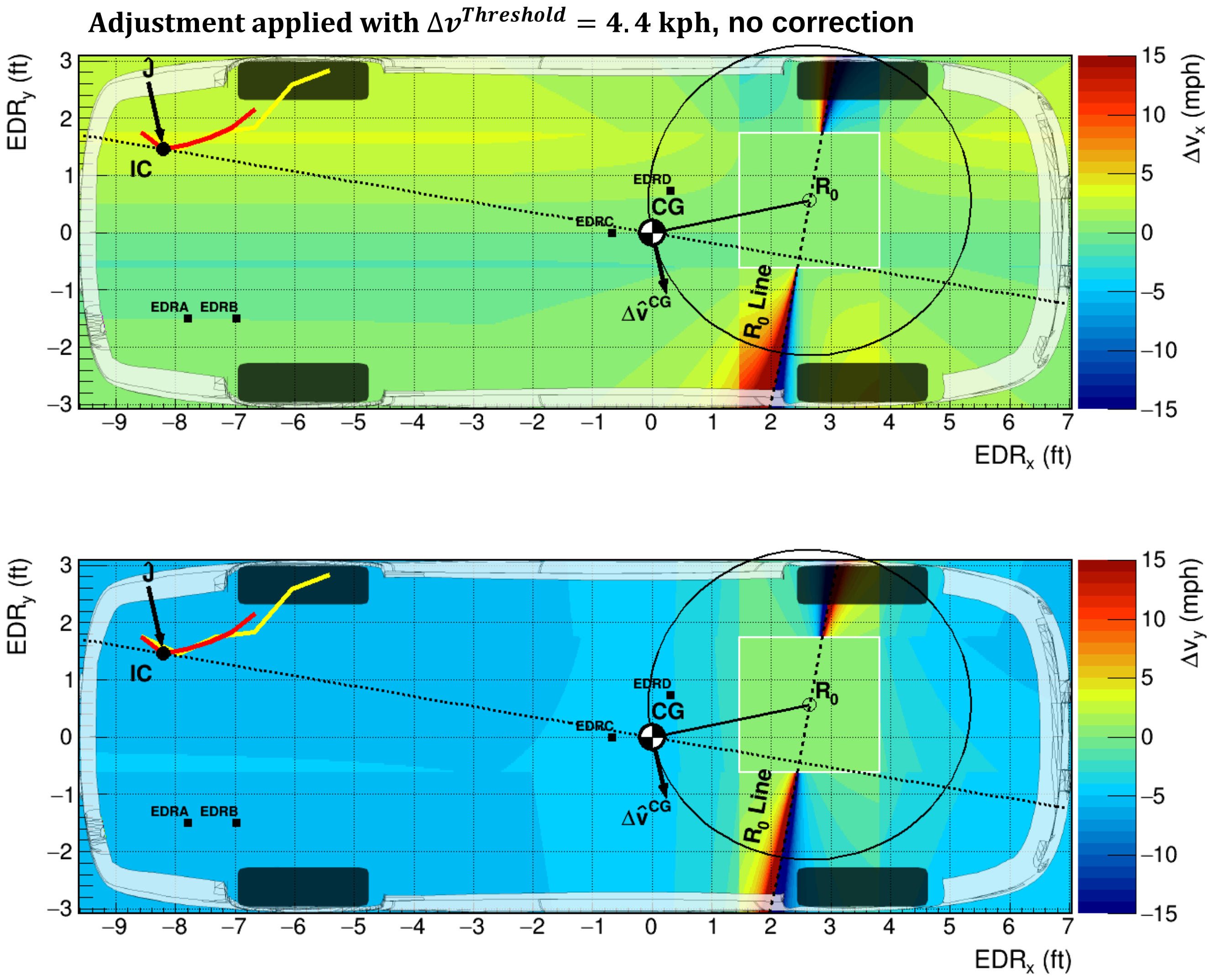

**Figure 29: (Top) Estimated $|\Delta v_x|$ at the center-of gravity as a function of EDR position. (Bottom) Estimated $|\Delta v_y|$ at the center-of gravity as a function of EDR position. In both cases, $\Delta v_x^{A\ Est}$ and $\Delta v_y^{A\ Est}$ were adjusted values following upper limit of the $\Delta v$ corridor. No corrections were applied. An EDR within the white box will result in $\Delta v_x^{A\ Est} = 0$ and $\Delta v_y^{A\ Est} = 0$. Note, view is from below vehicle looking up.**

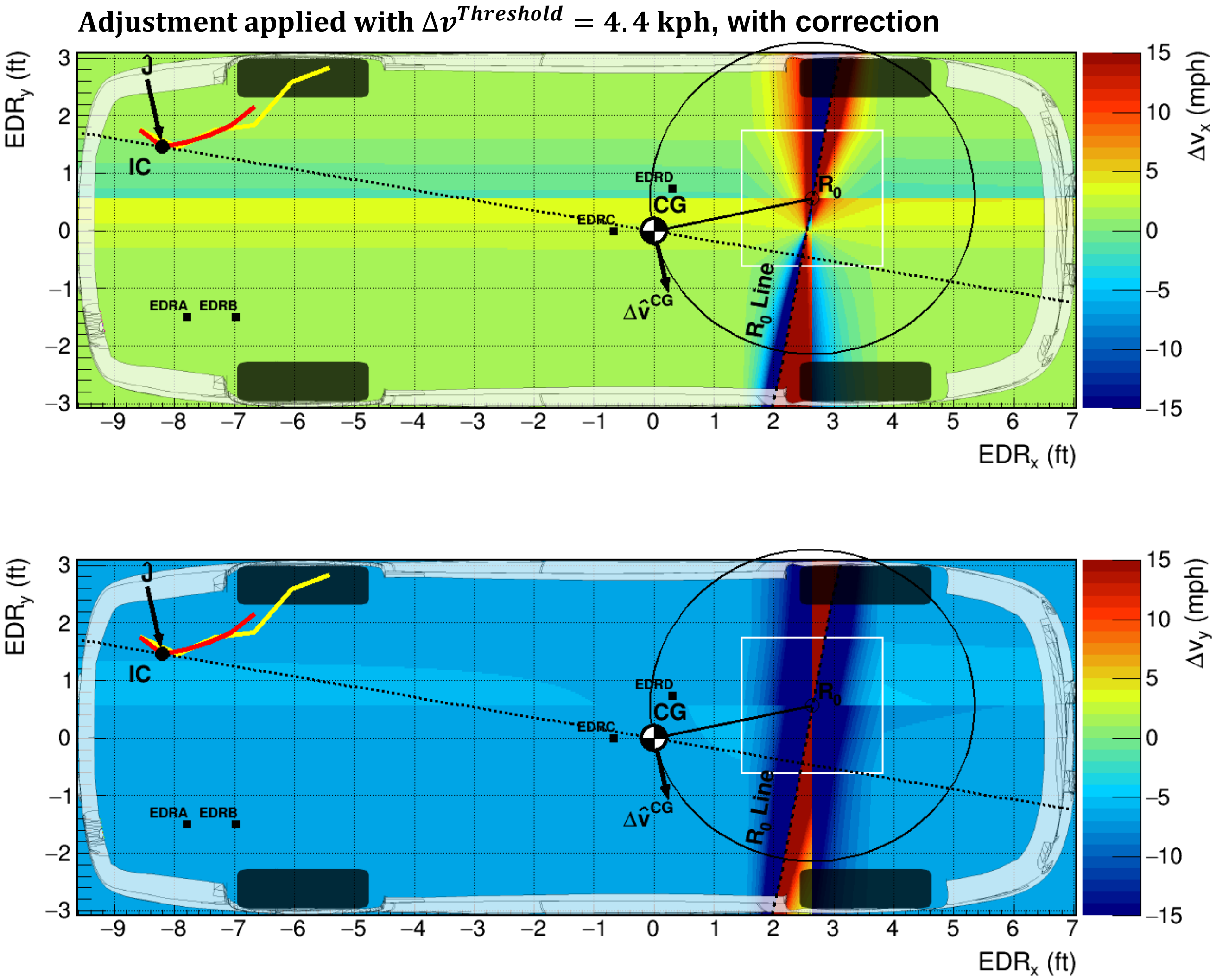

**Figure 30: (Top) Estimated $|\Delta v_x|$ at the center-of gravity as a function of EDR position. (Bottom) Estimated $|\Delta v_y|$ at the center-of gravity as a function of EDR position. In both cases, $\Delta v_x^{A\ Est}$ and $\Delta v_y^{A\ Est}$ were adjusted values following upper limit of the $\Delta v$ corridor. Corrections were applied. An EDR within the white box will result in $\Delta v_x^{A\ Est} = 2.9$ kph and $\Delta v_y^{A\ Est} = 2.9$ kph. Note, view is from below vehicle looking up.**

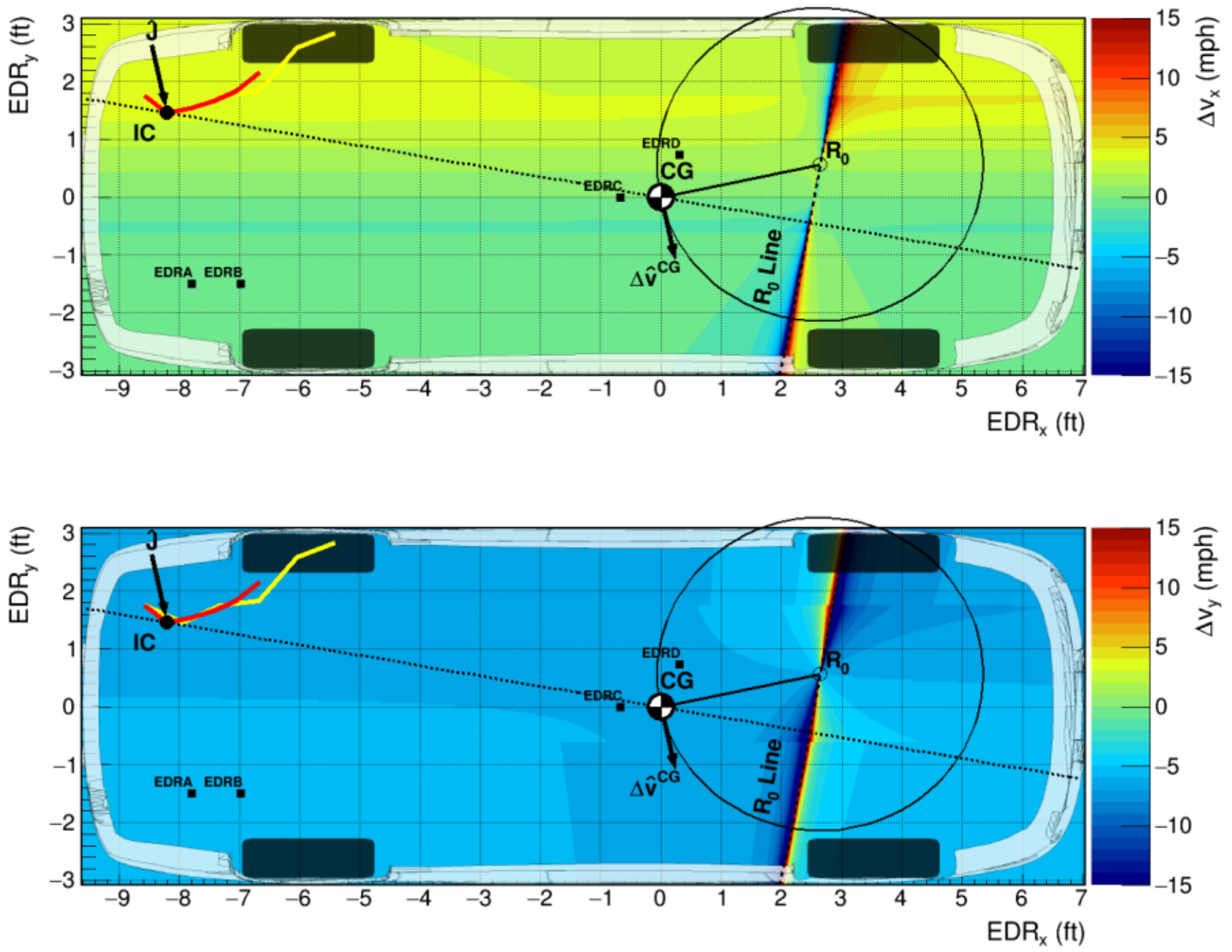

(a)

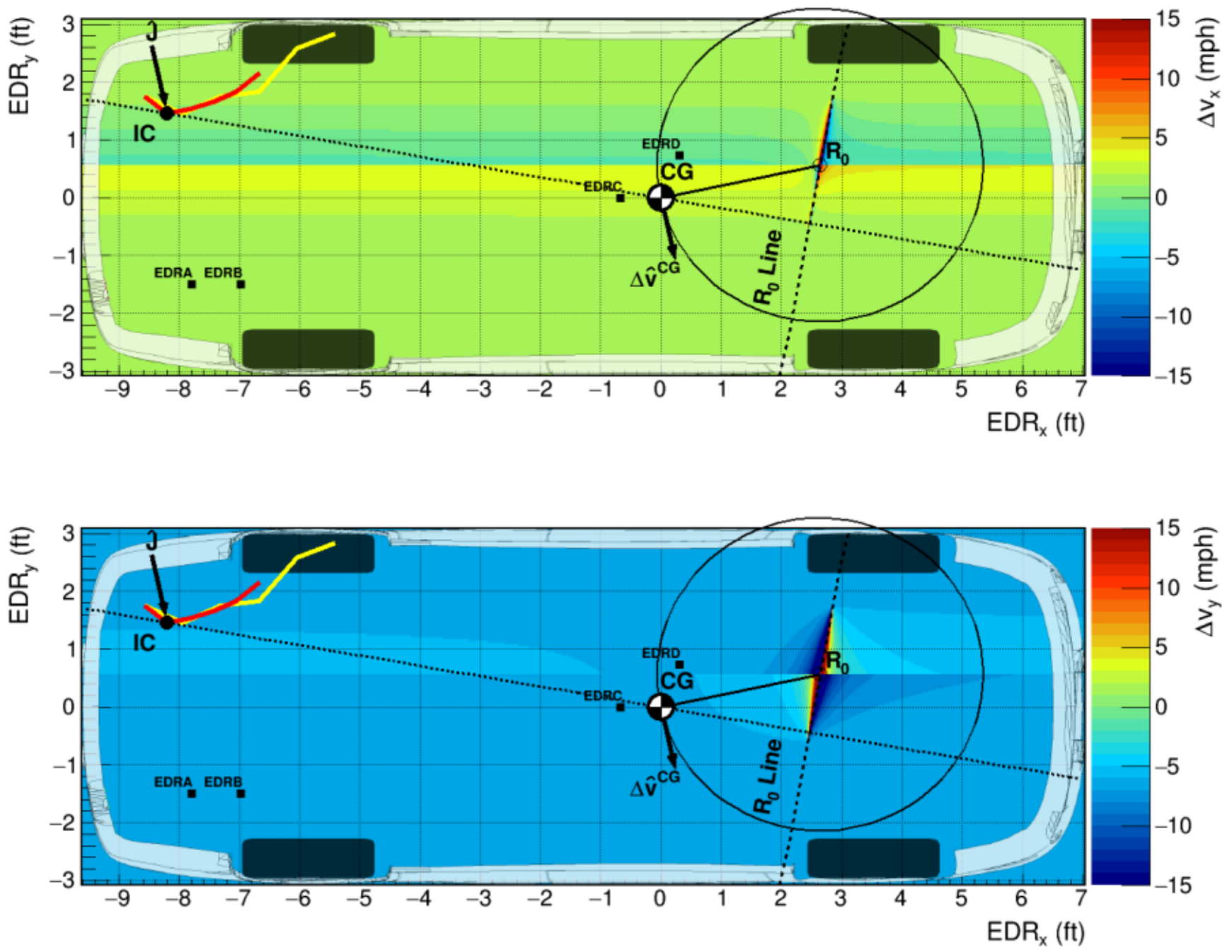

(b)

**Figure 31: Estimated $|\Delta v_x|$ and $|\Delta v_y|$ at the center-of gravity as a function of EDR position given: (a) longitudinal $\Delta v$ reduced using corridor and unmodified lateral $\Delta v$, (b) longitudinal $\Delta v$ reduced using corridor followed by inverse correction and unmodified lateral $\Delta v$.**

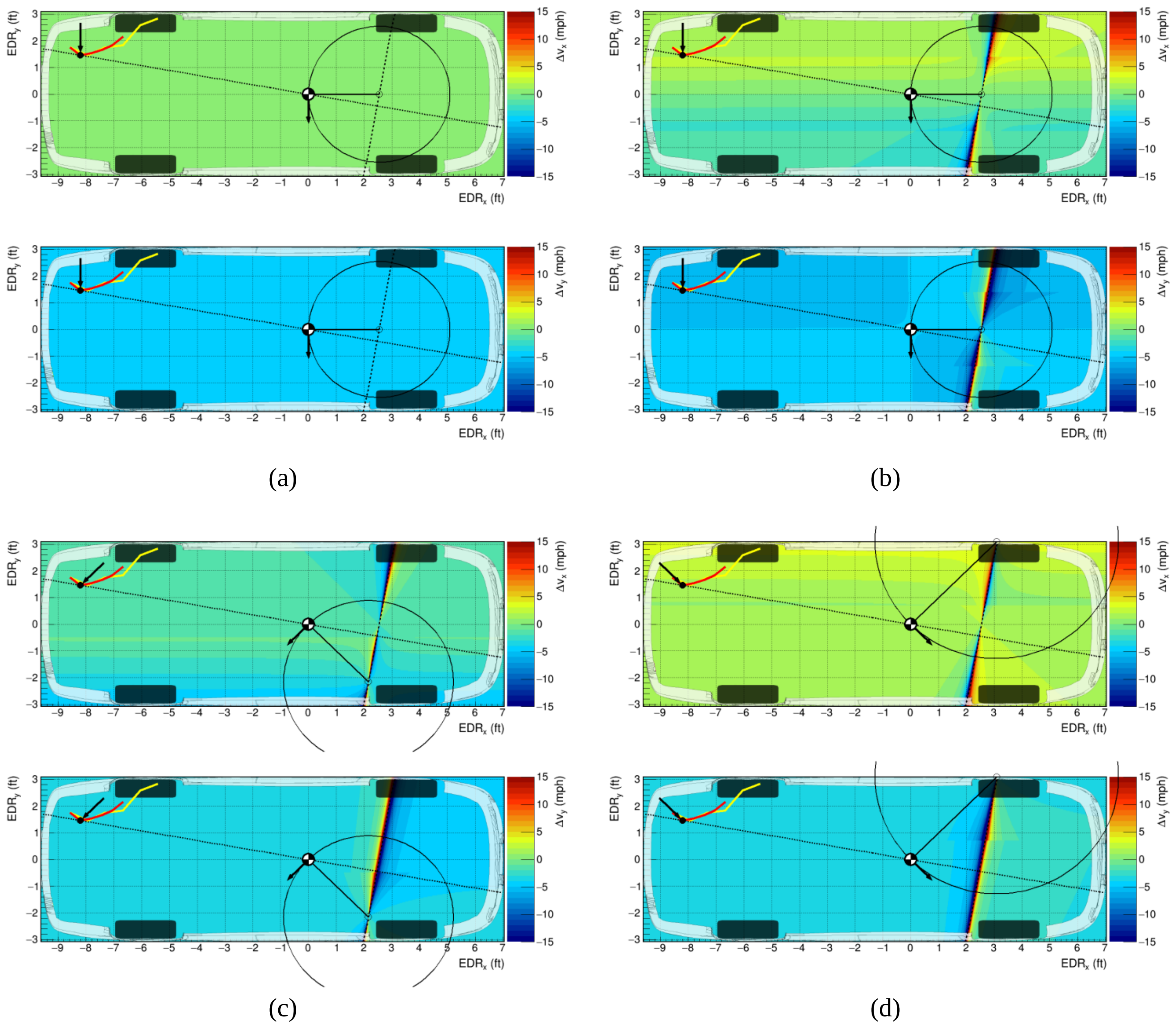

**Figure 32: Estimated $|\Delta v_x|$ and $|\Delta v_y|$ at the center-of-gravity as a function of EDR position given for $|\Delta\bar{v}|$ = 5 mph: (a) impulse at -90 degrees, unmodified longitudinal $\Delta v$ and unmodified lateral $\Delta v$, (b) impulse at -90 degrees, longitudinal $\Delta v$ reduced using corridor and unmodified lateral $\Delta v$, (c) impulse at -135 degrees, longitudinal $\Delta v$ reduced using corridor and unmodified lateral $\Delta v$, (d) impulse at -45 degrees, longitudinal $\Delta v$ reduced using corridor and unmodified lateral $\Delta v$.**

**Appendix 1**

*The closing-velocity ambiguity problem*

Typically, the analyst would like to use EDR $\Delta v$ to arrive at some estimates of pre-impact ground speeds for the interacting vehicles; however, our key result, given by (101), does not yield ground speeds, but rather the projection of the closing-velocity vector onto the axis given by the impulse vector for vehicle 1, thus leaving the component tangent to the vehicle 1 impulse vector axis undetermined. It is possible, in some cases, to resolve this ambiguity problem by either using more information, such as departure angles, or using a simplified system where one vehicle is initially at rest. Resolving the ambiguity problem is explored in this appendix[5]. Note, in the presentation below, it is assumed that the orientations at impact are known for the vehicles. Impact orientation is needed in order to use much of the formalism presented in this paper.

Starting with (100):

$$-(1+\varepsilon)\cdot\left(\bar{v}^{P}_{Rel,i}\cdot\Delta\hat{v}^{CG}_{1}\right)=\frac{\left|\Delta\bar{v}^{CG}_{1}\right|}{\gamma_1}+\frac{\left|\Delta\bar{v}^{CG}_{2}\right|}{\gamma_2}$$

From this we have:

$$-\bar{v}^{P}_{Rel,i}\cdot\Delta\hat{v}^{CG}_{1}=\frac{1}{1+\varepsilon}\cdot\left(\frac{\left|\Delta\bar{v}^{CG}_{1}\right|}{\gamma_1}+\frac{\left|\Delta\bar{v}^{CG}_{2}\right|}{\gamma_2}\right)$$

Next, we distribute dot product:

$$\begin{aligned}-\left(\bar{v}^{P}_{1,i}-\bar{v}^{P}_{2,i}\right)\cdot\Delta\hat{v}^{CG}_{1}&\\ &=-\bar{v}^{P}_{1,i}\cdot\Delta\hat{v}^{CG}_{1}+\bar{v}^{P}_{2,i}\cdot\Delta\hat{v}^{CG}_{1}\\ &=\frac{1}{1+\varepsilon}\cdot\left(\frac{\left|\Delta\bar{v}^{CG}_{1}\right|}{\gamma_1}+\frac{\left|\Delta\bar{v}^{CG}_{2}\right|}{\gamma_2}\right)\end{aligned}$$

Factoring out the magnitude of $\bar{v}^{P}_{1,i}$ and $\bar{v}^{P}_{2,i}$ gives:

$$\begin{aligned}-\bar{v}^{P}_{1,i}\cdot\Delta\hat{v}^{CG}_{1}+\bar{v}^{P}_{2,i}\cdot\Delta\hat{v}^{CG}_{1}&\\ &=-\left(\hat{v}^{P}_{1,i}\cdot\Delta\hat{v}^{CG}_{1}\right)\cdot\left|\bar{v}^{P}_{1,i}\right|+\left(\hat{v}^{P}_{2,i}\cdot\Delta\hat{v}^{CG}_{1}\right)\cdot\left|\bar{v}^{P}_{2,i}\right|\end{aligned}$$

Therefore, we have:

$$\begin{aligned}&-\left(\hat{v}^{P}_{1,i}\cdot\Delta\hat{v}^{CG}_{1}\right)\cdot\left|\bar{v}^{P}_{1,i}\right|+\left(\hat{v}^{P}_{2,i}\cdot\Delta\hat{v}^{CG}_{1}\right)\cdot\left|\bar{v}^{P}_{2,i}\right|\\ &=\frac{1}{1+\varepsilon}\cdot\left(\frac{\left|\Delta\bar{v}^{CG}_{1}\right|}{\gamma_1}+\frac{\left|\Delta\bar{v}^{CG}_{2}\right|}{\gamma_2}\right)\end{aligned}$$

Let us assume the special case that for each vehicle, the initial velocity vector at the effective point of contact, *P*, is the same as at the CG. Therefore:

$$\begin{aligned}-\left(\hat{v}^{CG}_{1,i}\cdot\Delta\hat{v}^{CG}_{1}\right)\cdot\left|\bar{v}^{CG}_{1,i}\right|+\left(\hat{v}^{CG}_{2,i}\cdot\Delta\hat{v}^{CG}_{1}\right)\cdot\left|\bar{v}^{CG}_{2,i}\right|&\\ &=\frac{1}{1+\varepsilon}\cdot\left(\frac{\left|\Delta\bar{v}^{CG}_{1}\right|}{\gamma_1}+\frac{\left|\Delta\bar{v}^{CG}_{2}\right|}{\gamma_2}\right)\end{aligned}$$

The above is in the form:

$$Ax+By=C$$

where *A*, *B*, and *C* are known:

$$A=-\hat{v}^{CG}_{1,i}\cdot\Delta\hat{v}^{CG}_{1}=\hat{v}^{CG}_{1,i}\cdot\Delta\hat{v}^{CG}_{2}$$

$$B=\hat{v}^{CG}_{2,i}\cdot\Delta\hat{v}^{CG}_{1}=-\hat{v}^{CG}_{2,i}\cdot\Delta\hat{v}^{CG}_{2}$$

$$C=\frac{1}{1+\varepsilon}\cdot\left(\frac{\left|\Delta\bar{v}^{CG}_{1}\right|}{\gamma_1}+\frac{\left|\Delta\bar{v}^{CG}_{2}\right|}{\gamma_2}\right)$$

and *x* and *y* are unknown:

$$x=\left|\bar{v}^{CG}_{1,i}\right|$$

$$y=\left|\bar{v}^{CG}_{2,i}\right|$$

Here, $x$ and $y$ are free parameters, thus causing an ambiguity in $\bar{v}^{P}_{Rel,i}$. This ambiguity can be solved if one vehicle is initially at rest; however, if both vehicles are initially in motion, the ambiguity can be solved if a departure angle is known for one of the vehicles.

From the law of sines, we know:

$$\frac{\left|\Delta\bar{v}^{CG}_{1}\right|}{\left|\sin(\theta_{1i})\right|}=\frac{\left|\bar{v}^{CG}_{1,f}\right|}{\left|\sin(\alpha_{1i})\right|}=\frac{\left|\bar{v}^{CG}_{1,i}\right|}{\left|\sin(\beta_{1i})\right|}$$

and

$$\frac{\left|\Delta\bar{v}^{CG}_{2}\right|}{\left|\sin(\theta_{2i})\right|}=\frac{\left|\bar{v}^{CG}_{2,f}\right|}{\left|\sin(\alpha_{2i})\right|}=\frac{\left|\bar{v}^{CG}_{2,i}\right|}{\left|\sin(\beta_{2i})\right|}$$

where,

$$\begin{aligned}\theta_{1i}&=\cos^{-1}\left|\hat{v}^{CG}_{1,i}\cdot\hat{v}^{CG}_{1,f}\right|\\ \alpha_{1i}&=\cos^{-1}\left|\hat{v}^{CG}_{1,i}\cdot\Delta\hat{v}^{CG}_{1}\right|\\ \beta_{1i}&=\cos^{-1}\left|\hat{v}^{CG}_{1,f}\cdot\Delta\hat{v}^{CG}_{1}\right|\end{aligned}$$

and

$$\begin{aligned}\theta_{2i}&=\cos^{-1}\left|\hat{v}^{CG}_{2,i}\cdot\hat{v}^{CG}_{2,f}\right|\\ \alpha_{2i}&=\cos^{-1}\left|\hat{v}^{CG}_{2,i}\cdot\Delta\hat{v}^{CG}_{2}\right|\\ \beta_{2i}&=\cos^{-1}\left|\hat{v}^{CG}_{2,f}\cdot\Delta\hat{v}^{CG}_{2}\right|\end{aligned}$$

Note, in the special case where, at the moment of impact, $\hat{v}^{CG}_{1,i}$ is aligned with vehicle 1's local *x* axis and $\hat{v}^{CG}_{2,i}$ is aligned with vehicle 2's local *x* axis, $\alpha_{1i}$ and $\alpha_{2i}$ are equivalent to each vehicle's principal direction of force (PDOF).

Again, we assume the vehicle orientations at impact are known. This implies that $\alpha_{1i}$ and $\alpha_{2i}$ are known. We are now ready to examine some special cases that help resolve the closing-velocity ambiguity problem.

*Known vehicle departure angle*

With known vehicle 1 departure angle, thus giving $\hat{v}^{CG}_{1,f}$, we then also have $\theta_{1i}$ and $\beta_{1i}$. This implies we can solve for both $\left|\bar{v}^{CG}_{1,f}\right|$ and $\left|\bar{v}^{CG}_{1,i}\right|$ using the law of sines above. With this, *x* is given by:

$$x=\left|\bar{v}^{CG}_{1,i}\right|=\left|\Delta\bar{v}^{CG}_{1}\right|\cdot\frac{\left|\sin(\beta_{1i})\right|}{\left|\sin(\theta_{1i})\right|}$$

Thus, we can now easily solve for $y$:

$$y=\left|\bar{v}^{CG}_{2,i}\right|=\frac{C}{B}-\frac{A}{B}x$$

If, on the other hand, the vehicle 2 departure angle is known, this implies we have $\hat{v}^{CG}_{2,f}$, and therefore $\theta_{2i}$ and $\beta_{2i}$. In this case, we know *y*:

$$y=\left|\bar{v}^{CG}_{2,i}\right|=\left|\Delta\bar{v}^{CG}_{2}\right|\cdot\frac{\left|\sin(\beta_{2i})\right|}{\left|\sin(\theta_{2i})\right|}$$

Thus, we can now easily solve for $x$:

$$x=\left|\bar{v}^{CG}_{1,i}\right|=\frac{C}{A}-\frac{B}{A}y$$

*Solving for non-zero initial speed when other vehicle is initially at rest*

If one vehicle is initially at rest, no departure angles are needed to resolve the closing-velocity ambiguity problem. Suppose we know vehicle 2 is initially at rest. With $y=0$, in this case, our above equation reduces to:

$$x=\left|\bar{v}^{CG}_{1,i}\right|=\frac{C}{A}$$

If, on the other hand, vehicle 1 is initially at rest, we have $x=0$, and therefore:

$$y=\left|\bar{v}^{CG}_{2,i}\right|=\frac{C}{B}$$

[5] Note, in this model, where $\Delta\hat{v}^{CG}_{1}$ is approximately aligned with the direction of crush, we define restitution as the point of contact negative ratio of separation-velocity to closing-velocity vector components directed along the $\Delta\hat{v}^{CG}_{1}$ axis. In cases where frictional effects are non-negligible, one may wish to use the more generally applicable formalism, where $\Delta\bar{v}^{CG}_{1}$ is represented in a normal "crush axis" and tangent "friction axis" basis, and where restitution is defined as the point of contact negative ratio of separation-velocity to closing-velocity vector components directed along the normal axis (see for example, R. Brach's planar impact mechanics). This topic will be considered in a future article.

**Appendix 2**

Example calculation using data from EDR A.

## 1998 Chevy Malibu

$W_1 := 2832\ lbf$

$m_1 := \frac{W_1}{g} = 88.021\ slug$

$OL_1 := 190\ in$

$WB_1 := 107\ in$

$I_1 := 0.1478 \cdot m_1 \cdot OL_1 \cdot WB_1 = \left(1.837 \cdot 10^3\right)\ slug \cdot ft^2$

$k_1 := \sqrt{\frac{I_1}{m_1}} = 4.568\ ft$

## 2002 Buick LeSabre

$W_2 := 3755\ lbf$

$m_2 := \frac{W_2}{g} = 116.709\ slug$

$OL_2 := 200\ in$

$WB_2 := 112\ in$

$I_2 := 0.1478 \cdot m_2 \cdot OL_2 \cdot WB_2 = \left(2.683 \cdot 10^3\right)\ slug \cdot ft^2$

$k_2 := \sqrt{\frac{I_2}{m_2}} = 4.795\ ft$

$FBtoCG_2 := 84.56\ in$

## EDR Data A

$\Delta v_{EDR1x} := 5.09\ mph$ $\quad r_{EDR1x} := -7.8\ ft$

$\Delta v_{EDR1y} := -24.19\ mph$ $\quad r_{EDR1y} := -1.5\ ft$

## Impulse Centroid

$r_{1x} := -8.2\ ft$

$r_{1y} := 1.46\ ft$

$r_{2x} := FBtoCG_2 = 7.047\ ft$

$r_{2y} := 1.3\ ft$

## Rotation Matrix

$a := 1 + \frac{r_{1y} \cdot r_{EDR1y}}{{k_1}^2} = 0.895$ $\quad c := -\frac{r_{1y} \cdot r_{EDR1x}}{{k_1}^2} = 0.546$

$b := -\frac{r_{1x} \cdot r_{EDR1y}}{{k_1}^2} = -0.589$ $\quad d := 1 + \frac{r_{1x} \cdot r_{EDR1x}}{{k_1}^2} = 4.065$

$R := \begin{bmatrix} a & b \\ c & d \end{bmatrix} = \begin{bmatrix} 0.895 & -0.589 \\ 0.546 & 4.065 \end{bmatrix}$ $\quad \|R\| = 3.960$

## Vehicle 1 Delta-V at Center-of-Gravity in Vehicle 1 Frame

$\Delta v_{CG1x} := \left(\frac{1}{\|R\|}\right) \cdot \left(d \cdot \Delta v_{EDR1x} - b \cdot \Delta v_{EDR1y}\right) = 1.624\ mph$

$\Delta v_{CG1y} := \left(\frac{1}{\|R\|}\right) \cdot \left(-c \cdot \Delta v_{EDR1x} + a \cdot \Delta v_{EDR1y}\right) = -6.169\ mph$

$\Delta v_{CG1} := \sqrt{{\Delta v_{CG1x}}^2 + {\Delta v_{CG1y}}^2} = 6.379\ mph$

$\Delta v_{CG1x_hat} := \frac{\Delta v_{CG1x}}{\Delta v_{CG1}} = 0.255$ $\quad \Delta v_{CG1y_hat} := \frac{\Delta v_{CG1y}}{\Delta v_{CG1}} = -0.967$

$\omega_1 := \frac{r_{1x} \cdot \Delta v_{CG1y} - r_{1y} \cdot \Delta v_{CG1x}}{{k_1}^2} = 3.389\ \frac{1}{s}$ $\quad \omega_1 = 194.155\ \frac{deg}{s}$

## Vehicle 2 Delta-V at Center-of-Gravity in Vehicle 1 Frame

$$\Delta v_{CG2x} := -\frac{m_1}{m_2} \cdot \Delta v_{CG1x} = -1.225\ mph$$

$$\Delta v_{CG2y} := -\frac{m_1}{m_2} \cdot \Delta v_{CG1y} = 4.652\ mph$$

$$\Delta v_{CG2} := \sqrt{{\Delta v_{CG2x}}^2 + {\Delta v_{CG2y}}^2} = 4.811\ mph$$

## Vehicle 2 Heading

$\theta_2 := -90.0\ deg$
$v_{Reli_x_hat} := -\cos\left(\theta_2\right) = -6.123 \cdot 10^{-17}$
$v_{Reli_y_hat} := -\sin\left(\theta_2\right) = 1.000$

## Rotation Matrix from Vehicle 1 frame to Vehicle 2 Frame

$M_{11} := \cos\left(\theta_2\right) = 6.123 \cdot 10^{-17}$ $\quad M_{12} := \sin\left(\theta_2\right) = -1.000$
$M_{21} := -\sin\left(\theta_2\right) = 1.000$ $\quad M_{22} := \cos\left(\theta_2\right) = 6.123 \cdot 10^{-17}$

## Vehicle 2 Delta-V at Center-of-Gravity in Vehicle 2 Frame

$$\Delta v'_{CG2x} := M_{11} \cdot \Delta v_{CG2x} + M_{12} \cdot \Delta v_{CG2y} = -4.652\ mph$$
$$\Delta v'_{CG2y} := M_{21} \cdot \Delta v_{CG2x} + M_{22} \cdot \Delta v_{CG2y} = -1.225\ mph$$

$$\Delta v'_{CG2} := \sqrt{{\Delta v'_{CG2x}}^2 + {\Delta v'_{CG2y}}^2} = 4.811\ mph$$

$$\omega_2 := \frac{r_{2x} \cdot \Delta v'_{CG2y} - r_{2y} \cdot \Delta v'_{CG2x}}{{k_2}^2} = -0.165\ \frac{1}{s}$$

$$\omega_2 = -9.447\ \frac{deg}{s}$$

## Lever-arms

$$h_1 := \frac{r_{1x} \cdot \Delta v_{CG1y} - r_{1y} \cdot \Delta v_{CG1x}}{\Delta v_{CG1}} = 7.558\ ft \qquad h_2 := \frac{r_{2x} \cdot \Delta v'_{CG2y} - r_{2y} \cdot \Delta v'_{CG2x}}{\Delta v'_{CG2}} = -0.537\ ft$$

## Gamma Factors

$$\gamma_1 := \frac{{k_1}^2}{{k_1}^2 + {h_1}^2} = 0.268 \qquad \gamma_2 := \frac{{k_2}^2}{{k_2}^2 + {h_2}^2} = 0.988$$

$$\varepsilon := 0.15$$

$$v_{c_fhat} := \frac{1}{1+\varepsilon} \cdot \left(\frac{\Delta v_{CG1}}{\gamma_1} + \frac{\Delta v'_{CG2}}{\gamma_2}\right) = 24.967\ mph$$

$$cos\varphi := \Delta v_{CG1x_hat} \cdot v_{Reli_x_hat} + \Delta v_{CG1y_hat} \cdot v_{Reli_y_hat} = -0.967$$

## Closing-speed

$$v_{c_nhat} := \frac{\left|v_{c_fhat}\right|}{\left|cos\varphi\right|} = 25.818\ mph$$